\documentclass[]{aa} 

\usepackage{epsfig}

\renewcommand{\d}{{\rm d}}
\newcommand{\pl}{\partial}
 
\newcommand{\beq}{\begin{equation}} 
\newcommand{\eeq}{\end{equation}} 
\newcommand{\beqa}{\begin{eqnarray}} 
\newcommand{\eeqa}{\end{eqnarray}} 
\newcommand{\bea}{\begin{array}} 
\newcommand{\ea}{\end{array}} 
\newcommand{\cH}{{\cal H}} 
\newcommand{\rhob}{\overline{\rho}} 
\newcommand{\lag}{\langle} 
\newcommand{\rag}{\rangle}
\newcommand{\bx}{{\bf x}}
\newcommand{\bv}{{\bf v}}
\newcommand{\bk}{{\bf k}}
\newcommand{\Om}{\Omega_{\rm m}}
\newcommand{\OL}{\Omega_{\Lambda}}
\newcommand{\cO}{{\cal O}}

\newcommand{\tKs}{{\tilde{K}_s}}

\newcommand{\bq}{{\bf q}}

\newcommand{\bu}{{\bf u}}
\newcommand{\bs}{{\bf s}}
\newcommand{\bw}{{\bf w}}
\newcommand{\xb}{\overline{x}} 
\newcommand{\cR}{{\cal R}}
\newcommand{\tcR}{\tilde{\cal R}}
\newcommand{\cS}{{\cal S}}
\newcommand{\tcS}{\tilde{\cal S}}
\newcommand{\tr}{\tilde{r}}
\newcommand{\tsig}{\tilde{\sigma}}
\newcommand{\hG}{\hat{G}}
\newcommand{\hPi}{\hat{\Pi}}
\newcommand{\hpi}{\hat{\pi}}
\newcommand{\rb}{\bar{r}}
\newcommand{\cD}{{\cal D}}
\newcommand{\hpsi}{\hat{\psi}}
\newcommand{\hdelta}{\hat{\delta}}
\newcommand{\tg}{\tilde{g}}
\newcommand{\trho}{\tilde{\rho}}

\begin{document} 
 
\title{Using the Zeldovich dynamics to test expansion schemes}    
\author{P. Valageas}   
\institute{Service de Physique Th\'eorique, CEA Saclay, 91191 Gif-sur-Yvette, 
France}  
\date{Received / Accepted } 
 
\abstract
{}
{We apply various expansion schemes that may be used to study gravitational 
clustering to the simple case of the Zeldovich dynamics.} 
{Using the well-known exact 
solution of the Zeldovich dynamics we can compare the predictions of
these various perturbative methods with the exact nonlinear result. We can also
study their convergence properties and their behavior at high orders.} 
{We find that most systematic expansions fail to recover the decay of the
response function in the highly nonlinear regime. ``Linear methods'' lead to 
increasingly fast growth in the nonlinear regime for higher orders, 
except for Pad\'{e} approximants that give a bounded response at any order. 
``Nonlinear methods'' manage to obtain some damping at one-loop order but they 
fail at higher orders.
Although it recovers the exact Gaussian damping, a resummation 
in the high-$k$ limit is not justified very well as the generation 
of nonlinear power does not originate from a finite range of wavenumbers 
(hence there is no simple separation of scales). No method is able to recover 
the relaxation of the matter power spectrum on highly nonlinear scales. It is 
possible to impose a Gaussian cutoff in a somewhat ad-hoc fashion to reproduce 
the behavior of the exact two-point functions for two different times. However, 
this cutoff is not directly related to the clustering of matter and disappears 
in exact equal-time statistics such as the matter power spectrum.
On a quantitative level, on weakly nonlinear scales, the usual perturbation theory, 
and the nonlinear scheme to which one
adds an ansatz for the response function with such a Gaussian cutoff, are the two
most efficient methods. We can expect these results to hold for the gravitational
dynamics as well (this has been explicitly checked at one-loop order), since the 
structure of the equations of motion is identical for both dynamics.}
{}


\maketitle

\section{Introduction} 
\label{Introduction}

The growth of large-scale structures in the Universe through the amplification
of small primordial fluctuations by gravitational instability is a key ingredient
of modern cosmology (Peebles 1980). This process can be used to constrain 
cosmological 
parameters through the dependence of the matter power spectrum on scale and redshift.
For instance, observations of highly nonlinear objects such as galaxy clusters
can help constrain the normalization of the matter power spectrum and the average
matter density (Oukbir \& Blanchard 1992; Younger et al. 2005). 
Although theoretical predictions are not very accurate on
these small scales, one can still derive useful constraints because they are very 
rare objects, so that their dependence on cosmological parameters is very steep.
Alternative probes, such as baryonic acoustic oscillations 
(Eisenstein et al. 1998, 2005) or weak lensing surveys (Munshi et al. 2007;
Massey et al. 2007), focus on weakly nonlinear scales where they aim at measuring 
the matter distribution through its power spectrum or its two-point correlation, 
which are sensitive to typical fluctuations. 
In such cases, one needs an accurate theoretical prediction
to derive useful constraints on cosmology. This problem is usually tackled through
N-body simulations or perturbative expansions for scales not too far from the linear
regime (Bernardeau et al. 2002).
However, numerical simulations are rather costly and analytical methods
may have the advantage of leading to a better understanding of the physics at work.
Moreover, on such large scales, a hydrodynamical description should be
sufficient (i.e. one neglects shell crossing). This facilitates analytical approaches
as one can use the continuity and Euler equations instead of the Vlasov equation.
Therefore, there has recently been renewed interest in perturbation theory
to devise analytical methods that could exhibit good accuracy on weakly nonlinear
scales to be used for such observational probes 
(Crocce \& Scoccimarro 2006a,b; Valageas 2007; McDonald 2007; 
Matarrese \& Pietroni 2007a,b).

Thus, Crocce \& Scoccimarro (2006a,b) find that one can perform a partial resummation
of the diagrammatic series that appears in the standard perturbative expansion
to obtain a response function that decays into the nonlinear regime as expected 
(whereas the standard expansion grows as a polynomial of increasing order as we 
truncate the series at higher order). 
Moreover, this result agrees well with numerical simulations
and can be used as an intermediate tool for obtaining a more accurate prediction of
the matter power spectrum than with the usual perturbation theory
(Crocce \& Scoccimarro 2007). On the other hand, Valageas (2007) present a
path-integral formalism, so that the system is fully defined by its action $S$
(or its partition function $Z$). Then, one can apply the usual tools of field 
theory, such as large-$N$ expansions (similar to a semi-classical expansion over
powers of $\hbar$ or a generalization to $N$ fields), to derive the matter
power spectrum (Valageas 2007). Note that this method also applies to the highly
nonlinear scales described by the Vlasov equation (Valageas 2004). Next,
Matarrese \& Pietroni (2007a) have recently proposed an alternative method based on 
the path-integral formalism where one considers the dependence of the system on
a large-wavenumber cutoff $\Lambda$. This gives rise to a new set of equations 
and, by taking the limit $\Lambda\rightarrow\infty$, one recovers the original 
system.
These various methods may be seen as different reorganizations of the standard
perturbative expansion. Of course, they all involve some truncation at some stage
(otherwise the problem would be solved exactly), and they are all consistent
up to that order (i.e. they only differ by higher-order terms).   

To check the range of validity of such expansions, one must compare their
predictions with N-body simulations (and assume for observational purposes that 
the accuracy remains the same for close cosmological parameters). This is not very 
convenient, since simulations themselves may be of limited accuracy. Along with
this problem, the
behaviors of other two-point functions than the power spectrum $P(k)$, such
as the response function and self-energies and especially
different-time functions such as $\lag\delta(\bx_1,t_1)\delta(\bx_2,t_2)\rag$
with $t_1\neq t_2$, have not been analyzed in detail from N-body simulations
(which do not give direct access to self-energies either). Therefore, it is
interesting to investigate these theoretical methods applied to a simpler dynamics 
that can be solved exactly. Then, one can do a detailed comparison of the 
predictions of various expansion schemes with the exact nonlinear results. 
Moreover, one can reconstruct such expansion schemes in a direct manner from the
exact two-point functions without computing high-order diagrams that involve 
many integrals, simply
by expanding back the exact nonlinear result. In this way one can more easily 
investigate the convergence properties of these expansions and the behavior of
high-order terms. A simple dynamics that can be solved exactly but that remains
close to the gravitational dynamics (at least up to weakly nonlinear scales)
is provided by the Zeldovich approximation (Zeldovich 1970; Gurbatov et al. 1989;
Shandarin \& Zeldovich 1989). The latter was originally devised as
an approximation to the gravitational dynamics. Here we take a different point of 
view as we modify the equations of motion so that the system is exactly given
by the simple Zeldovich dynamics. Then, we apply to these new equations of motion
various methods which can be applied to both dynamics (and to other stochastic
dynamics such as the Navier-Stokes equations). Taking advantage of the
exact results which can be obtained for the Zeldovich dynamics and its simpler
properties we study the accuracy and the general properties of these 
expansion schemes in detail. This should shed light on the behavior of these 
methods applied to the gravitational dynamics, because both dynamics exhibit 
similar equations of motion, and these expansions apply in identical manner to 
both systems.  

This article is organized as follows. First, in Sect.~\ref{Equations-of-motion}
we derive the equations of motion associated with the Zeldovich dynamics and their
linear solution. Next, in Sect.~\ref{Path-integral} we obtain the path-integral 
formulation of this system, starting from either the differential form or the
integral form of the equations of motion, in order to make the connection with
the different approaches used in the literature. Then, we briefly describe how 
some expansion schemes can be built from this path-integral formalism, such as 
the large-$N$ expansions in Sect.~\ref{Large-N-expansions} and the evolution 
equations with respect to a high-$k$ cutoff in Sect.~\ref{Running}. Before 
investigating such methods, we first derive the exact nonlinear two-point 
functions which can be obtained from the well-known exact solution of the Zeldovich 
dynamics in Sect.~\ref{Exact}. Then, we describe the behavior of the standard 
perturbation theory in Sect.~\ref{Standard-perturbative-expansions} and of the 
steepest-descent method (built from a large-$N$ expansion) in 
Sect.~\ref{Direct-steepest-descent}.
Next, we discuss in Sect.~\ref{High-k-limit} the high-$k$ resummation proposed
by Crocce \& Scoccimarro (2006b) to improve the behavior of such expansion
schemes. We turn to the 2PI effective action method in 
Sect.~\ref{2PI-effective-action-method2} (a second approach built from a large-$N$ 
expansion) and to simple nonlinear schemes associated with this expansion
in Sect.~\ref{nonlinear-schemes}. We investigate simple nonlinear schemes 
associated with the dependence on a high-$k$ cutoff in 
Sect.~\ref{nonlinear-schemes-2}. Finally, in Sect.~\ref{Weakly-nonlinear-scales} 
we study the quantitative predictions on weakly nonlinear scales of these methods 
at one-loop order and we conclude in Sect.~\ref{Conclusion}.

\section{Equations of motion}
\label{Equations-of-motion}

\subsection{Zeldovich approximation}
\label{Zeldovich}

On scales much larger than the Jeans length, both the cold dark matter 
and the baryons can be described as a pressureless dust. Then, we can
neglect orbit crossings and use a hydrodynamical description governed 
by the equations of motion (Peebles 1980):
\beq
\frac{\pl\delta}{\pl\tau} + \nabla.[(1+\delta) \bv] = 0,
\label{continuity1}
\eeq
\beq
\frac{\pl\bv}{\pl\tau} + \cH \bv + (\bv .\nabla) \bv = - \nabla \phi,
\label{Euler1}
\eeq
\beq
\Delta \phi = \frac{3}{2} \Om \cH^2 \delta , 
\label{Poisson1}
\eeq
where $\tau=\int \d t/a$ is the conformal time (and $a$ the scale factor), 
$\cH=\d\ln a/\d\tau$ the
conformal expansion rate, and $\Om$ the matter density cosmological parameter.
Here, $\delta$ is the matter density contrast and $\bv$ the peculiar velocity.
Since the vorticity field decays within linear theory (Peebles 1980), we take
the velocity to be a potential field so that $\bv$ is fully specified by its 
divergence $\theta$ or by its potential $\chi$ with
\beq
\theta=\nabla.\bv , \;\;\; \bv= - \nabla \chi \;\;\; \mbox{whence} \;\;\; 
\theta=-\Delta \chi .
\label{thetachi}
\eeq
In the linear regime, one finds that the linear growing mode satisfies
\beq
\theta_L = - f \cH \delta_L \;\;\; \mbox{whence} \;\;\;
\phi_L = \frac{3\Om\cH}{2f} \chi_L ,
\label{thetaLdeltaL}
\eeq
where $f(\tau)$ is defined from the linear growing rate $D_+(\tau)$ 
of the density contrast by
\beq
f=\frac{\d\ln D_+}{\d\ln a}
= \frac{1}{\cH}\frac{\d\ln D_+}{\d\tau} ,
\label{fdef}
\eeq
and $D_+(\tau)$ is the growing solution of
\beq
\frac{\d^2 D_+}{\d\tau^2}+\cH\frac{\d D_+}{\d\tau} = \frac{3}{2}\Om\cH^2 D_+ .
\label{D+}
\eeq
If we make the approximation that relation (\ref{thetaLdeltaL}) remains 
valid in the nonlinear regime, that is, we replace the Poisson equation 
(\ref{Poisson1}) by the second Eq.(\ref{thetaLdeltaL}):  
$\phi=3\Om\cH\chi/2f$, then we obtain for the Euler equation (\ref{Euler1}):
\beq
\frac{\pl\bv}{\pl\tau} + \left(1-\frac{3}{2} \frac{\Om}{f}\right) \cH \bv 
+ (\bv .\nabla) \bv = 0.
\label{Euler2}
\eeq
Obviously, as shown by Eq.(\ref{Euler2}), within this approximation the 
velocity field now evolves independently of the density field.
As is well known (Gurbatov et al. 1989), approximation (\ref{Euler2}) 
is actually identical to the Zeldovich approximation. Indeed, a change of 
variables for the velocity field yields
\beq
\frac{\pl\bu}{\pl D_+} + (\bu .\nabla) \bu = 0 \;\;\; \mbox{with} \;\;\; 
\bv= \left(\frac{\d D_+}{\d\tau}\right) \bu .
\label{Euler3}
\eeq
Equation (\ref{Euler3}) is the equation of motion of free particles, 
$\d\bu/\d D_+=0$, hence the trajectories are given by
\beq
\bx= \bq + D_+(\tau) \bs_{L0}(\bq) , \;\;\;
\bv= \frac{\d D_+}{\d\tau} \, \bs_{L0}(\bq) ,
\label{xq}
\eeq
where $\bq$ is the Lagrangian coordinate and $\bs=\bs_L=D_+ \bs_{L0}$ is the 
displacement field that is exactly given by the linear theory. 
Equation (\ref{xq}) is the usual definition of the Zeldovich approximation (i.e.
setting $\bs=\bs_L$).

Thus, the Zeldovich approximation corresponds to a change in the linear term
of the Euler equation, keeping the quadratic term and the continuity equation
unchanged. Therefore, the analysis presented in Valageas (2007) for the case
of the exact gravitational dynamics applies to the Zeldovich dynamics up to
minor modifications. First, the equations of motion (\ref{continuity1})
and (\ref{Euler2}) read in Fourier space as
\beqa
\frac{\pl\delta(\bk,\tau)}{\pl\tau} + \theta(\bk,\tau) & = & 
- \int\d\bk_1\d\bk_2 \; \delta_D(\bk_1+\bk_2-\bk) \nonumber \\
&& \times \alpha(\bk_1,\bk_2) \theta(\bk_1,\tau) \delta(\bk_2,\tau) 
\label{continuity2}
\eeqa
\beqa
\lefteqn{ \frac{\pl\theta(\bk,\tau)}{\pl\tau} + 
\left(1-\frac{3\Om}{2f}\right) \cH \theta(\bk,\tau) = } \nonumber \\
&& \!\!\!\! - \!\! \int\d\bk_1\d\bk_2 \; \delta_D(\bk_1+\bk_2-\bk) 
\beta(\bk_1,\bk_2) \theta(\bk_1,\tau) \theta(\bk_2,\tau) 
\label{Euler4}
\eeqa
where $\delta_D$ is the Dirac distribution. The coupling functions $\alpha$ 
and $\beta$ are given by
\beq
\!\! \alpha(\bk_1,\bk_2)= \frac{(\bk_1+\bk_2).\bk_1}{k_1^2} ,
\beta(\bk_1,\bk_2)= \frac{|\bk_1+\bk_2|^2 (\bk_1.\bk_2)}{2 k_1^2 k_2^2} ,
\label{alphabeta}
\eeq
and we defined the Fourier transforms as
\beq
\delta(\bk) = \int\frac{\d\bx}{(2\pi)^3} e^{-i\bk.\bx} \delta(\bx) .
\label{deltak}
\eeq
As in Crocce \& Scoccimarro (2006a,b), let us define the two-component vector
$\psi$ as
\beq
\psi(\bk,\eta) = \left(\bea{c} \psi_1(\bk,\eta) \\ \psi_2(\bk,\eta)\ea \right)
= \left( \bea{c} \delta(\bk,\eta) \\ -\theta(\bk,\eta)/f\cH 
\ea \right) ,
\label{psidef}
\eeq
where we have introduced the time coordinate $\eta$ defined from the linear 
growing rate $D_+$ of the density contrast (normalized to unity today):
\beq
\eta= \ln D_+(\tau)  \;\;\; \mbox{with} \;\;\; D_+(z=0)=1 .
\label{eta}
\eeq
Then, the equations of motion (\ref{continuity2})-(\ref{Euler4}) can be
written as
\beq
\cO(x,x').\psi(x') = K_s(x;x_1,x_2) . \psi(x_1) \psi(x_2) ,
\label{OKsdef}
\eeq
where we have introduced the coordinate $x=(\bk,\eta,i)$ where $i=1,2$ is the 
discrete index of the two-component vectors. In Eq.(\ref{OKsdef}) and in the 
following, we use the convention that repeated coordinates are integrated over
as
\beq
\cO(x,x').\psi(x') = \!\! \int\!\d\bk'\d\eta'\sum_{i'=1}^2 
\cO_{i,i'}(\bk,\eta;\bk',\eta') \psi_{i'}(\bk',\eta') .
\label{defproduct}
\eeq
The matrix $\cO$ reads
\beq
\cO(x,x') = \left( \bea{cc} \frac{\pl}{\pl\eta} & -1 \\ 0
& \; \frac{\pl}{\pl\eta} -1 \ea \right) \delta_D(\bk-\bk') \,
\delta_D(\eta-\eta')
\label{Odef}
\eeq
whereas the symmetric vertex $K_s(x;x_1,x_2)=K_s(x;x_2,x_1)$ writes as
\beqa
K_s(x;x_1,x_2) & = & \delta_D(\bk_1+\bk_2-\bk) \delta_D(\eta_1-\eta) 
\delta_D(\eta_2-\eta) \nonumber \\ && \times \gamma^s_{i;i_1,i_2}(\bk_1,\bk_2)
\label{Ksdef}
\eeqa
with
\beq
\gamma^s_{1;1,2}(\bk_1,\bk_2)= \frac{\alpha(\bk_2,\bk_1)}{2} , \;\;
\gamma^s_{1;2,1}(\bk_1,\bk_2)= \frac{\alpha(\bk_1,\bk_2)}{2} ,
\label{gamma1}
\eeq
\beq
\gamma^s_{2;2,2}(\bk_1,\bk_2)= \beta(\bk_1,\bk_2) ,
\label{gamma2}
\eeq
and zero otherwise (Crocce \& Scoccimarro 2006a). We can note that all 
the dependence on cosmology is contained in the time-redshift relation
$\eta\leftrightarrow z$. Indeed, the equation of motion (\ref{OKsdef})
written in terms of the coordinate $\eta$ no longer involves 
time-dependent factors such as $\Om/f^2$. Therefore, the evolution of
the density field only depends on cosmology through the time-coordinate 
$\eta(z)$. In this article we study the system defined by the the
equation of motion  (\ref{OKsdef}), which shows the exact solution
(\ref{xq}). This will allow us to compare various expansion methods with exact
nonlinear results.

\subsection{Linear regime}
\label{Linear}

On large scales or at early times where the density and velocity fluctuations
are small, one can linearize the equation of motion (\ref{OKsdef}), which yields
$\cO.\psi_L=0$. This gives the two linear modes:
\beq
\psi_+ = e^{\eta} \left(\bea{c} 1 \\ 1 \ea\right) , \;\;\;
\psi_-= \left(\bea{c} 1 \\ 0 \ea\right) .
\label{psilinear}
\eeq
Of course we recover the linear growing mode $\psi_+$ of the gravitational
dynamics, since approximation (\ref{thetaLdeltaL}) is valid in this case.
However, the usual decaying mode $\psi_-$ has been changed to a constant mode.
As seen in Eq.(\ref{psilinear}), it corresponds to a mere perturbation of the
density field that is transported by the unchanged velocity field.
Indeed, since the velocity field is now decoupled from the density field,
it obeys a first-order differential equation in the linear regime (rather than
a second-order differential equation), which only admits one linear mode.
As usual we define the initial conditions by the linear growing mode
$\psi_L$:
\beq
\psi_L(x) = e^{\eta} \delta_{L0}(\bk) \left(\bea{c} 1 \\ 1 \ea\right) ,
\label{psiL}
\eeq
where $\delta_{L0}(\bk)$ is the linear density contrast today at redshift
$z=0$. In this fashion the system (\ref{continuity2})-(\ref{Euler4}) that 
we study here agrees with the gravitational dynamics in the linear regime. 
Besides, from Eqs.(\ref{thetaLdeltaL}) and (\ref{xq}), we see that the 
displacement field $\bs_{L0}(\bq)$ obeys
\beq
\nabla_{\bq} . \bs_{L0} = - \delta_{L0} .
\label{s0delta}
\eeq
Moreover, we assume Gaussian homogeneous and isotropic initial conditions 
defined by the linear power spectrum $P_{L0}(k)$:
\beq
\lag \delta_{L0}(\bk_1) \delta_{L0}(\bk_2) \rag =  \delta_D(\bk_1+\bk_2)
\; P_{L0}(k_1) .
\label{PL0}
\eeq
Then, as for the gravitational dynamics studied in Valageas (2007),
the linear two-point correlation function $G_L(x_1,x_2)$ reads as:
\beqa
G_L(x_1,x_2) & = & \lag \psi_L(x_1) \psi_L(x_2) \rag \nonumber \\
& = & \delta_D(\bk_1+\bk_2) \, e^{\eta_1+\eta_2} P_{L0}(k_1) 
\left(\bea{cc} 1 & 1 \\ 1 & 1 \ea\right) .
\label{GL}
\eeqa
As in Valageas (2004), it is convenient to introduce the response function
$R(x_1,x_2)$ (related to the propagator used in Crocce \& Scoccimarro 
(2006a,b), see Sect.~\ref{intetaI} below) defined by the functional derivative
\beq
R(x_1,x_2) = \lag \frac{\delta \psi(x_1)}{\delta\zeta(x_2)} \rag_{\zeta=0} ,
\label{Rdef}
\eeq
where $\zeta(x)$ is a ``noise'' added to the r.h.s. in Eq.(\ref{OKsdef}).
Thus, $R(x_1,x_2)$ measures the linear response of the system to an external 
source of noise. Because of causality it contains an Heaviside factor
$\theta(\eta_1-\eta_2)$ since the field $\psi(x_1)$ can only depend on the
values of the ``noise'' at earlier times $\eta_2\leq\eta_1$. Moreover,
it satisfies the initial condition:
\beq
\eta_1\rightarrow\eta_2^+ : \; R(x_1,x_2) \rightarrow \delta_D(\bk_1-\bk_2) 
\delta_{i_1,i_2} .
\label{Requaltimes}
\eeq
In the linear regime, the response $R_L$ can be obtained from 
the initial condition (\ref{Requaltimes}) and the linear dynamics 
$\cO.R_L=0$ for $\eta_1>\eta_2$ (as implied by the definition (\ref{Rdef}) 
and $\cO.\psi_L =0$). This yields (Crocce et al. 2006)
\beqa
\lefteqn{R_L(x_1,x_2) = \delta_D(\bk_1-\bk_2) \, \theta(\eta_1-\eta_2) } 
\nonumber \\
&& \times \left\{ e^{\eta_1-\eta_2} \left(\bea{cc} 0 & 1 \\ 0 & 1 \ea\right)
+ \left(\bea{cc} 1 & -1 \\ 0 & 0 \ea\right)
\right\} .
\label{RL}
\eeqa
This expression holds for any cosmology, whereas for the case of the
gravitational dynamics factors, such as $\Om/f^2$, lead to a small explicit
dependence on cosmological parameters.
Note that by symmetry the two-point correlation $G$ has the form
\beq
G(x_1,x_2) = \delta_D(\bk_1+\bk_2) G_{i_1,i_2}(k_1;\eta_1,\eta_2)
\label{Gfac}
\eeq
with
\beq
G_{i_1,i_2}(k;\eta_1,\eta_2) = G_{i_2,i_1}(k;\eta_2,\eta_1)
\label{Gsym}
\eeq
whereas the response function has the form
\beq
R(x_1,x_2) = \delta_D(\bk_1-\bk_2) \, \theta(\eta_1-\eta_2) \, 
R_{i_1,i_2}(k_1;\eta_1,\eta_2) .
\label{Rsym}
\eeq
On the other hand, as noticed above, the linear two-point functions obey
\beq
\cO(x,z).G_L(z,y) = 0 , \;\;\; \cO(x,z).R_L(z,y) = \delta_D(x-y) .
\label{OGLORL}
\eeq
This can also be checked from the explicit expressions (\ref{GL}),(\ref{RL}).
Finally, it is convenient to define the power per logarithmic wavenumber 
$\Delta^2(k)$ by
\beq
\Delta^2(k) = 4\pi k^3 P(k) , \;\;
\Delta^2(k;\eta_1,\eta_2)= 4\pi k^3 G_{11}(k;\eta_1,\eta_2)
\label{Delta2def}
\eeq
where the second expression generalizes $\Delta^2(k)$ at different times.
Note that for $\eta_1\neq\eta_2$ we can have $\Delta^2(k;\eta_1,\eta_2)<0$,
whereas at equal times we have $\Delta^2(k;\eta,\eta)\geq 0$.
Then we have, for instance,
\beq
\lag\delta(\bx_1)\delta(\bx_2)\rag = \int_0^{\infty} \frac{\d k}{k} 
\Delta^2(k) \frac{\sin k|\bx_1-\bx_2|}{k|\bx_1-\bx_2|} .
\label{xir}
\eeq
Thus, for a CDM cosmology the linear power $\Delta^2_{L0}(k)$
grows as $k^4$ at low $k$ and as $\ln k$ at high $k$.

\section{Path-integral formalism}
\label{Path-integral}

\subsection{Differential form}
\label{Differential}

As in Valageas (2004, 2007) we can apply a path-integral approach to the 
hydrodynamical system described in Sect.~\ref{Zeldovich}. Let us briefly
recall how this can be done from the differential equation (\ref{OKsdef})
(see also Martin et al. 1973; Phythian 1977). In order to explicitly include
the initial conditions, we rewrite Eq.(\ref{OKsdef}) as
\beq
\cO . \psi = K_s . \psi \psi + \mu_I
\label{OKsmu}
\eeq
with $\psi=0$ for $\eta<\eta_I$ and
\beq
\! \mu_I(x) \! = \delta_D(\eta-\eta_I) e^{\eta_I} \delta_{L0}(\bk) 
\left(\bea{c} 1 \\ 1 \ea\right) = \delta_D(\eta-\eta_I) \psi_I(\xb)  ,
\label{mui}
\eeq
where we have introduced the coordinate $\xb$:
\beq
\xb=(\bk,i) \;\;\; \mbox{and} \;\;\; \psi_I(\xb) = \psi_L(\xb,\eta_I) .
\label{xb}
\eeq
Thus, the source $\mu_I$ (which formally plays the role of some external noise)
merely provides the initial conditions at time $\eta_I$, obtained from the 
linear growing mode (\ref{psiL}). We shall eventually take the limit 
$\eta_I\rightarrow -\infty$. Next, we define the generating functional $Z[j]$ by
\beq
Z[j] = \lag e^{j.\psi} \rag = \int [\d\mu_I] \;
e^{j.\psi[\mu_I]-\frac{1}{2} \mu_I . \Delta_I^{-1} . \mu_I} ,
\label{Zjmui}
\eeq
where we took the average over the Gaussian initial conditions
\beq
\lag\mu_I\rag=0, \;\;\; \lag\mu_I(x_1)\mu_I(x_2)\rag= \Delta_I(x_1,x_2) ,
\label{Deltai}
\eeq
with
\beqa
\Delta_I(x_1,x_2) & = &\delta_D(\eta_1-\eta_I) \delta_D(\eta_2-\eta_I) 
G_I(\xb_1,\xb_2) , \\
G_I(\xb_1,\xb_2) & = & G_L(\xb_1,\eta_I;\xb_2,\eta_I) .
\eeqa
All statistical properties of the field $\psi$ may be obtained from $Z[j]$.
It is convenient to write Eq.(\ref{Zjmui}) as
\beqa
Z[j] & = &\int [\d\mu_I] [\d\psi] \; |\det M| \;
\delta_D(\mu_I-\cO.\psi+K_s.\psi\psi) \nonumber \\
&& \times e^{j.\psi-\frac{1}{2} \mu_I . \Delta_I^{-1} . \mu_I} ,
\label{Zjmuipsi}
\eeqa
where the Jacobian $|\det M|$ is defined by the functional derivative 
$M=\delta\mu_I/\delta\psi$. As in Valageas (2007), a simple computation shows 
that this Jacobian is equal to
an irrelevant constant. Then, introducing an imaginary ghost field $\lambda$ 
to express the Dirac as an exponential and performing the Gaussian 
integration over $\mu_I$, we obtain
\beq
Z[j] = \int [\d\psi] [\d\lambda] \; e^{j.\psi 
+ \lambda.(-\cO.\psi+K_s.\psi\psi) + \frac{1}{2}\lambda.\Delta_I.\lambda} .
\label{Zjpsilambda}
\eeq
Thus, the statistical properties of the system (\ref{OKsdef}) are described
by the action $S[\psi,\lambda]$ defined by
\beq
S[\psi,\lambda] = \lambda.(\cO.\psi-K_s.\psi\psi) 
- \frac{1}{2}\lambda.\Delta_I.\lambda.
\label{Spsilambda}
\eeq
Moreover, we can note that adding a ``noise'' $\zeta$ to the r.h.s. of 
Eq.(\ref{OKsdef}) amounts to changing $\mu_I\rightarrow\mu_I+\zeta$, which
translates into $S\rightarrow S-\lambda.\zeta$. Therefore, functional 
derivatives with respect to $\zeta$ are equivalent to insertions of the ghost
field $\lambda$. In particular, we have
\beq
R(x_1,x_2) = \lag \psi(x_1) \lambda(x_2) \rag, \;\;\; \lag\lambda\rag=0 , 
\;\;\; \lag\lambda\lambda\rag=0 .
\label{Rpsilambda}
\eeq
The response function $R$ is also related to the correlation with the initial
conditions $\mu_I$ through
\beqa
\lefteqn{\lag\psi\mu_I\rag = \lag\psi(\cO.\psi-K_s.\psi\psi)\rag } \nonumber \\
&& = \int [\d\psi] [\d\lambda] \psi 
\left[ - \frac{\delta}{\delta\lambda} + \Delta_I.\lambda \right] 
e^{\lambda.(-\cO.\psi+K_s.\psi\psi) + \frac{1}{2}\lambda.\Delta_I.\lambda}
\nonumber \\
&& = \lag\psi (\Delta_I.\lambda)\rag = R . \Delta_I
\eeqa
since the integral of a total derivative vanishes, and we have used the symmetry 
of $\Delta_I$. This also reads as
\beq
\lag\psi(x_1)\psi_I(\xb_2)\rag = R(x_1;\xb,\eta_I) \times G_I(\xb;\xb_2) ,
\label{psipsiI0}
\eeq
where we define the cross-product $\times$ as the dot product (\ref{defproduct})
without integration over time, such as:
\beq
R\times\psi_I=\int\d\bk'\sum_{j=1}^2 
R_{ij}(\bk,\eta;\bk',\eta_I) \psi_{Ij}(\bk') .
\label{deftimeprod}
\eeq

\subsection{Integral form}
\label{Integral}

In order to make the connection with the approach developed in 
Crocce \& Scoccimarro (2006a,b), we describe here how the same path-integral
method can be applied to the equation of motion (\ref{OKsdef}) written
in integral form.

\subsubsection{Letting $\eta_I\rightarrow-\infty$}
\label{etaIinfty}

First, as in Valageas (2001) (see also Scoccimarro 2000), we can integrate the 
equation of motion (\ref{OKsdef}) as
\beq
\psi(x)=\psi_L(x)+\tKs(x;x_1,x_2) . \psi(x_1) \psi(x_2)
\label{OtKs}
\eeq
with
\beq
\cO.\tKs=K_s \;\;\; \mbox{or} \;\;\; \tKs=R_L.K_s
\label{tKs}
\eeq
as seen from Eq.(\ref{OGLORL}). Here the initial time $\eta_I$ no longer appears,
because we have already taken the limit $\eta_I\rightarrow-\infty$.
Then, following the same procedure as in Sect.~\ref{Differential} we can write
\beq
Z[j]=\lag e^{j.\psi} \rag = \int [\d\psi_L] \;
e^{j.\psi[\psi_L]-\frac{1}{2} \psi_L . G_L^{-1} . \psi_L} .
\label{ZjpsiL}
\eeq
Introducing again an imaginary field $\chi$ to impose the constraint associated
with the equation of motion (\ref{OtKs}), we finally obtain (the Jacobian is 
equal to unity):
\beq
Z[j] = \int [\d\psi] [\d\chi] \; e^{j.\psi 
+ \chi.(-\psi+\tKs.\psi\psi) + \frac{1}{2}\chi.G_L.\chi} .
\label{Zjpsichi}
\eeq
Thus, the statistical properties of the system (\ref{OtKs}) are now described
by the action $\cS[\psi,\chi]$ defined by
\beq
\cS[\psi,\chi] = \chi.(\psi-\tKs.\psi\psi) 
- \frac{1}{2}\chi.G_L.\chi.
\label{cSpsichi}
\eeq
Note that this formulation is equivalent to the one described in 
Sect.~\ref{Differential} except that we have already taken the limit 
$\eta_I\rightarrow-\infty$ directly into the equation of motion (\ref{OtKs}).
From the response field $\chi$, we can again obtain a new response function
$\cR$ associated with Eq.(\ref{OtKs}). From the comparison with 
Eq.(\ref{Zjpsilambda}), we see that we have the relations between both approaches:
\beq
\chi=\lambda.\cO , \;\;\; \cR=\lag\psi\chi\rag=R.\cO
=\lag\frac{\delta\psi}{\delta\psi_L}\rag ,
\label{cR}
\eeq
where in the last expression we recall that from Eq.(\ref{OtKs}) a variation
with respect to an external noise $\zeta$ can be seen as a variation with respect
to $\psi_L$. In the linear regime we simply have 
$\cR_L(x,y)=\delta_D(x-y)$. Moreover, in a fashion similar to Eq.(\ref{psipsiI0}) 
we have the property
\beqa
\lefteqn{\lag\psi\psi_L\rag = \lag\psi(\psi-\tKs.\psi\psi)\rag } \nonumber \\
&& = \int [\d\psi] [\d\chi] \psi \left[ - \frac{\delta}{\delta\chi} 
+ G_L.\chi \right]
e^{\chi.(-\psi+\tKs.\psi\psi) + \frac{1}{2}\chi.G_L.\chi} 
\nonumber \\
&& = \lag\psi (G_L.\chi)\rag
\eeqa
which yields the relation
\beq
\lag\psi(x_1)\psi_L(x_2)\rag = \cR(x_1,x) . G_L(x,x_2) ,
\label{psipsiL}
\eeq
where we use the symmetry of $G_L$.

\subsubsection{Integral form with finite $\eta_I$}
\label{intetaI}

Finally, as in Crocce \& Scoccimarro (2006a,b), it is possible to apply the 
initial conditions at some finite time $\eta_I$, as in Sect.~\ref{Differential}.
Thus, we may write the linear growing mode $\psi_L$ at times $\eta>\eta_I$ as
\beq
\psi_L(x)=R_L(x;\xb',\eta_I)\times\psi_I(\xb') 
\label{psiipsiL}
\eeq
where $\psi_I$ was defined in Eq.(\ref{xb}) and the cross-product $\times$
in Eq.(\ref{deftimeprod}).
Then, following the same procedure as in Eqs.(\ref{OtKs})- (\ref{Zjpsichi}),
where the Gaussian average is now taken over the field $\psi_I$ with two-point
correlation $G_I$, we now obtain the generating functional:
\beqa
\!\!\lefteqn{Z[j] \!\! = \!\!\! \int \! [\d\psi_I] [\d\psi] [\d\chi] e^{j.\psi 
+ \chi.(R_L\times\psi_I-\psi+\tKs.\psi\psi) 
- \frac{1}{2} \psi_I\!\times G_I^{-1} \!\!\times\!\psi_I} }\nonumber \\
&& = \!\! \int \! [\d\psi] [\d\chi] \; e^{j.\psi 
+ \chi.(-\psi+\tKs.\psi\psi) 
+ \frac{1}{2} \chi.(R_L \times G_I \times R_L^T).\chi} .
\label{ZjpsichiI}
\eeqa
Of course, we can check that, by taking the limit $\eta_I\rightarrow-\infty$ 
in Eq.(\ref{ZjpsichiI}), we recover Eq.(\ref{Zjpsichi}) since we have
\beq
\eta_1,\eta_2\!>\!\eta_I \! : \;
G_L(\eta_1,\eta_2)= R_L(\eta_1,\eta_I) \times G_I \times R_L^T(\eta_2,\eta_I) .
\label{GLGI}
\eeq
The system is now described by the action $\tcS[\psi,\chi]$ defined by
\beq
\tcS[\psi,\chi] = \chi.(\psi-\tKs.\psi\psi) 
- \frac{1}{2} \chi.(R_L \times G_I \times R_L^T).\chi .
\label{tcSpsichi}
\eeq
Next, we can define a response function with respect to the initial conditions by
\beq
\tcR(x_1,\xb_2) = \lag\frac{\delta\psi(x_1)}{\delta\psi_I(\xb_2)}\rag 
= \lag\psi(x_1) \chi(x) . R_L(x;\xb_2,\eta_I)\rag.
\label{tcR}
\eeq
From the comparison of (\ref{tcSpsichi}) with (\ref{Spsilambda}), we obtain 
$\chi=\lambda.\cO$ and
\beq
\tcR(x_1,\xb_2)= R.\cO.R_L=R(x_1;\xb_2,\eta_I) .
\label{tcRR}
\eeq
Thus, the response $\tcR(x_1,\xb_2)$, which is called the ``propagator 
$G_{i_1 i_2}(k_1,\eta_1)\delta_D(\bk_1-\bk_2)$'' in Crocce \& Scoccimarro (2006a,b)
is equal to the response function $R$ of Sect.~\ref{Differential} restricted
to time $\eta_2=\eta_I$, without taking the limit $\eta_I\rightarrow-\infty$.
Finally, we can note that from Eq.(\ref{psipsiI0}) we have
\beq
\lag\psi(x_1)\psi_I(\xb_2)\rag = \tcR(x_1;\xb) \times G_I(\xb,\xb_2) .
\label{psipsiI}
\eeq
This relation was 
obtained in Crocce \& Scoccimarro (2006b) from a diagrammatic approach.
Thus, we see that the three approaches (\ref{Spsilambda}), (\ref{cSpsichi}),
and (\ref{tcSpsichi}) are closely related. In the integral method we simply
absorb the matrix $\cO$ into the response field $\chi$. Next, we can
either take the limit $\eta_I\rightarrow-\infty$ from the start, as for
the action $\cS$, or keep a finite $\eta_I$ in the computation, as for $\tcS$. 
Then, we can take $\eta_I\rightarrow-\infty$ in the final results for the
nonlinear two-point correlation.
Note, however, that for the approach of Crocce \& Scoccimarro (2006a,b),
which corresponds to the action $\tcS$, it is not possible to take this limit
in a practical manner, since one needs to keep track of the response
$\tcR$, which has no finite limit for $\eta_I\rightarrow-\infty$. This leads
to somewhat more complicated expressions than for the approaches based on the
actions $S$ and $\cS$ of Eqs.(\ref{Spsilambda}),(\ref{cSpsichi}), where the 
response functions $R$ and $\cR$ remain well-defined for 
$\eta_I\rightarrow-\infty$.
Of course, the analysis described above also applies to the case of the
gravitational dynamics.

\section{Large-$N$ expansions}
\label{Large-N-expansions}

The path integrals (\ref{Zjpsilambda}), (\ref{Zjpsichi}), and (\ref{ZjpsichiI}) 
can be computed by expanding
over powers of the non-Gaussian part (i.e. over powers of $K_s$ or $\tKs$). 
This actually gives the usual perturbative expansion over powers of the
linear power spectrum $P_L$ (see also Valageas (2001, 2004) for the
case of the Vlasov equation of motion). 
On the other hand, these path integrals may also be studied
through large-$N$ expansions as in Valageas (2004). We focus below on the 
differential form (\ref{Zjpsilambda}), but the formalism also applies to the
integral forms (\ref{Zjpsichi}) and (\ref{ZjpsichiI}).
Thus, one considers the generating functional $Z_N[j,h]$ defined by
\beq
Z_N[j,h] = \int [\d\psi] [\d\lambda] \; e^{N[j.\psi+h.\lambda-S[\psi,\lambda]]} ,
\label{ZN}
\eeq
and one looks for an expansion over powers of $1/N$, taking eventually $N=1$
into the results.
As discussed in Valageas (2004), the large-$N$ expansions may also be derived
from a generalization of the dynamics to $N$ fields $\psi^{(\alpha)}$. 
This yields the same results once we deal with the long-wavelength 
divergences that constrain which subsets of diagrams need to be gathered.
 
The interest of such large-$N$ expansions is to provide new systematic
expansion schemes that may show improved convergence properties as compared
with the standard perturbation theory. Besides, it is clear from Eq.(\ref{ZN})
that the symmetries of the system (e.g. invariance through translations)
are automatically conserved at any order. These methods have been applied to 
many fields of theoretical physics, such as quantum field theory 
(e.g. Zinn-Justin 1989; 
Berges 2002), statistical physics (e.g. study of growing interfaces described
by the Kardar-Parisi-Zhang equation, Doherty et al. 1994), and turbulence 
(e.g. Mou \& Weichman 1993). They are closely related
at lowest order to the so-called ``mode-coupling approximations'' used for 
critical dynamics, liquids, or glassy systems (Bouchaud et al. 1996), and to the 
``direct interaction approximation'' used for turbulent flows (Kraichnan 1961).
Therefore, it is natural to investigate their application to the
cosmological gravitational dynamics described by 
Eqs.(\ref{continuity1})-(\ref{Poisson1}), which are similar to the Navier-Stokes
equations. In some cases (e.g. Berges 2002), it has been found that, whereas the
simplest perturbative expansions give rise to secular terms
(which grow as powers of time), the 2PI effective action method derived from
such a large-$N$ method (discussed below in Sect.~\ref{2PI-effective-action})
could achieve a non-secular behavior and display relaxation processes.
Of course, the actual behavior of such schemes depends on the specific problem.
It has already been shown in Valageas (2007) that, for the case of the gravitational
dynamics in the expanding Universe, the large-$N$ expansions indeed show
a qualitative improvement over standard perturbation theory at one-loop order,
as they display bounded oscillations (for the steepest-descent method) 
or decaying oscillations (for the 2PI effective action method) for the response 
functions instead of the secular terms encountered in the standard perturbative 
expansion (which gives increasingly large powers of time $D(\tau)^p$ at higher 
orders). In this article, we investigate whether this good behavior extends to 
higher orders in the case of the Zeldovich dynamics.

We discuss below both ``linear schemes'', such as the standard perturbation
theory or the steepest-descent method of Sect.~\ref{Steepest-descent}, which
involve expansions over linear two-point functions, and ``nonlinear schemes'', 
such as the 2PI effective action method of Sect.~\ref{2PI-effective-action},
which involve expansions over nonlinear two-point functions themselves.
By expanding different intermediate quantities or different equations (derived from
the same equation of motion), one obtains different methods
that also correspond to different partial resummations.

\subsection{Steepest-descent method}
\label{Steepest-descent}

A first approach to handle the large-$N$ limit of 
Eq.(\ref{ZN}) is to use a steepest-descent method (also called a 
semi-classical or loopwise expansion in the case of usual quantum field 
theory with $\hbar=1/N$). For auxiliary correlation and response
functions $G_0$ and $R_0$, this yields the equations (Valageas 2004)
\beqa
\cO(x,z).G_0(z,y) &=& 0 \label{G0eq} \\
\cO(x,z).R_0(z,y) &=& \delta_D(x-y) \label{R0forward} \\
R_0(x,z).\cO(z,y) &=& \delta_D(x-y) , \label{R0eq} 
\eeqa
whereas the actual correlation and response functions obey
\beqa
\cO(x,z).G(z,y) &=& \Sigma(x,z).G(z,y) + \Pi(x,z).R^T(z,y) \label{Geq} \\
\cO(x,z).R(z,y) &=& \delta_D(x-y) + \Sigma(x,z).R(z,y) \label{Rforward} \\
R(x,z).\cO(z,y) &=& \delta_D(x-y) + R(x,z).\Sigma(z,y) \label{Req}  
\eeqa
where the self-energy terms $\Sigma$ and $\Pi$ are given at one-loop order by
\beqa
\!\!\!\Sigma(x,y) \!\!\! & = & \!\!\! 4 K_s(x;x_1,x_2) K_s(z;y,z_2) 
R_0(x_1,z) G_0(x_2,z_2) \label{Seq} \\
\!\!\!\Pi(x,y) \!\!\! & = & \!\!\! 2 K_s(x;x_1,x_2) K_s(y;y_1,y_2) 
G_0(x_1,y_1) G_0(x_2,y_2) .\label{Peq} 
\eeqa
Note that Eqs.(\ref{G0eq})-(\ref{Req}) are exact and that the expansion over 
powers of $1/N$ only enters the expression of the self-energy 
(\ref{Seq})-(\ref{Peq}). Here we only kept the lowest-order terms 
(see Valageas 2004 for the next-order terms). We also took the limit 
$\eta_I\rightarrow -\infty$ so that terms involving $\Delta_I$ vanish.
The comparison of Eqs.(\ref{G0eq})-(\ref{R0forward}) with Eqs.(\ref{OGLORL}) 
shows that the auxiliary matrices $G_0$ and $R_0$ are actually equal to their 
linear counterparts:
\beq
G_0= G_L, \;\;\; R_0 = R_L .
\label{G0GLR0RL}
\eeq
Next, substituting $G_0$ and $R_0$ into Eqs.(\ref{Seq})-(\ref{Peq}), we obtain
the self-energies at one-loop order. First, we can note that $\Sigma$ has the 
same form (\ref{Rsym}) as the response $R$, whereas $\Pi$ is symmetric and
has the same form (\ref{Gfac}) as the two-point correlation $G$. Then, a simple
calculation gives
\beqa
\Sigma_0(x_1,x_2) & = & - \omega_1^2 \theta(\eta_1-\eta_2) \delta_D(\bk_1-\bk_2) 
\nonumber \\
&& \times \left[ e^{2\eta_1} \left(\bea{cc} 0 & 1 \\ 0 & 1 \ea\right) 
+ e^{\eta_1+\eta_2} \left(\bea{cc} 1 & -1 \\ 0 & 0 \ea\right) \right],
\label{S0}
\eeqa
where we define $\omega_1=\omega(k_1)$ as
\beq
\omega(k) = k \sigma_v \;\; \mbox{with} \;\; 
\sigma_v^2 = \frac{1}{3} \lag s_{L0}^2 \rag 
= \frac{4\pi}{3} \int_0^{\infty} \d w \, P_{L0}(w) .
\label{sigvdef}
\eeq
Here $\sigma_v^2$ is the variance of the one-dimensional displacement 
field $\bs_{L0}$ (or of the one-dimensional velocity dispersion up to a normalization 
factor). On the other hand, $\Pi$ is given at one-loop order by
\beq
\Pi_0(x_1,x_2) = \delta_D(\bk_1+\bk_2) e^{2\eta_1+2\eta_2} \Pi_0(k_1) ,
\label{Pi0}
\eeq
with
\beqa
\Pi_0(k) & = & 2 \int\d\bk_1\d\bk_2 \delta_D(\bk_1+\bk_2-\bk) P_{L0}(k_1) P_{L0}(k_2)
\nonumber \\
&& \times \left(\bea{cc} \pi_1^2 & \pi_1 \pi_2 \\ \pi_1 \pi_2 & \pi_2^2 \ea\right)
\label{Pi0k}
\eeqa
and
\beq
\pi_1 = \frac{\alpha(\bk_1,\bk_2)+\alpha(\bk_2,\bk_1)}{2} , \;\;\;
\pi_2 = \beta(\bk_1,\bk_2) .
\eeq
Then, the response $R$ and the correlation $G$ can be obtained by integrating
Eqs.(\ref{Geq})-(\ref{Rforward}).

\subsection{The 2PI effective action method}
\label{2PI-effective-action}

As described in Valageas (2004), a second approach is to first introduce the
double Legendre transform $\Gamma[\psi,G]$ of the functional $W=\ln Z$ (with
respect to both the field $\psi$ and its two-point correlation $G$) and next 
to apply the $1/N$ expansion to $\Gamma$. 
This ``2PI effective action''method yields the
same equations (\ref{Geq})-(\ref{Req}), and the self-energy shows the same
structure at one-loop order as (\ref{Seq})-(\ref{Peq}) where $G_0$ and $R_0$ are 
replaced by $G$ and $R$:
\beqa
\lefteqn{\!\Sigma(x,y) \! = \! 4 K_s(x;x_1,x_2) K_s(z;y,z_2) R(x_1,z) G(x_2,z_2)} 
\label{S2PI} \\
&& \!\!\!\!\!\!\!\Pi(x,y) \! = \! 2 K_s(x;x_1,x_2) K_s(y;y_1,y_2) G(x_1,y_1) 
G(x_2,y_2) .
\label{P2PI} 
\eeqa
Thus, the direct steepest-descent method yields a series of
linear equations that can be solved directly, whereas the 2PI effective 
action method gives a system of nonlinear equations (through the dependence
on $G$ and $R$ of $\Sigma$ and $\Pi$) that usually must be solved numerically 
by an iterative scheme. However, thanks to the Heaviside factors appearing
in the response $R$ and the self-energy $\Sigma$, these equations can be solved
directly by integrating forward over the time $\eta_1$.

\subsection{Role of self-energy terms}
\label{Role-of-self-energy-terms}

From Eq.(\ref{Rforward}) we can see that the self-energy $\Sigma$ plays the
role of a damping term. Indeed, Eq.(\ref{Rforward}) has the form 
$\pl R/\pl\eta_1=\Sigma.R$ so that large negative values of $\Sigma$ are associated
with a strong damping of the response function (Exact details are somewhat more
intricate since Eq.(\ref{Rforward}) is actually an integro-differential equation.)
This agrees with Eq.(\ref{S0}) which shows that the one-loop self-energy $\Sigma_0$
becomes large and negative at high $k$ as $\Sigma_0 \propto -k^2$.
Thus, the self-energy $\Sigma$ encodes the loss of memory associated with the 
nonlinear dynamics. 

On the other hand, we can see from Eq.(\ref{Geq}) that the self-energy $\Pi$
is associated with the continuous production of power due to nonlinear
mode couplings. Indeed, we can see from Eqs.(\ref{Geq})-(\ref{Req}) that the 
correlation $G$ can also be written in terms of the response $R$ as
\beq
G(x_1,x_2) = R \times G_I \times R^T + R.\Pi.R^T ,
\label{GPi}
\eeq
and we let $\eta_I\rightarrow -\infty$. 
The physical meaning of Eq.(\ref{GPi}) is clear. The first term on the right hand 
side 
means that the fluctuations at the initial time $\eta_I$ are merely 
transported forward in time through the response $R$. This is the only nonzero 
term in the linear regime (with $R=R_L$ hence $G=G_L$). The effect of the 
nonlinear dynamics is to modify the transport matrix $R$ and to add a second 
term to the right hand side of Eq.(\ref{GPi}). 
The latter has the meaning of a source term that produces fluctuations with
two-point correlation $\Pi(\eta_1',\eta_2')$ at the times $(\eta_1',\eta_2')$ 
that are next transported 
forward to later times $(\eta_1,\eta_2)$ by the matrices $R(\eta_1,\eta_1')$ 
and $R^T(\eta_2',\eta_2)$.

\section{Running with a high-$k$ cutoff}
\label{Running}

\begin{figure}[htb]
\begin{center}
\epsfxsize=5.3 cm \epsfysize=1.7 cm {\epsfbox{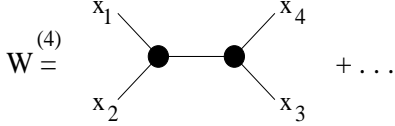}}
\end{center}
\caption{The first diagram of the expansion of the fourth derivative $W^{(4)}$
as in Eq.(\ref{W41}). The big dots are the vertices $K_s$ and the lines are
the two-point functions $R$ or $G$.}
\label{figdiag3}
\end{figure}

In a recent paper, Matarrese \& Pietroni (2007a) introduce another approach to
studying the gravitational dynamics within the hydrodynamical framework. It is
also based on a path-integral formulation. Although they use the integral form
of the equations of motion, as in Sect.~\ref{Integral}, we briefly describe in 
this section how this method may be applied to the path integral 
(\ref{Zjpsilambda}). First, from Eq.(\ref{Zjpsilambda}) we define the generating
functional $Z[j,h]$ as
\beq
Z[j,h] = \int [\d\psi] [\d\lambda] \; e^{j.\psi+h.\lambda - S[\psi,\lambda]} ,
\label{Zjh}
\eeq
where we have introduced the external source $h$. This allows us to obtain the
correlations of the response field $\lambda$ through derivatives with respect 
to $h$. Next, following Matarrese \& Pietroni (2007a), we add a high-$k$ cutoff
$\Lambda$ to the linear power spectrum $P_{L0}(k)$ by changing the kernel
$\Delta_I$, which appears in the action $S$ of Eq.(\ref{Spsilambda}) as
\beq
\Delta_I \rightarrow \Delta_{\Lambda} = \theta(\Lambda-k_1) \Delta_I(x_1,x_2) .
\label{DeltaLambda}
\eeq
Thus, the Heaviside factor $\theta(\Lambda-k_1)$ removes the linear power
at high wavenumbers $k_1>\Lambda$, and we recover the full system in the limit
$\Lambda\rightarrow\infty$. Then, the idea proposed in Matarrese \& Pietroni (2007a)
is to study the evolution of the system as a function of the cutoff $\Lambda$.
Therefore, one first looks for equations that describe the dependence on 
$\Lambda$. Second, one derives some approximation for these equations, for
instance by a truncation of some expansion, and finally solves these
approximate equations from $\Lambda=0$ up to $\Lambda=\infty$.
First, the dependence on $\Lambda$ may obviously be described through the
derivative of $Z$ with respect to $\Lambda$, which reads
\beq
\!\!\! \frac{\pl Z}{\pl\Lambda} \!= \frac{e^{2\eta_I} \! P_{L0}(\Lambda)}{2} \!\!\! 
\int\!\!\d\bk 
\delta_D(\Lambda-k) \!\! \sum_{i,j} \! \frac{\delta^2 Z}{\delta h_i(\bk,\eta_I) 
\delta h_j(-\bk,\eta_I)} .
\label{dZdLambda}
\eeq
Next, introducing the generating functional $W$ of the connected correlation
functions,
\beq
W=\ln Z , \;\;\; R(x_1,x_2) = \left. \frac{\delta^2 W}{\delta j(x_1)\delta h(x_2)} 
\right|_{j=h=0} ,
\label{Wdef}
\eeq
we obtain from Eq.(\ref{dZdLambda}) the evolution of the response $R$ with
the cutoff $\Lambda$ as
\beqa
\lefteqn{\frac{\pl R}{\pl\Lambda}(x_1,x_2) = \frac{e^{2\eta_I}P_{L0}(\Lambda)}{2} 
\int\d\bw \, \delta_D(\Lambda-w) } \nonumber \\
&& \;\;\;\;\;\;\;\;\; \times \sum_{i,j} \frac{\delta^4 W}{\delta h_i(\bw,\eta_I) 
\delta h_j(-\bw,\eta_I) \delta j(x_1)\delta h(x_2)} .
\label{dRdLambda}
\eeqa
Here we use the property (\ref{Rpsilambda}): 
$\lag\lambda\rag=\lag\lambda\lambda\rag=0$.
Next, to make some progress, one needs to obtain an expression for
the fourth-derivative $W^{(4)}$. Of course, in generic cases this quantity
is not known exactly and one must introduce some approximations. The usual
procedure is to write a diagrammatic expansion for $W$, using the path integral
expression (\ref{Zjh}), and to truncate at some finite order.
For the cubic action (\ref{Spsilambda}), the lowest-order contribution is associated
with the diagram of Fig.~\ref{figdiag3}, which gives

\beqa
\frac{\delta^4 W}{\delta j_1\delta h_2\delta h_3\delta h_4} 
& = & \lag \psi_1\lambda_2\lambda_3\lambda_4\rag_c \nonumber \\
& = & 12 R R ( K_s R  K_s ) R R  + ..
\label{W41}
\eeqa
Using expression (\ref{Ksdef}) of the vertex $K_s$, this gives:
\beqa
\lefteqn{ \frac{\pl R_{i_1i_2}}{\pl\Lambda}(k;\eta_1,\eta_2) \! = 4 e^{2\eta_I} 
\! P_{L0}(\Lambda) \!\!\int\!\!\d\bw \, \delta_D(\Lambda-w) \! 
\int_{\eta_2}^{\eta_1}\!\!\!\!\d\eta \!\int_{\eta_2}^{\eta}\!\!\!\d\eta'} 
\nonumber \\
&& \!\!\!\!\!\times R_{i_1i_1'}(k;\eta_1,\eta) R_{i_2'i_2}(k;\eta',\eta_2) 
R_{i_3'i_3}(w;\eta,\eta_I) R_{i_4'i_4}(w;\eta',\eta_I) \nonumber \\
&& \!\!\!\!\!\times R_{ij}(\bk-\bw;\eta,\eta') 
\gamma^s_{i_1';i_3'i}(\bw,\bk-\bw) \gamma^s_{j;i_2'i_4'}(\bk,-\bw) .
\label{dRdLambda1}
\eeqa
Following Matarrese \& Pietroni (2007a,b), we note that at lowest-order we may 
replace the response functions on the right hand side in Eq.(\ref{dRdLambda1}) 
by the linear 
responses, which do not depend on $\Lambda$ since they do not depend on the 
linear power spectrum, see Eq.(\ref{RL}). Then, in the limit 
$\eta_I\rightarrow-\infty$, only the linear growing modes of $R_{i_3'i_3}$ and
$R_{i_4'i_4}$ give a non-vanishing contribution:
\beqa
\lefteqn{ \frac{\pl R_{i_1i_2}}{\pl\Lambda} \! = 4 P_{L0}(\Lambda) \int\d\bw \, 
\delta_D(\Lambda-w) \int_{\eta_2}^{\eta_1}\d\eta \int_{\eta_2}^{\eta}\d\eta'
e^{\eta+\eta'} } \nonumber \\
&& \times R_{Li_1i_1'}(\eta_1,\eta) R_{Li_2'i_2}(\eta',\eta_2) 
R_{Lij}(\eta,\eta') \nonumber \\
&& \times \gamma^s_{i_1';i_3'i}(\bw,\bk-\bw) \gamma^s_{j;i_2'i_4'}(\bk,-\bw) .
\label{dRdLambdaRL0}
\eeqa
Using Eqs.(\ref{gamma1})-(\ref{gamma2}) for the vertices $\gamma^s$ 
we obtain:
\beq
\frac{\pl R}{\pl\Lambda} = - \frac{4\pi}{3} k^2 P_{L0}(\Lambda) 
\frac{(D_1-D_2)^2}{2} R_L .
\label{dRdLambdaRL}
\eeq
Starting from the initial condition $R(\Lambda=0)=R_L$, this yields at 
$\Lambda=\infty$,
\beq
R= R_L \left[ 1 -  \frac{\omega^2(D_1-D_2)^2}{2} \right] ,
\label{RRL1}
\eeq
which agrees with the usual perturbative result at order $P_{L0}$ 
(see Eq.(\ref{Rp1}) below).
The running with $\Lambda$ of the two-point correlation $G$ (whence of
the nonlinear power spectrum) is then obtained by taking the derivative with respect
to $\Lambda$ of Eq.(\ref{GPi}) and again using a loopwise expansion for the
self-energy $\Pi$. In practice, Matarrese \& Pietroni (2007a,b) use some ansatz
for $G$ to derive a linear equation that can be solved up to $\Lambda=\infty$.
Although they refer to this approach as a renormalization group method, we can note 
that it is somewhat different from the usual renormalization group techniques.
Although considering the evolution with a cutoff $\Lambda$, one does
not look for a fixed point of renormalization group equations that would govern 
the properties of the system in some large-scale limit.

Matarrese \& Pietroni (2007a,b) notice that, if we promote the linear response $R_L$ 
on the right hand side of Eq.(\ref{dRdLambdaRL}) to the nonlinear response $R$, 
we obtain the response (\ref{RNL}) with a Gaussian decay at high $k$ as the 
solution of this 
linear equation. Note that all the previous steps apply identically 
to the case of the gravitational dynamics, where this procedure again leads to 
the response (\ref{RNL}). In this case, this expression is no longer exact, but 
it does show the expected damping into the nonlinear regime. 
This remark suggests that this procedure may provide a very efficient expansion
scheme for the response function. However, this is somewhat artificial.
Indeed, to derive Eq.(\ref{dRdLambdaRL}) from Eq.(\ref{dRdLambda1}),
one makes use of the properties of the linear response to simplify the right 
hand side, 
so that substituting back the nonlinear response $R$ is somewhat ad-hoc (although
it is correct at lowest-order it makes the procedure not systematic).
Moreover, it is clear that one can apply the same procedure to any other scheme 
that gives an equation of the form $\pl R/\pl\alpha=F[P_{L0},R,G]$ where 
$\alpha$ can be any variable among $\{k,\eta_1,\eta_2,\Lambda,..\}$. 
(For the large-$N$ 
expansions of Sect.~\ref{Large-N-expansions}, it would be $\eta_1$.) 
Indeed, at lowest order one can always simplify $F$ as a linear functional of $R_L$
such that one obtains the exact response (\ref{RNL}) by substituting 
$R_L\rightarrow R$, 
since the right hand side must be consistent at lowest-order with 
$\pl R/\pl\alpha$ evaluated for the exact response (\ref{RNL}). Thus, the latter 
satisfies the equation
\beq
\frac{\pl R}{\pl\alpha} = \left[ \frac{\pl \ln R_L}{\pl\alpha} 
- \frac{1}{2} \frac{\pl}{\pl\alpha} (D_1-D_2)^2\omega^2 \right] R ,
\label{dRdalpha}
\eeq
which implies that at lowest order one can always write 
$\pl R/\pl\alpha=F[P_{L0},R,G]$ as
\beq
\left[\frac{\pl}{\pl\alpha} - \frac{\pl \ln R_L}{\pl\alpha} \right] R = 
- \frac{1}{2} \left(\frac{\pl}{\pl\alpha} (D_1-D_2)^2\omega^2\right) R_L + ...
\label{dRdalphaRL}
\eeq
where the dots stand for higher-order terms over $\{(D_1-D_2),\omega\}$. 
For the specific case $\alpha=\Lambda$, Eq.(\ref{dRdalphaRL}) leads back to 
Eq.(\ref{dRdLambdaRL}). For the large-$N$ expansion schemes, Eq.(\ref{dRdalphaRL})
would be Eq.(\ref{Rforward}) with $\alpha\rightarrow\eta_1$, the left hand side 
corresponding to $\cO.R$ and the right hand side to $\Sigma.R$ at lowest order.
Then, substituting $R$ to $R_L$ on the right hand side of 
Eq.(\ref{dRdalphaRL}), one recovers Eq.(\ref{dRdalpha}) and the nonlinear 
response (\ref{RNL}) with the Gaussian cutoff. This reasoning also applies to
the gravitational dynamics, to which one adds high-$k$ approximations so that
the response (\ref{RNL}) (or a variant) still applies; see also 
Sect.~\ref{High-k-limit} below. Therefore, recovering the response (\ref{RNL})
in this manner does not imply that the underlying expansion scheme is very efficient.
However, Matarrese \& Pietroni (2007b) argue that for the running with
$\Lambda$ it is possible to derive a stronger justification of this procedure
which still applies at higher orders. Then, for cases where additional arguments can
be obtained, such techniques based on the evolution of the system with respect to some
parameter $\alpha$ may provide useful alternative expansion schemes. 
We discuss this method, based on the dependence of the system on a high-$k$ 
cutoff $\Lambda$, in Sect.~\ref{nonlinear-schemes-2} below
within the framework of a simple systematic expansion, and we find that it
actually gives similar results to the 2PI effective action method.

\section{Exact two-point functions}
\label{Exact}

For the Zeldovich dynamics, all quantities of interest can be computed exactly,
since we know the solution (\ref{xq}) of the equations of motion.
This makes the Zeldovich dynamics an interesting test of approximation schemes,
since we can compare their predictions with the exact results. As the equations of
motion in the form (\ref{continuity2})-(\ref{Euler4}) are very similar to those 
associated with the exact gravitational dynamics, we can expect that the behavior
of various approximation schemes will be similar for both dynamics.
Therefore, we compute the exact two-point functions associated
with the Zeldovich dynamics in this section.

\subsection{Two-point correlation}
\label{Two-point-correlation}

As is well known, the two-point correlation $G$ for the Zeldovich dynamics can be
computed exactly from the solution (\ref{xq}) of the equations of motion
(e.g. Schneider \& Bartelmann 1995; Taylor \& Hamilton 1996).
Indeed, starting from the uniform density $\rhob$ at $t\rightarrow 0$, 
the conservation of matter gives, before orbit-crossing,
\beq
\rho(\bx) \d\bx = \rhob \d\bq \;\;\; \mbox{whence} \;\;\; 
1+\delta(\bx) = \left|\det\left(\frac{\pl\bx}{\pl\bq}\right)\right|^{-1} .
\label{rhox}
\eeq
This also reads from Eq.(\ref{xq}) as
\beq
\delta(\bx,\eta) = \int\d\bq \; \delta_{D}[\bx-\bq-\bs(\bq,\eta)] - 1 ,
\label{deltax}
\eeq
where $\bs(\bq,\eta)= D_+(\eta) \bs_{L0}(\bq)$ is the displacement field.
Note that this expression remains valid after shell crossing: all particles
of Lagrangian coordinate $\bq$ that happen to be at location $\bx$ at the time
of interest contribute to the right hand side. In Fourier space we obtain, 
for $k\neq 0$,
\beq
\delta(\bk) = \int\frac{\d\bq}{(2\pi)^3} \; e^{-i\bk.(\bq+\bs)} .
\label{deltakq}
\eeq
Therefore, the density-density two-point correlation reads as
\beqa
\lefteqn{\delta_D(\bk_1+\bk_2) G_{11}(k_1;\eta_1,\eta_2) = } \nonumber \\
&&  \int\frac{\d\bq_1\d\bq_2}{(2\pi)^6} \, e^{-i(\bk_1.\bq_1+\bk_2.\bq_2)}
\lag e^{-i(\bk_1.\bs_1+\bk_2.\bs_2)}\rag .
\label{G11Z1}
\eeqa
Since the displacement field $\bs_{L0}$ given by Eq.(\ref{s0delta}) is Gaussian,
the average in Eq.(\ref{G11Z1}) reads as
\beq
\lag e^{-i(\bk_1.\bs_1+\bk_2.\bs_2)}\rag = 
e^{-\frac{1}{2} \lag (k_{1i} s_{1i}+ k_{2i} s_{2i})
(k_{1j} s_{1j}+ k_{2j} s_{2j}) \rag} ,
\label{G11Z2}
\eeq
where we sum over the 3D components $i,j=1,2,3$. Let us define the displacement 
correlation $\Psi_0$:
\beq
\Psi_{0;ij}(\bq_1,\bq_2) = \lag s_{L0;i}(\bq_1) s_{L0;j}(\bq_2) \rag .
\label{Psi0def}
\eeq
Thanks to statistical homogeneity, it obeys
\beq
\Psi_{0;ij}(\bq_1,\bq_2) = \Psi_{0;ij}(\bq_1-\bq_2) ,
\eeq
and from Eq.(\ref{s0delta}) it is given by
\beq
\Psi_{0;ij}(\bq) = \int\d\bk \; e^{i\bk.\bq} \; \frac{k_i k_j}{k^4} \; P_{L0}(k) .
\label{Psi0}
\eeq
Then, Eq.(\ref{G11Z1}) writes as
\beqa
\lefteqn{ G_{11}(k;\eta_1,\eta_2) = \int\frac{\d\bq}{(2\pi)^3} \; e^{-i\bk.\bq} }
\nonumber \\
&& \times \; e^{e^{\eta_1+\eta_2} k_i k_j [\Psi_{0;ij}(\bq) - \cosh(\eta_1-\eta_2) 
\Psi_{0;ij}(0)]} .
\label{G11Z3}
\eeqa
Using Eq.(\ref{Psi0}) this can also be written as
\beqa
\lefteqn{ G_{11}(k;\eta_1,\eta_2) = \int\frac{\d\bq}{(2\pi)^3} \; e^{-i\bk.\bq} }
\nonumber \\
&& \times \; e^{e^{\eta_1+\eta_2} \int\d\bw \, \frac{(\bk.\bw)^2}{w^4} P_{L0}(w) 
[\cos(\bw.\bq)-\cosh(\eta_1-\eta_2)]} .
\label{G11Z4}
\eeqa
The integration over angles in the exponent of Eq.(\ref{G11Z4}) can be performed
analytically. Thus, let us define the quantity $I(\bq;\bk)$ by
\beq
I(\bq;\bk) = k_i k_j \Psi_{0;ij}(\bq) =  \int\d\bw \,  e^{i \bw.\bq} \, 
\frac{(\bk.\bw)^2}{w^4} \, P_{L0}(w) .
\label{Idef}
\eeq
Then, by expanding the exponential over spherical harmonics, we obtain
\beq
I(\bq;\bk) = k^2 I_0(q) + k^2 (1-3\mu^2) I_2(q) , \;\;\;\; 
\mu=\frac{\bk.\bq}{kq} , 
\label{Iq}
\eeq
where we introduce
\beq
I_n(q) = \frac{4\pi}{3} \int_0^{\infty} \d w \, P_{L0}(w) j_n(wq) ,
\label{Indef}
\eeq
and $j_n$ is the spherical Bessel function of order $n$. Note that the
variance $\sigma_v^2$ of the one-dimensional displacement field, defined in
Eq.(\ref{sigvdef}), also satisfies $\sigma_v^2=I_0(0)$. Then, Eq.(\ref{G11Z4}) 
reads (using the linear growth factor $D=e^{\eta}$ as the time-coordinate) as
\beqa
G_{11}(k;D_1,D_2) & = & e^{-\frac{D_1^2+D_2^2}{2} \, k^2\sigma_v^2} 
\int\frac{\d\bq}{(2\pi)^3} \; \cos(kq\mu)  \nonumber \\
&& \times \; e^{D_1D_2k^2[I_0+(1-3\mu^2) I_2]} .
\label{G11Z5}
\eeqa
Following Schneider \& Bartelmann (1995), we can perform the integration over
angles by expanding the exponential and using the property
\beq
\int_0^1\d\mu \, \cos(kq\mu) (1-\mu^2)^n = n! \left(\frac{2}{kq}\right)^n 
j_n(kq) .
\eeq
This gives
\beqa
G_{11}(k;D_1,D_2) & \!\! = \!\! & e^{-\frac{D_1^2+D_2^2}{2} \, k^2\sigma_v^2} 
\int_0^{\infty}\frac{\d q \, q^2}{2\pi^2} \; e^{D_1D_2k^2(I_0-2I_2)} \nonumber \\
&& \times \; \sum_{n=0}^{\infty} \left( D_1D_2 \frac{6 k^2 I_2}{kq}\right)^n
j_n(kq) .
\label{G11Z6}
\eeqa

\subsection{Asymptotic behavior}
\label{Asymptotic-behavior}

From the explicit expression (\ref{G11Z5}), we can obtain the asymptotic behavior
of the two-point correlation function in the highly nonlinear regime. 
Thus, we can formally write for a power-law linear power spectrum
\beq
G_{11}= e^{[D_1D_2-\frac{D_1^2+D_2^2}{2}]k^2\sigma_v^2} \; \frac{1}{4\pi k^3} \;
F\left[ \Delta_L^2(k;D_1,D_2) \right] 
\label{G11Z7}
\eeq
with
\beq
F(x) \! = \!\!\int\!\frac{\d\bq}{2\pi^2} \cos(q\mu) 
e^{\frac{x}{3} \int\!\d w w^n [j_0(qw)-1+(1-3\mu^2)j_2(qw)]} 
\label{F1}
\eeq
where we make the change of variables $q\rightarrow q/k, w\rightarrow kw$, for
\beq
P_{L0}(k) = \frac{1}{4\pi k_0^3} \left(\frac{k}{k_0}\right)^n ,
\label{P0kn}
\eeq
whence
\beq
\Delta_L^2(k;D_1,D_2) =  D_1D_2 \left(\frac{k}{k_0}\right)^{n+3} .
\label{D0kn}
\eeq
First, we note that infrared (IR) divergences appear in the one-dimensional 
velocity dispersion $\sigma_v^2$ (defined in Eq.(\ref{sigvdef})) for $n\leq-1$ 
at low $k$, and in the integral over $w$ in Eq.(\ref{F1}) for $n\leq-3$. 
As is well-known, the IR divergence at $n\leq-1$ should
disappear for equal-time statistics (Vishniac 1983; Jain \& Bertschinger 1996)
because of Galilean invariance. This is explicitly checked in Eq.(\ref{G11Z7})
since for $D_1=D_2$ the prefactor of $k^2\sigma^2$ vanishes so that the
contribution associated with $\sigma_v$ cancels out. Thus the equal-time 
nonlinear power spectrum is well-defined for $n>-3$.
Second, we can see that both $\sigma_v^2$ and the integral over $w$ in 
Eq.(\ref{F1}) diverge if $n\geq-1$ at high $k$. Thus, this UV divergence 
remains untamed in the full non-perturbative result (\ref{G11Z7}).
This is a qualitative difference with the true gravitational dynamics where 
such UV divergences are expected to disappear in the exact nonlinear 
power spectrum for $-3<n<1$. However, this may require going beyond the 
single-stream approximation.
Therefore, in the following we assume $-3<n<-1$. 
After performing the integral over $w$ and making a change of variable, we obtain
\beqa
\lefteqn{F(x) = x^{\frac{3}{n+1}} \frac{2}{\pi}\int_0^{\infty} \d q \, q^2 
\int_0^1 \d\mu \cos\left(x^{\frac{1}{n+1}} q \mu\right) } \nonumber \\
&& \!\!\!\!\! \times \!\exp\!\left[ -q^{-n-1} \frac{\pi^{1/2}2^{n-1}\Gamma[(n+3)/2]}
{(-n-1)\Gamma[(4-n)/2]} [1-(n+1)\mu^2] \right] ,
\label{F2}
\eeqa
which shows that $F(x)$ is well-defined for $-3<n<-1$ and obeys the asymptotic
behavior
\beq
F(x) \sim x^{\frac{3}{n+1}} \;\;\; \mbox{for} \;\;\; x\gg 1.
\label{F3}
\eeq
Thus, the equal-time power $\Delta^2(k;D)$ decreases in the highly 
nonlinear regime as
\beq
\Delta^2(k;D) \sim \Delta^2_L(k;D)^{\frac{3}{n+1}}  \;\;\; \mbox{for} 
\;\;\; \Delta^2_L\gg 1.
\label{Delta2asymp}
\eeq
Therefore, if $P_{L0}(k) \sim k^n$ at high $k$ the nonlinear power decreases
as a power law $P(k) \sim k^{-3+3(n+3)/(n+1)}$ in the highly nonlinear regime.

\subsection{Response function}
\label{Response-function}

Using the exact solution (\ref{xq}) we can also compute the exact response 
function $R$. First, we note that, since the velocity field is decoupled from
the density field, we have the simple exact result:
\beq
R_{21}= 0 .
\label{R21}
\eeq
Of course, the linear solution (\ref{RL}) is consistent with Eq.(\ref{R21}).
Next, we can compute the response $R_{12}$ of the density to a velocity perturbation
as follows. At time $\eta_2^-$, before the velocity perturbation localized at time
$\eta_2$, the location and velocity of the particle of Lagrangian coordinate 
$\bq$ are from Eq.(\ref{xq}),
\beq
\bx_2^-= \bq+D_2\bs_{L0}(\bq) , \;\;\; \bv_2^-= \dot{D}_2 \bs_{L0}(\bq) ,
\label{x2m}
\eeq
whereas at time $\eta_2^+$, after the velocity perturbation $\zeta_2(\bx)$, we have:
\beq
\bx_2^+=\bx_2^- , \;\;\; \bv_2^+=\bv_2^- - f_2\cH_2 \nabla_{\bx_2}^{-1}.\zeta_2
\label{x2p}
\eeq
where we have used the definition of $\psi_2$ in Eq.(\ref{psidef}). Therefore, the 
location of the particle at time $\eta_1>\eta_2$ is
\beq
\bx_1= \bq+D_1\bs_{L0}(\bq) - \frac{D_1-D_2}{D_2} \nabla_{\bx_2}^{-1}.\zeta_2 .
\label{x1}
\eeq
The density contrast is again given by expressions of the form 
(\ref{rhox})-(\ref{deltakq}) so that $\psi_1(\bk_1,\eta_1)=\delta(\bk_1,\eta_1)$
reads for $k_1\neq 0$ as
\beq
\psi_1(\bk_1,\eta_1)= \int\frac{\d\bq}{(2\pi)^3} \, 
e^{-i\bk_1.[\bq+D_1\bs_{L0}-\frac{D_1-D_2}{D_2} \nabla_{\bx_2}^{-1}.\zeta_2]} .
\label{psi1zeta2}
\eeq
Definition (\ref{Rdef}) of the response function reads here as
\beq
R_{12}(\bk_1,\eta_1;\bk_2,\eta_2) = 
\lag \left. \frac{\delta\psi_1(\bk_1)}{\delta\zeta_2(\bk_2)}\right|_{\zeta_2=0} 
\rag .
\eeq
Then, using the expression
\beq
-\nabla_{\bx_2}^{-1}.\zeta_2 = \int\d\bk \, e^{i\bk.\bx_2} \, i \frac{\bk}{k^2} ,
\, \zeta_2(\bk) 
\eeq
we obtain from Eq.(\ref{psi1zeta2})
\beqa
R_{12} & = & \frac{D_1-D_2}{D_2} \, \frac{\bk_1.\bk_2}{k_2^2} 
\int\frac{\d\bq}{(2\pi)^3} \, e^{i(\bk_2-\bk_1).\bq} \nonumber \\
&& \times \; \lag e^{-i(D_1\bk_1-D_2\bk_2).\bs_{L0}(\bq)} \rag .
\label{R12_1}
\eeqa
Because of homogeneity, the average in Eq.(\ref{R12_1}) does not depend on $\bq$
so that the integral over $\bq$ yields a Dirac factor $\delta_D(\bk_1-\bk_2)$.
On the other hand, since $\bs_{L0}$ is Gaussian the average can be easily performed
as in Sect.~\ref{Two-point-correlation}, which gives for $D_1>D_2$:
\beqa
R_{12}(k;D_1,D_2) & = & \frac{D_1-D_2}{D_2} \, 
e^{-\frac{1}{2} (D_1-D_2)^2 k^2 \sigma_v^2} \\
& = & R_{L12} \, e^{-\frac{1}{2} (D_1-D_2)^2 \omega^2} ,
\label{R12_2}
\eeqa
where $R_{L12}$ is the linear response from Eq.(\ref{RL}) and we have introduced
$\omega(k)=k\sigma_v$ as in Eq.(\ref{sigvdef}).
The computation of $R_{11}$ proceeds along the same lines. A perturbation 
$\zeta_1(\bx_2)$ of the density field at time $\eta_2$ does not modify the
velocity field, and we obtain, for $k_1\neq 0$,
\beqa
\lefteqn{ \psi_1(\bk_1,\eta_1) = \int\frac{\d\bx_2}{(2\pi)^3} \left[ 1 
+ \delta(\bx_2,\eta_2^+)\right] e^{-i\bk_1.\bx_1} } \nonumber \\
&& \!\!\!\!\!\! = \!\! \int\!\! \frac{\d\bx_2}{(2\pi)^3} e^{-i\bk_1.\bx_1} \!\!
\left[ \!\int\!\!\d\bq \delta_D[\bx_2\!-\!\bq\!-\!D_2\bs_{L0}(\bq)] 
+\zeta_1(\bx_2) \right]
\label{psi1zeta1}
\eeqa
where we have used Eq.(\ref{deltax}) and $\bx_1, \bx_2$ are the locations at times
$\eta_1,\eta_2$ of the particle of Lagrangian coordinate $\bq$. This gives
\beq
R_{11} = \lag \det\left(\frac{\pl\bx_2}{\pl\bq}\right) 
e^{-i (D_1-D_2) \bk.\bs_{L0}(\bq)} \rag .
\label{R11_1}
\eeq
Expanding the determinant, we find that most terms cancel out, and we obtain the
simple result:
\beq
R_{11} = R_{L11} \, e^{-\frac{1}{2} (D_1-D_2)^2 \omega^2} .
\label{R11_2}
\eeq
In a similar fashion, for $R_{22}$ we can write:
\beqa
\lefteqn{ \psi_2(\bk_1,\eta_1) = \int\frac{\d\bq}{(2\pi)^3} 
\det\left(\frac{\pl\bx_1}{\pl\bq}\right) e^{-i\bk_1.\bx_1} \int\d\bk 
\frac{\bk_1.\bk}{k^2} } \nonumber \\
&& \times \left[ D_1 e^{i\bk.\bq} \delta_{L0}(\bk) + \frac{D_1}{D_2} 
e^{i\bk.\bx_2}\zeta_2(\bk) \right]
\label{psi2zeta2}
\eeqa
The computation is slightly more intricate than for $R_{11}$, since $\bx_1$ 
also depends on $\zeta_2$; however, most terms cancel out and
we recover again the same form as in Eqs.(\ref{R12_2}), (\ref{R11_2}).
Thus, the exact nonlinear response function is merely given by
\beq
R(k;D_1,D_2) = R_L \, e^{-\frac{1}{2} (D_1-D_2)^2 \omega^2} ,
\label{RNL}
\eeq
that is, all linear components are multiplied by the same damping factor.
We can see from Eq.(\ref{RNL}) that the response function only depends on
the linear power spectrum through the linear velocity dispersion $\sigma_v^2$,
and on scale through $\omega^2=k^2\sigma_v^2$. This property extends
to the self-energy $\Sigma$ which is related to $R$ through Eq.(\ref{Rforward}).
This is a big simplification with respect to the gravitational dynamics, where 
the response $R$ and the self-energy $\Sigma$ depend on the detailed shape of
$P_{L0}(k)$, see Valageas (2007). However, even in that case it appears that
the behavior of the response function is mostly governed by 
$\omega^2=k^2\sigma_v^2$; see for instance the analysis in Sect.~5.2 of 
Valageas (2007). This also shows that both dynamics share important features.

\subsection{Damping self-energy $\Sigma$}
\label{Self-energy-Sigma}

The self-energy $\Sigma$ introduced in Sect.~\ref{Large-N-expansions} is usually
obtained as a series of diagrams from the path integral (\ref{ZN}). However,
since we know the exact response function $R$, we can directly compute $\Sigma$
from Eq.(\ref{Rforward}). From the simple result (\ref{RNL}), we can see that the
matrix structure of $R$, hence of $\Sigma$, is not changed by the nonlinear
corrections. Therefore, from Eq.(\ref{S0}) and Eq.(\ref{RNL}), we write the exact 
self-energy $\Sigma$ as
\beq
\Sigma(k;D_1,D_2)= \Sigma_0 \, \sigma[\omega(D_1-D_2)] ,
\label{SNL}
\eeq
where the matrix $\Sigma_0$ was obtained in Eq.(\ref{S0}). To generalize the 
calculation for future use, we consider a response of the form
\beq
R(k;D_1,D_2) \! = R_L \, r(t), \;\;\;  t=\omega(D_1-D_2) , \;\;\; r(0)= 1 ,
\label{RLr}
\eeq
where the constraint $r(0)= 1$ comes from Eq.(\ref{Requaltimes}).
Substituting Eqs.(\ref{SNL})-(\ref{RLr}) into Eq.(\ref{Rforward}) yields
\beq
r'(t)= - \int_0^t \d t' \; \sigma(t-t') r(t') .
\label{rsig2}
\eeq
The fact that the system (\ref{Rforward}) can be reduced to Eq.(\ref{rsig2})
shows that the scalings (\ref{SNL})-(\ref{RLr}) are indeed self-consistent.
Note that the functions $r(t)$ and $\sigma(t)$ are defined for $t\geq 0$,
because of the Heaviside factors $\theta(\eta_1-\eta_2)$ in $R$ and $\Sigma$. Then,
introducing the Laplace transform
\beq
\tr(s)= \int_0^{\infty} \d t \; e^{-s t} r(t) ,
\label{Laplace}
\eeq
we obtain from Eq.(\ref{rsig2})
\beq
s\tr(s) - 1 = - \tsig(s) \tr(s) .
\label{trtsig}
\eeq
For the exact nonlinear response (\ref{RNL}), we have:
\beq
r(t)=e^{-t^2/2} , \;\;\; \tr(s)= \sqrt{\frac{\pi}{2}} \, e^{s^2/2} 
\mbox{erfc}\left(\frac{s}{\sqrt{2}}\right) ,
\label{rtrsNL}
\eeq
which gives
\beq
\tsig(s)=  \sqrt{\frac{2}{\pi}} \frac{e^{-s^2/2}}{\mbox{erfc}(s/\sqrt{2})} - s ,
\label{tsigNL}
\eeq
where $\mbox{erfc}(x)$ is the complementary error function:
\beq
\mbox{erfc}(x) = \frac{2}{\sqrt{\pi}} \int_x^{\infty} \d t \; e^{-t^2} .
\eeq

\begin{figure}[htb]
\begin{center}
\epsfxsize=4.34 cm \epsfysize=4.34 cm {\epsfbox{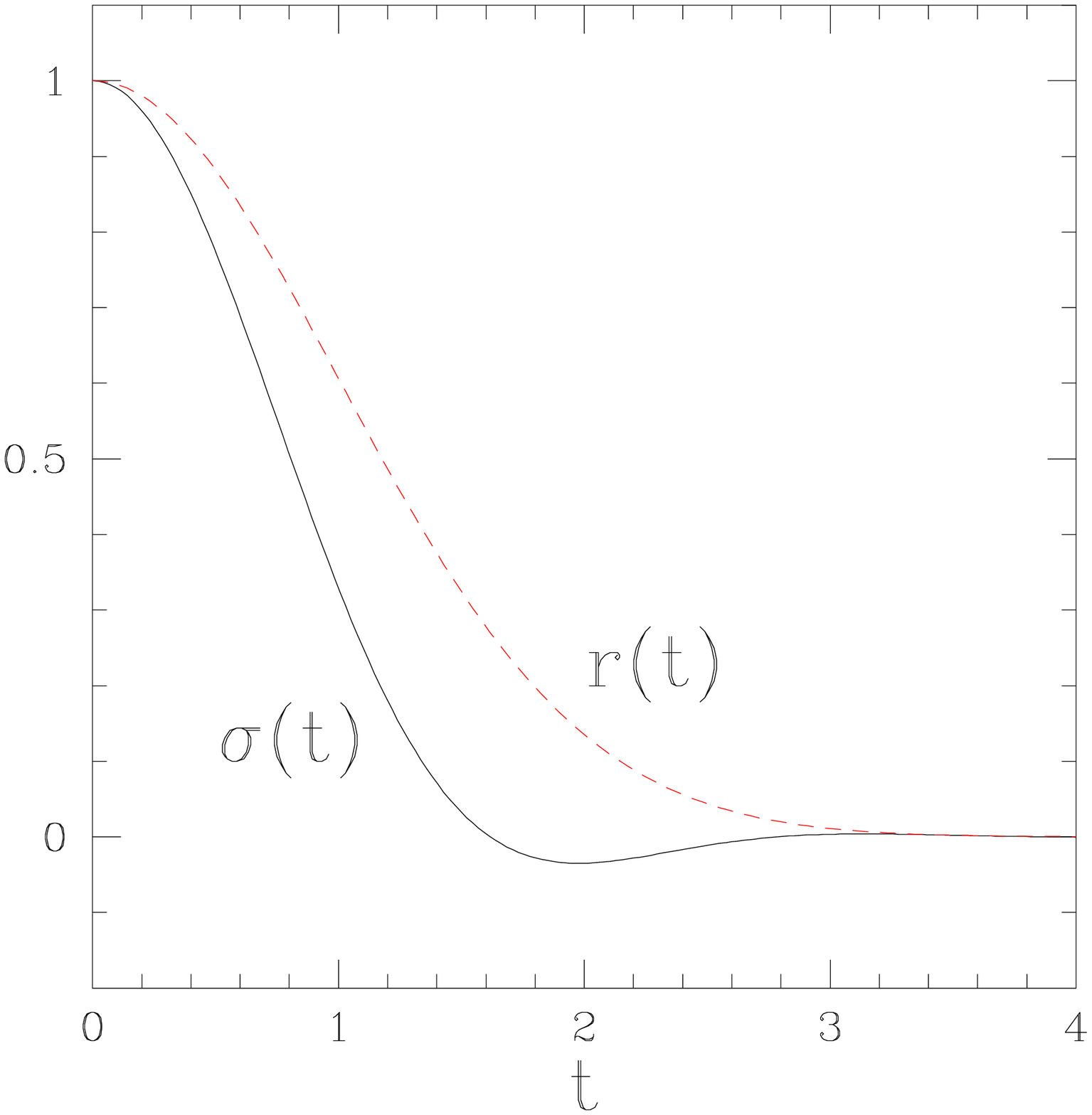}}
\epsfxsize=4.34 cm \epsfysize=4.34 cm {\epsfbox{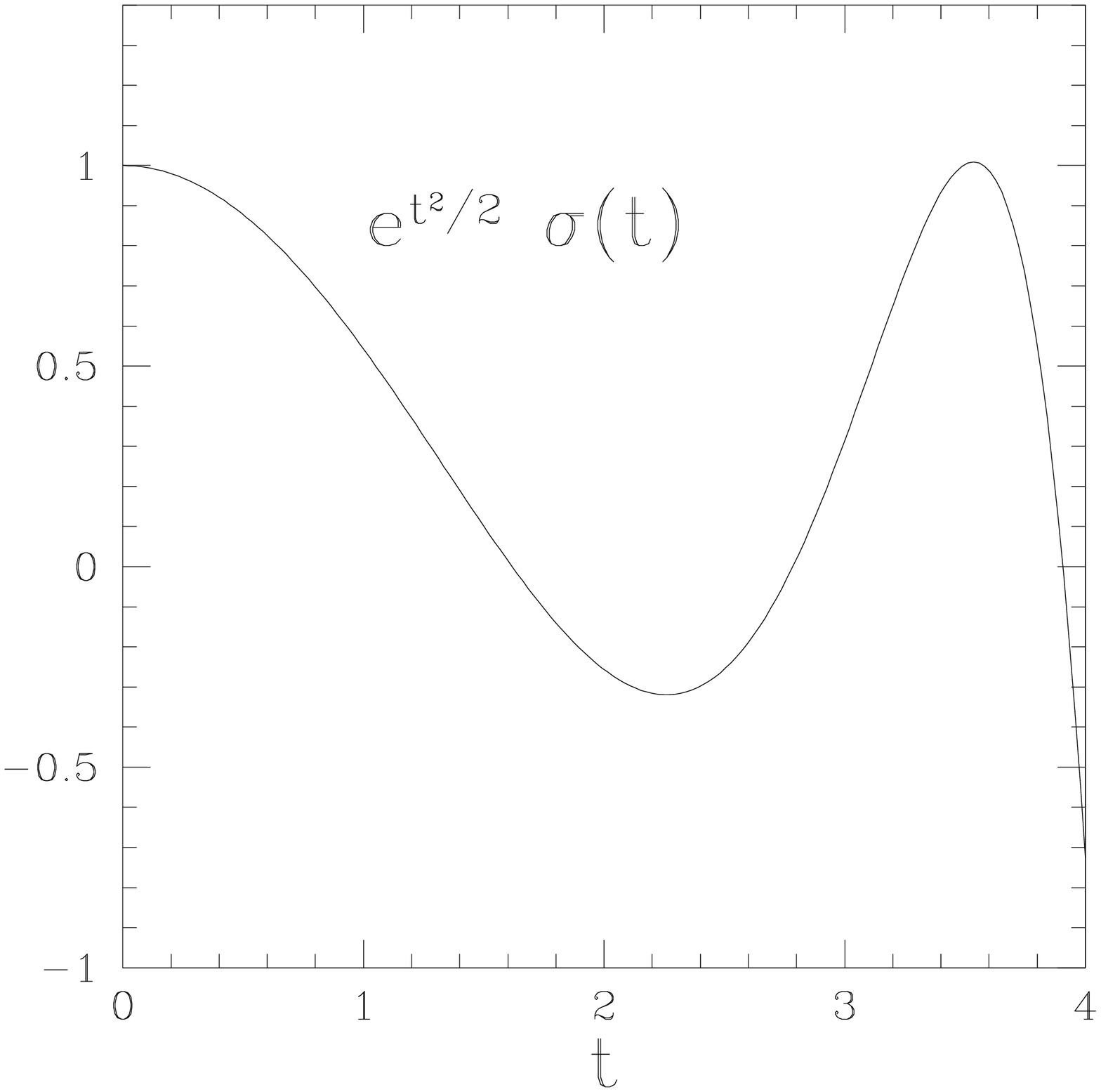}}
\end{center}
\caption{{\it Left panel}: the functions $r(t)$ and $\sigma(t)$.
{\it Right panel}: the function $\sigma(t)$ multiplied by a factor $e^{t^2/2}$.}
\label{figrsigt}
\end{figure}

On the other hand, if we write the expansion of $\sigma(t)$ around $t=0$ as
\beq
t\geq 0 : \;\; \sigma(t) = \sum_{p=0}^{\infty} (-1)^p \sigma_p \frac{t^{2p}}{(2p)!} ,
\label{sigtexp}
\eeq
we obtain by substituting into Eq.(\ref{rsig2})
\beq
\sigma_p= (2p+1)!! - \sum_{m=1}^p \sigma_{p-m} (2m-1)!! ,
\label{sigmap}
\eeq
and the first few coefficients are
\beq
\sigma_0=1, \; \sigma_1=2, \; \sigma_2=10, \; \sigma_3=74, \; \sigma_4=706,...
\label{sigmap04}
\eeq 
We note that this series also appears in other problems of field theory
as the number of Feynman diagrams associated for instance to a cubic complex action
with two fields (Cvitanovic et al. 1978). This is not surprising since in our case 
we also have a cubic action (\ref{Spsilambda}) with two fields $\psi,\lambda$.
We show in Fig.~\ref{figrsigt} the behavior of the self-energy function $\sigma(t)$
computed from the expansion (\ref{sigtexp}). We can see that it shows a fast decay,
together with oscillations (but the numerical range is too small to check whether
the asymptotic is of the form $e^{-t^2/2}\cos(t)$). Thus, the self-energy exhibits
a more intricate behavior than the response $r(t)$. This may explain why 
path-integral
methods based on the Schwinger-Dyson equation (\ref{Rforward}) have difficulty
reproducing the response $R$ from simple approximations to $\Sigma$, as we shall
see below.

\subsection{Self-energy $\Pi$}
\label{Self-energy-Pi}

We can also compute the self-energy $\Pi$ from Eq.(\ref{GPi}), which gives
\beq
\Pi(x_1,x_2)= (\cO-\Sigma).G.(\cO-\Sigma)^T
\label{PiG}
\eeq
using Eqs.(\ref{Rforward})-(\ref{Req}). Then, using the exact expressions
of the two-point correlation $G$ and of the self-energy $\Sigma$,
we can obtain $\Pi$. However, using the exact two-point correlation $G_{ij}$
would give intricate expressions, as the velocity correlations involve a few
prefactors up to order $P_{L0}^8$ in front of the exponential of Eq.(\ref{G11Z4}).
Here we are mostly interested in the qualitative behavior of the self-energy $\Pi$,
therefore, we use the simple approximation
\beq
G \simeq \hG \;\;\; \mbox{with} \;\;\;
\hG \equiv G_{11} \left(\bea{cc} 1 & 1 \\ 1 & 1 \ea\right) ,
\label{hG}
\eeq
where $G_{11}$ is the exact density-density correlation derived in 
Sect.~\ref{Two-point-correlation}. The simple form (\ref{hG}) is also consistent
with the linear regime limit (\ref{GL}). Then, expanding the exponential of 
Eq.(\ref{G11Z5}), we can write
\beqa
\hG(k;D_1,D_2) & = & \! \int \! \frac{\d\bq}{(2\pi)^3} \; e^{-i\bk.\bq} 
\sum_{n=1}^{\infty} \frac{1}{n!} I(\bq;\bk)^n \nonumber \\
&& \times \hG_n(k,D_1) \hG_n(k,D_2)^T
\label{hGhGn}
\eeqa
where $I(\bq;\bk)$ is defined in Eq.(\ref{Iq}), and we introduce the vectors 
$\hG_n$ defined by
\beq
\hG_n(k,D) = e^{-\omega^2 D^2/2} D^n \left(\bea{c} 1 \\ 1 \ea\right) .
\label{hGn}
\eeq
In Eq.(\ref{hGhGn}) we use the fact that the term $n=0$ does not contribute to
$\hG(k)$. Then, the self-energy $\hPi$ associated with $\hG$
through Eq.(\ref{PiG}) reads as
\beqa
\hPi(k;D_1,D_2) & = & \! \int \! \frac{\d\bq}{(2\pi)^3} \; e^{-i\bk.\bq} 
\sum_{n=1}^{\infty} \frac{1}{n!} I(\bq;\bk)^n \nonumber \\
&& \times \hPi_n(k,D_1) \hPi_n(k,D_2)^T
\label{hPhPn}
\eeqa
with
\beq
\hPi_n(k,D) = (\cO-\Sigma).\hG_n = \hpi_n(k,D) \left(\bea{c} 1 \\ 1 \ea\right) .
\label{hPihG}
\eeq
Using Eq.(\ref{SNL}) we obtain
\beqa
\lefteqn{ \hpi_n(k,D) = (n-1-\omega^2 D^2) D^n e^{-\omega^2 D^2/2} } \nonumber \\
&& + \omega^2 D^2 \int_0^D \d D' \sigma[\omega(D-D')] D'^{n-1} e^{-\omega^2 D'^2/2} .
\label{hpin}
\eeqa
Then, using Eq.(\ref{rsig2}) 
we can check that $\hpi_1=0$. Moreover, if we apply
the response $R$ to $\hPi$ as in Eq.(\ref{GPi}), we obtain obviously from 
Eq.(\ref{hPihG}) and Eq.(\ref{Req})
\beqa
R.\hPi.R^T & = & \! \int \! \frac{\d\bq}{(2\pi)^3} \; e^{-i\bk.\bq} 
\sum_{n=2}^{\infty} \frac{1}{n!} I(\bq;\bk)^n \nonumber \\
&& \times \hG_n(k,D_1) \hG_n(k,D_2)^T ,
\label{RPiR}
\eeqa
and we recover the full two-point correlation $\hG$ by noticing that the term $n=1$
in Eq.(\ref{hGhGn}) is equal to $R\times G_I \times R^T$ 
(with $\eta_I\rightarrow-\infty$) in agreement with Eq.(\ref{GPi}).

\section{Standard perturbative expansions}
\label{Standard-perturbative-expansions}

\begin{figure}[htb]
\begin{center}
\epsfxsize=8 cm \epsfysize=7 cm {\epsfbox{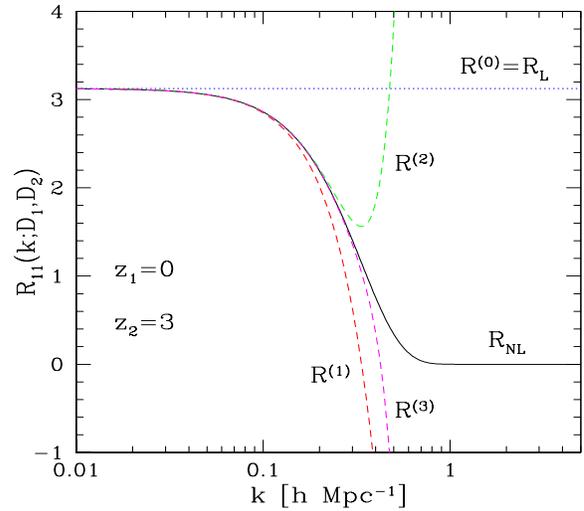}}
\end{center}
\caption{The standard perturbative expansion of the response function over powers
of $P_{L0}$ as in Eq.(\ref{Rp1}). We only show the density-density component $R_{11}$
for clarity. The solid curve $R_{\rm{NL}}$ is the exact response (\ref{RNL}), whereas
curves labeled $R^{(p)}$ are the expansion of the response function up to order $p$
over $P_{L0}$. Here we consider a $\Lambda$CDM Universe with ``$n=-2$'' normalized
as in Eq.(\ref{k0}), but the results are identical for any CDM cosmology up to a 
rescaling of $k$.}
\label{figR_P0}
\end{figure}

In this section, we describe the usual perturbative expansion over powers of
the linear power spectrum $P_{L0}$ applied to the Zeldovich dynamics.
In this article we are considering a $\Lambda$CDM universe with
$\Om=0.3, \OL=0.7, \Omega_b=0.046, \sigma_8=0.9$, a reduced Hubble constant 
$h=0.7$, and we use the linear power spectrum given by the CAMB Boltzmann code 
(Lewis et al. 2000). This gives at $z=0$ for a smooth linear power spectrum
taken from Eisenstein \& Hu (1998):
\beq
\Delta^2_{L0}(k_0)=1 , \;\;\; \omega(k_0)=1.3, \;\;\; \mbox{for} \;\;\;  
k_0=0.21 h \mbox{Mpc}^{-1} .
\label{k0}
\eeq
However, until Sect.~\ref{Weakly-nonlinear-scales} where we focus on weakly 
nonlinear scales and baryonic acoustic oscillations, we use a 
power-law linear power spectrum (\ref{P0kn}) with $n=-2$ (except in 
Sect.~\ref{2PI-effective-action-method2}), normalized as in Eq.(\ref{k0}):
\beq
n=-2 : \;\;\; \Delta^2_{L0}(k) = k/k_0 ,
\label{n2}
\eeq
where $k_0$ is given in Eq.(\ref{k0}). This is not important for the
response function $R$, which only depends on $\omega(k)=k\sigma_v$, whatever the
linear power spectrum, but this will allow us to simplify the analysis for
the nonlinear two-point correlation $G$.
First, we consider the response function $R$. Expanding Eq.(\ref{RNL}) over powers
of $P_{L0}$ is equivalent to expanding over powers of $\omega^2$, since 
$\omega^2\propto P_{L0}$ from Eq.(\ref{sigvdef}). Therefore, the response function
$R^{(p)}$ expanded up to order $P_{L0}^p$ is:
\beq
R^{(p)}(k;D_1,D_2)= R_L \sum_{m=0}^p \frac{(-1)^m}{m!} 
\left[ \frac{\omega^{2}}{2} (D_1-D_2)^2 \right]^m .
\label{Rp1}
\eeq
This expansion converges absolutely at all times and on all scales, but the 
convergence
is not uniform. Thus, the convergence rate is very slow for $\omega(D_1-D_2)\ga 1$ 
and for any finite order $R^{(p)}$ grows without bound at large times or 
wavenumbers instead of decreasing. From Eq.(\ref{Rp1}) we can see that in order
to obtain a reliable prediction at a given scale we need to go at least up to order
$p \sim \omega^{2}(D_1-D_2)^2$.
We display the first few terms in Fig.~\ref{figR_P0}, which clearly shows that
increasing the order $p$ improves the agreement with the exact result on weakly
nonlinear scales but worsens the prediction in the highly nonlinear regime.

\begin{figure}[htb]
\begin{center}
\epsfxsize=8 cm \epsfysize=7 cm {\epsfbox{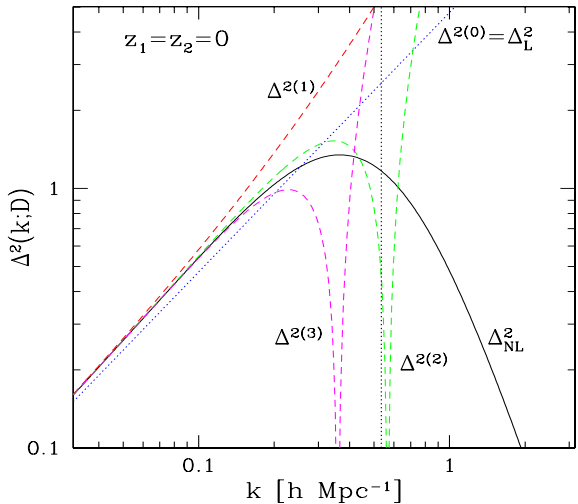}}
\end{center}
\caption{The standard perturbative expansion of the two-point correlation over 
powers of $P_{L0}$ from Eq.(\ref{Dp1}). We only show the density-density equal-time
logarithmic power $\Delta^2(k;D)$ at redshift $z=0$, for the case $n=-2$. 
The solid curve $\Delta^2_{\rm{NL}}$ is 
the exact power from Eqs.(\ref{FtF})-(\ref{tF}), whereas curves labeled 
$\Delta^{2(p)}$ are the expansion up to order $p+1$ over $P_{L0}$. Higher-order
terms grow as $k^{p+1}$ at high wavenumbers. The perturbative 
expansion diverges beyond the vertical dotted line, for $k>0.53h$Mpc$^{-1}$.}
\label{figlG_P0}
\end{figure}

Next, we turn to the standard perturbative expansion of the two-point 
density-density correlation $G_{11}$. For illustrative purposes we consider the 
power-law linear power spectrum (\ref{n2}).
In this case we can perform the integrals in Eq.(\ref{F2}) and 
for the equal-time nonlinear power we obtain:
\beq
\Delta^2= F(\Delta_L^2) \;\;\; \mbox{with} \;\;\; F(x)=x \tilde{F}(x\pi/8)
\label{FtF}
\eeq
and
\beqa
\lefteqn{ \!\!\! \tilde{F}(y) \! = \! \frac{1}{(1+y^2)^2} 
+ \frac{3 y (1+\sqrt{1+y^2}) {\rm Arctan} \frac{y}{\sqrt{2+y^2+2\sqrt{1+y^2}}}}
{4(1+y^2)^{5/2}\sqrt{2+y^2+2\sqrt{1+y^2}}} } \nonumber \\
&& + \frac{3 y (-1+\sqrt{1+y^2}) 
{\rm Arctan} \frac{y}{\sqrt{2+y^2-2\sqrt{1+y^2}}}}
{4(1+y^2)^{5/2}\sqrt{2+y^2-2\sqrt{1+y^2}}} .
\label{tF}
\eeqa
Note that the last two terms have not been correctly written in 
Taylor \& Hamilton (1996). 
The expansion over powers of $P_{L0}$ corresponds to the expansion of $F(x)$ over 
powers of $x$, and we obtain
\beq
\Delta^2= \Delta_L^2 + \frac{3\pi^2}{64} \Delta_L^4 
- \frac{\pi^2}{32} \Delta_L^6 - \frac{15\pi^4}{8192} \Delta_L^8 + ...
\label{Dp1}
\eeq
Contrary to the response function, we can see from Eq.(\ref{tF}) that this expansion
diverges for $\Delta_L^2>8/\pi$. Therefore, one cannot describe nonlinear scales
from this perturbative expansion, and going to higher orders only improves the 
predictions for weakly nonlinear scales where $\Delta_L^2<8/\pi$. We can see from
Eqs.(\ref{F1})-(\ref{F2}) that the radius of convergence of the perturbative series 
is zero for $n<-2$. (Since at large $q$ we encounter an integrand of the form 
$\int \d q e^{-x q^{-n-1}}$, which gives rise to a singularity at $x<0$.) For a 
$\Lambda$CDM cosmology, the slope of the linear power spectrum goes to $n=1$ on
large scales; therefore the perturbative series should always converge. However, on
small scales where $n\leq-2$ the perturbative expansion is likely to be useless
(except for the quasi-linear regime) since the series only converges because of 
the behavior of the linear power spectrum on unrelated large scales. We compare
the first few terms $\Delta^{2(p)}$ of this perturbative expansion with the exact
nonlinear power in Fig.~\ref{figlG_P0}. In agreement with the analysis above, 
we can check that the perturbative predictions provide a good match on 
quasi-linear scales $\Delta_L^2 \ll 8/\pi, k \ll 0.53h$Mpc$^{-1}$ and becomes 
useless deeper into the nonlinear regime.

\begin{figure}[htb]
\begin{center}
\epsfxsize=8 cm \epsfysize=7 cm {\epsfbox{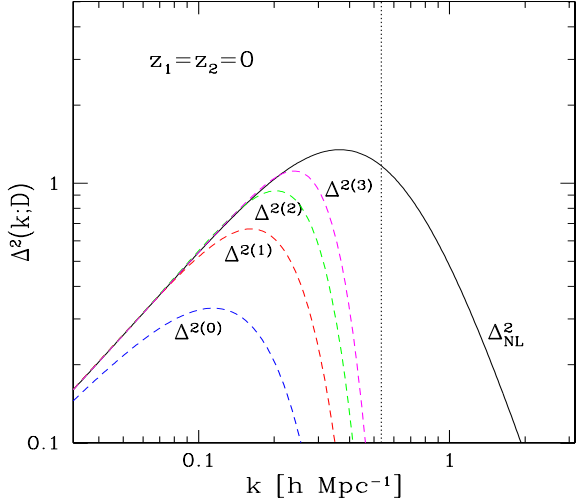}}
\end{center}
\caption{The perturbative expansion of the two-point correlation over powers
of $P_{L0}$ from Eq.(\ref{Dp2}) where we keep the exponential factor 
$e^{-D^2\omega^2}$ apart. The solid curve $\Delta^2_{\rm{NL}}$ is 
the exact power from Eqs.(\ref{FtF})-(\ref{tF}), whereas curves labeled 
$\Delta^{2(p)}$ are the expansion up to order $p+1$ over $P_{L0}$. 
All terms decay as $e^{-D^2\sigma_v^2k^2}$ at high $k$. The perturbative 
expansion again diverges beyond the vertical dotted line, for $k>0.53h$Mpc$^{-1}$.}
\label{figlG_P0exp}
\end{figure}

Crocce \& Scoccimarro (2006a) notice that instead
of the standard perturbative expansion (\ref{Dp1}), one could keep the exponential
factor $e^{-(D_1^2+D_2)^2\omega^2/2}$ in expression (\ref{G11Z5}) and only expand
the last exponential $e^{D_1D_2I}$. In this fashion, all terms of the new series
are positive and damped by the exponential factor at small scales, which gives 
a seemingly better behaved expansion. We display the results we obtain for the 
case $n=-2$ in Fig.~\ref{figlG_P0exp}, when we consider the series
\beqa
\Delta^2 & = & e^{-D^2\omega^2} \left[ e^{D^2\omega^2} F(\Delta_L^2) \right] 
\nonumber \\
& = & e^{-D^2\omega^2} \left[ \Delta_L^2 + \left( \frac{3\pi^2}{64} \Delta_L^4 + 
D^2\omega^2 \Delta_L^2 \right) + ... \right]
\label{Dp2}
\eeqa
where we use the normalization of Eq.(\ref{k0}) for $\omega(k)$. Of course, for 
a power-law linear power spectrum, we actually have $\omega=\infty$, but 
Eq.(\ref{Dp2}) describes a CDM-like power spectrum, such that $n=-2$ on the 
weakly nonlinear scales of interest and $\omega$ is made finite and normalized 
to Eq.(\ref{k0}) through IR and UV cutoffs. Fig.~\ref{figlG_P0exp} shows that indeed 
the terms at each order over $P_{L0}$ are positive, and the series looks better 
behaved. However, it is clear that the series still 
diverges beyond $k \simeq 0.53h$Mpc$^{-1}$ as for (\ref{Dp1}). Moreover, we note that 
increasing $\omega(k_0)$ (by moving the IR and UV cutoffs) would move the Gaussian 
cutoff towards smaller $k$ and would worsen the agreement with the exact 
nonlinear power.
In fact, for general power-spectra where the linear velocity dispersion $\sigma_v^2$
is not necessarily governed by the weakly nonlinear scales of interest, the
rewriting associated with Eq.(\ref{Dp2}) does not always improve the agreement with 
the exact power. Power-law linear power spectra are a clear example
of such cases, since $\omega=\infty$ so that the expansion (\ref{Dp2}) is not 
well-defined (unless one absorbs IR divergences into a renormalized 
$\sigma_v$\footnote{The procedure (\ref{Dp2}) could be modified to apply 
to cases such as $n=-2$ without any IR cutoff, as the exact expression (\ref{G11Z7}) 
shows that the IR divergences are fully absorbed into $\omega=k\sigma_v$. Then, by 
splitting all IR-divergent integrals into a $\sigma_v$ part and a finite part and 
using a given value for $\sigma_v$, one obtains finite expansions. The 
value of $\sigma_v$ is irrelevant for exact equal-time statistics (but would remain 
in approximate quantities obtained as in Eq.(\ref{Dp2}) by truncation at some 
order).}), even though the equal-time power remains finite, and the usual
perturbative expansion (\ref{Dp1}) is well-defined. Note that the expansion 
schemes based on
Eq.(\ref{GPi}), through path-integral or diagrammatic methods, do not exactly 
correspond to the expansion (\ref{Dp2}). Although the first term 
$R \times G_I \times R^T$ may exhibit the factor $e^{-D^2\omega^2}$, this is no 
longer true for the second term (because of the integrations over time 
in $R.\Pi.R^T$), as shown by the discussion in 
Sect.~\ref{Correlation-G-and-self-energy-Pi}. We note, however,
that since the response function $R$ also involves the quantity $\omega$ we can 
expect such schemes to fail in cases where the velocity dispersion $\sigma_v^2$ 
is governed by scales that are very far from those of interest.

\section{Direct steepest-descent method}
\label{Direct-steepest-descent}

\subsection{Response $R$ and damping self-energy $\Sigma$}
\label{Response-and-damping}

\begin{figure}[htb]
\begin{center}
\epsfxsize=8 cm \epsfysize=7 cm {\epsfbox{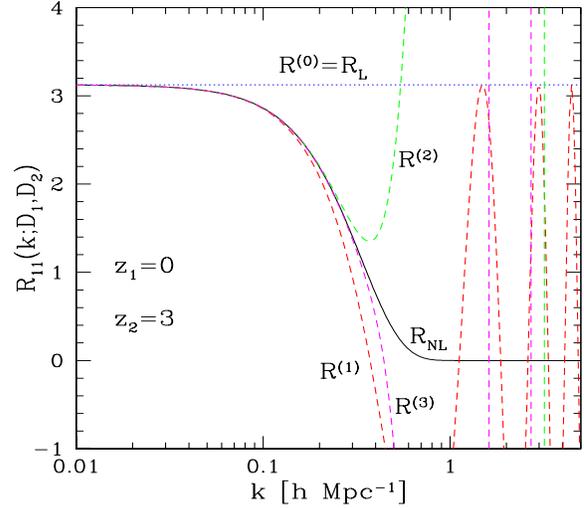}}
\end{center}
\caption{The expansion of the response function defined by the direct 
steepest-descent
method, from Eq.(\ref{trp}). We only show the density-density component $R_{11}$
for clarity. The solid curve $R_{\rm{NL}}$ is the exact response (\ref{RNL}), whereas
curves labeled $R^{(p)}$ are the expansion of the response function up to order $p$.}
\label{figR_dsd}
\end{figure}

We now investigate the properties of the steepest-descent method described in
Sect.~\ref{Steepest-descent}. We first consider the damping self-energy $\Sigma$
and the response $R$. The self-energy $\Sigma$ can be obtained at one-loop order
from Eq.(\ref{Seq}), which gives Eq.(\ref{S0}), and at higher orders by including
higher-order diagrams. However, for the Zeldovich dynamics, it can be directly
obtained from the exact expression derived in Sect.~\ref{Self-energy-Sigma}.
From Eq.(\ref{SNL}) it is clear that series (\ref{sigtexp}) corresponds to
the perturbative expansion of the self-energy $\Sigma$ over powers of $P_{L0}$
(through powers of $\omega^2$). The one-loop result (\ref{S0}) simply corresponds to
the first term $\sigma^{(1)}(t)=\sigma_0$ (whereas the linear regime corresponds
to $\sigma=0$) because of the prefactor $\omega^2$ in $\Sigma_0$ defined in
Eq.(\ref{S0}). In this fashion we obtain the self-energy $\Sigma$ up to order $p$ as
\beq
\Sigma^{(p)}=\Sigma_0 \, \sigma^{(p)} = \Sigma_0 \sum_{m=0}^{p-1} (-1)^m \sigma_m 
\frac{[\omega(D_1-D_2)]^{2m}}{(2m)!} .
\eeq
This in turn determines the response function through the steepest-descent method 
described in Sect.~\ref{Steepest-descent} as
\beq
R^{(p)}(k;D_1,D_2) = R_L \, r^{(p)}[\omega(D_1-D_2)] ,
\label{Rprp}
\eeq
where $r^{(p)}(t)$ is obtained from $\sigma^{(p)}(t)$ through Eq.(\ref{rsig2}).
From Eq.(\ref{trtsig}) this gives
\beq
\tr^{(p)}(s) = \frac{s^{2p-1}}{s^{2p}
+\sum_{m=0}^{p-1} (-1)^m \sigma_m s^{2(p-1-m)}} .
\label{trp}
\eeq
For the first few terms this gives
\beq
r^{(0)}=1, \; r^{(1)}=\cos t , \; 
r^{(2)}= \frac{1}{3} \cosh t + \frac{2}{3} \cos\sqrt{2}t .
\label{rpt}
\eeq
Indeed, the perturbative expansion of the self-energy $\Sigma$ gives $\sigma(t)$
as a series over powers of $t$, whence $\tsig(s)$ as a series over powers of $1/s$.
This yields a rational function for $\tr^{(p)}(s)$ and a sum of exponentials for
$r^{(p)}(t)$. At order $p=1$, the arguments of the exponentials are imaginary 
(they are given by the roots of the denominator of $\tr^{(p)}(s)$), but real parts 
appear at order $p=2,3$ (and presumably at higher orders).
Since the denominator is even, see Eq.(\ref{trp}), the roots appear by pairs 
$\pm s_i$
so that both decaying and growing exponentials appear. Therefore, the direct 
steepest-descent method cannot reproduce the decay of the response function at
large $t$ (i.e. in the highly nonlinear regime). Of course, at a given order $p$,
the response $r^{(p)}(t)$ agrees with the expansion of the exact response (\ref{Rp1})
up to order $t^{2p}$. We display the first few terms $R^{(p)}$
in Fig.~\ref{figR_dsd}. 
The behavior is actually rather close to the standard perturbative 
expansion shown in Fig.~\ref{figR_P0}, as the higher orders slowly improve the 
agreements over weakly nonlinear scales but explode increasingly fast into the 
highly nonlinear regime (but their amplitude grows even faster as exponentials 
instead of power laws).

\subsection{Pad\'{e} approximants}
\label{Pade}

\begin{figure}[htb]
\begin{center}
\epsfxsize=8 cm \epsfysize=7 cm {\epsfbox{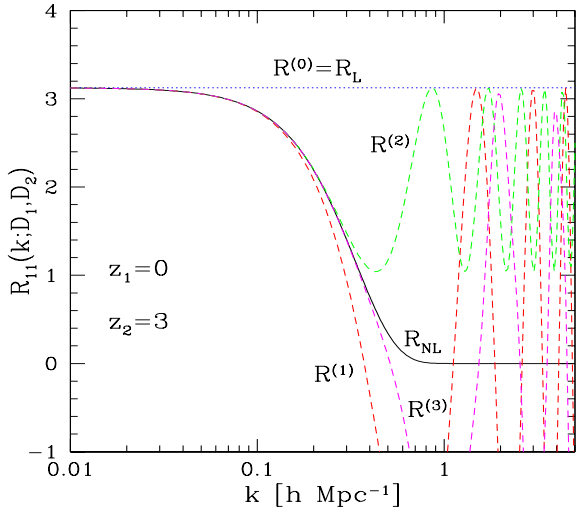}}
\end{center}
\caption{The expansion of the response function defined by the Pad\'{e} approximants
from Eqs.(\ref{r11}). At each order, one obtains a sum of cosines, with an 
offset for even $p$.}
\label{figR_Pade}
\end{figure}

The remarks above suggest that we may improve the expansion (\ref{trp}) by looking
for a Pad\'{e} approximant to the rational function $\tr^{(p)}(s)$, which has the
same expansion over $1/s$ up to $1/s^{2p+1}$. In fact, since we know the exact
response function, we can directly obtain the series of Pad\'{e} approximants from
Eq.(\ref{rtrsNL}). First, from the expansion of $e^{-t^2/2}$ at $t=0$, we obtain
formally the expansion of $\tr(s)$ over $1/s$ as
\beq
\tr(s)= \sum_{p=0}^{\infty} \, (-1)^p \; (2p-1)!! \; s^{-2p-1} .
\label{trsp}
\eeq
Note that this only provides an asymptotic series for $\tr(s)$ in the limit
$s\rightarrow\infty$. Here it is convenient to make the change of variable $y=2/s^2$
and to define $\rb(y)$ by
\beq
\tr(s)= \frac{1}{s} \, \rb(y=2/s^2) , \;\; \rb(y)= \sum_{p=0}^{\infty} \, (-1)^p 
\, \rb_p \, y^p .
\label{rbdef}
\eeq
From Eq.(\ref{trsp}) we obtain for the coefficients $\rb_p$
\beq
\rb_p=\frac{(2p-1)!!}{2^p}=\frac{\Gamma[p+1/2]}{\sqrt{\pi}}=
\int_0^{\infty}\frac{\d t}{\sqrt{\pi t}} \, e^{-t} \, t^p .
\label{rbp}
\eeq
This shows that the expansion (\ref{rbdef}) is a Stieltjes series since the 
coefficients $\rb_p$ are the moments of a real positive function defined over 
$t\geq 0$ (here of the function $e^{-t}/\sqrt{\pi t}$), see Bender \& Orszag (1978).
Since Carleman's condition is fulfilled, $\sum \rb_p^{-1/2p}=\infty$, the function
$\rb(y)$ is uniquely determined by its asymptotic expansion (\ref{rbdef}), which can
be resummed as
\beq
\rb(y) = \int_0^{\infty} \frac{\d t}{1+y t} \, \frac{e^{-t}}{\sqrt{\pi t}} .
\label{rby}
\eeq
Then, all coefficients of the continued-fraction representation of $\rb(y)$ are 
nonnegative, and both Pad\'{e} sequences $P_p^p$ and $P_{p+1}^p$ converge 
monotonically to $\rb(y)$ as
\beqa
P_1^0(y) \leq P_2^1(y) \leq P_3^2(y) \leq ... \leq \rb(y) && \label{P10} \\
\rb(y) \leq .. \leq P_2^2(y) \leq P_1^1(y) \leq P_0^0(y) && \label{P00}
\eeqa
and
\beq
\rb(y) = \lim_{p\rightarrow\infty} P_{p+1}^p(y) = \lim_{p\rightarrow\infty} 
P_p^p(y) .
\label{limPade}
\eeq
By contrast, the usual perturbative expansion (\ref{Rp1}), which is associated
with expansion (\ref{trsp}), amounts to approximate $\rb(y)$ by a polynomial,
that is, by the sequence $P^p_0(y)$, whereas the steepest-descent method (\ref{trp})
amounts to approximate $1/\rb(y)$ by a polynomial, that is, $\rb(y)$ by the sequence 
$P^0_p(y)$. As we have seen above, these two sequences do not converge very well
since they give a response $r^{(p)}(t)$ that grows without bound for 
$t\rightarrow\infty$. On the other hand, the sequence associated with (\ref{limPade})
gives
\beqa
\rb^{(0)}(y) & = & P_0^0=1, \; \rb^{(1)}(y)=P_1^0=\frac{1}{1+y/2} , \nonumber \\
\rb^{(2)}(y) & = & P_1^1=\frac{1+y}{1+3y/2} , ... \label{r11}
\label{rbyPade}
\eeqa
The first two terms give the same results $r^{(0)}$ and $r^{(1)}$ as Eq.(\ref{rpt})
but the next two terms give:
\beqa
r^{(2)}(t) & = & \frac{2}{3} + \frac{1}{3} \cos\sqrt{3} t  , \label{r2t}\\
r^{(3)}(t) & = & \frac{\sqrt{6}+2}{2\sqrt{6}} \cos\sqrt{3-\sqrt{6}} \, t \nonumber \\
&& + \frac{\sqrt{6}-2}{2\sqrt{6}} \cos\sqrt{3+\sqrt{6}} \, t \label{r3t} ,
\eeqa
which do not grow exponentially at large $t$ any more. Because both the
sequences $P_p^p$ and $P_{p+1}^p$ and $\rb(y)$ are Stieltjes functions, they are
analytic in the cut plane $|\arg(y)|<\pi$, and all poles of the Pad\'{e} approximants
$P_p^p$ and $P_{p+1}^p$ lie on the negative real axis (Bender \& Orszag 1978).
Therefore, Eq.(\ref{rbdef}) shows that all poles of the associated rational function
$\tr^{(p)}(s)$ lie on the imaginary axis and appear by pairs $\pm i s_j$ together
with a pole at $s=0$ for even $p$. Then, the response factor $r^{(p)}(t)$
is a sum of cosines $\cos(s_j t)$ plus a constant for even $p$, in agreement
with Eqs.(\ref{r2t})-(\ref{r3t}). This is a clear improvement over the standard
expansion (\ref{Rp1}) and the direct steepest-descent expansion 
(\ref{trp})-(\ref{rpt}). However, even this expansion cannot recover the Gaussian
decay $e^{-t^2/2}$, which is replaced by fast oscillations. Nevertheless, this 
gives rise to an effective damping (for odd $p$) once the response function is 
integrated with some weight function.
We compare the first few terms to the exact nonlinear response in 
Fig.~\ref{figR_Pade}. Of course, we recover the same agreement at low $k$ as
with the other expansions displayed in Figs.~\ref{figR_P0} and \ref{figR_dsd}, 
but as explained above, the response remains bounded at high $k$ with fast 
oscillations.

\subsection{Correlation $G$ and self-energy $\Pi$}
\label{Correlation-G-and-self-energy-Pi}

We now consider the predictions of the direct steepest-descent method for the
two-point correlation $G$ and self-energy $\Pi$. As in 
Sect.~\ref{Standard-perturbative-expansions}, we consider the case of
a power-law linear power spectrum $n=-2$, which simplifies the calculations, since
integrals over wavenumber $k$ have already been performed following 
Eqs.(\ref{FtF})-(\ref{tF}). Moreover, as in Eq.(\ref{Dp2}) and 
Fig.~\ref{figlG_P0exp},
we consider a finite linear velocity dispersion $\omega(k)=k\sigma_v$ normalized
in Eq.(\ref{k0}). As in Sect.~\ref{Self-energy-Pi}, in order to simplify the
computations, we approximate the matrix form of the correlation $G_{ij}$ by
Eq.(\ref{hG}). Thus, we keep the exact nonlinear, density-density correlation
$G_{11}$, and make approximation $G_{ij}=G_{11}$ for all $\{i,j\}$.
Using Eq.(\ref{G11Z7}) and expanding the exponential $e^{D_1D_2\omega^2}$ and $F(x)$,
we write $G_{11}$ as
\beq
G_{11}\!=\!\frac{1}{4\pi k^3} \!\! \sum_{p,m=0}^{\infty} \frac{F_p}{p! m!} 
\Delta_{L0}^{2p} \omega^{2m} (D_1D_2)^{p+m} e^{-\frac{D_1^2+D_2^2}{2}\omega^2} 
\label{G11pm}
\eeq
where $F_p$ is the $p$-th derivative of $F(x)$ at $x=0$:
\beq
F(x)= \sum_{p=0}^{\infty} \frac{F_p}{p!} x^p .
\label{Fp}
\eeq
The first few coefficients $F_p$ are given in Eq.(\ref{Dp1}). Note, however, 
that for 
the case $n=-2$, the Taylor series (\ref{Fp}) diverges for $x>8/\pi$ as seen in 
Sect.~\ref{Standard-perturbative-expansions}. 
Then, we can write the matrix two-point correlation $\hG$ as
\beq
\hG= \frac{1}{4\pi k^3} \! \sum_{p,m} \frac{F_p}{p! m!} \Delta_{L0}^{2p} \omega^{2m}
\hG_{p+m}(D_1) \hG_{p+m}(D_2)^T ,
\label{hGFp}
\eeq
where the vectors $\hG_n$ have been defined in Eq.(\ref{hGn}). Thus, as compared with
Eq.(\ref{hGhGn}), the use of a power-law linear power spectrum has simply replaced
the integral over $\bq$ by a discrete sum. Then Eqs.(\ref{hPhPn})-(\ref{RPiR})
still apply, once we replace the integral over $\bq$ by this discrete sum. We
now need to compute the functions $\hpi_n(D)$ defined in Eq.(\ref{hpin}). Using
the expansion (\ref{sigtexp}), we can perform the integrations over $D'$ and
obtain
\beqa
\lefteqn{\hpi_n(D)= (n-1) D^n + D^n \sum_{m=0}^{\infty} \frac{(-1)^m}{m!} 
\left(\frac{\omega^2D^2}{2}\right)^{m+1} } \nonumber \\
&& \!\!\!\!\!\!\!   \times \! \left[ - \frac{n-1}{m+1} -2 + 
\frac{m! \; 2^{m+1}}{(n+2m)!} \sum_{\ell=0}^m \sigma_{m-\ell} 
\frac{(n+2\ell-1)!}{\ell! \; 2^{\ell}} \right]
\label{hpinsign}
\eeqa
Of course, from Eq.(\ref{sigmap}) we can check that $\hpi_1=0$, in agreement with the
result already obtained in Sect.~\ref{Self-energy-Pi}. Substituting 
Eq.(\ref{hpinsign})
into the analog of Eq.(\ref{hPhPn}), which reads here (as Eq.(\ref{hGFp}) 
for $\hG$) as
\beq
\hPi= \frac{1}{4\pi k^3} \sum_{p,m} \frac{F_p}{p! m!} \Delta_{L0}^{2p} \omega^{2m}
\hPi_{p+m}(D_1) \hPi_{p+m}(D_2)^T ,
\label{hGPipm}
\eeq
we obtain the self-energy $\hPi$ up to order $P_{L0}^{p+1}$ as
\beq
\hPi^{(p)}(k;D_1,D_2) = \frac{1}{4\pi k^3} \sum_{n=1}^p \hPi_n(k;D_1,D_2)  
\left(\bea{cc} 1 & 1 \\ 1 & 1 \ea\right) ,
\label{hPindef}
\eeq
where $\hPi_n \propto P_{L0}^{n+1}$ and the first few terms are
\beqa
\hPi_1 & = & \Delta_L^2 \left( \frac{3\pi^2}{64} \Delta_L^2 
+ \omega_1\omega_2 \right)
\label{hPi1} \\
\hPi_2 & = & - \Delta_L^2 (\omega_1^2+\omega_2^2) \left( \frac{3\pi^2}{64} 
\Delta_L^2 + \omega_1\omega_2 \right) \nonumber \\ 
&& + \Delta_L^2 \left( - \frac{\pi^2}{8} \Delta_L^4 + \frac{3\pi^2}{16} 
\Delta_L^2 \omega_1 \omega_2 + 2 \omega_1^2 \omega_2^2\right)
\label{hPi2}
\eeqa
where we have introduced (we recall that $\Delta_L^2=D_1D_2\Delta_{L0}^2$)
\beq
\omega_1 = D_1 \omega(k)=D_1 k \sigma_v, \;\;\; \omega_2 = D_2 \omega(k) .
\label{omega12}
\eeq
As for the standard perturbative expansion (\ref{Dp1}) applied to the correlation
function $G$, the series expansion (\ref{hPindef}) for the self-energy $\hPi$
gives higher-order terms that grow increasingly fast into the highly nonlinear 
regime. Moreover, we can expect that it diverges at equal times 
for $\Delta_L^2>8/\pi$ as for $G$.

\begin{figure}[htb]
\begin{center}
\epsfxsize=8 cm \epsfysize=7 cm {\epsfbox{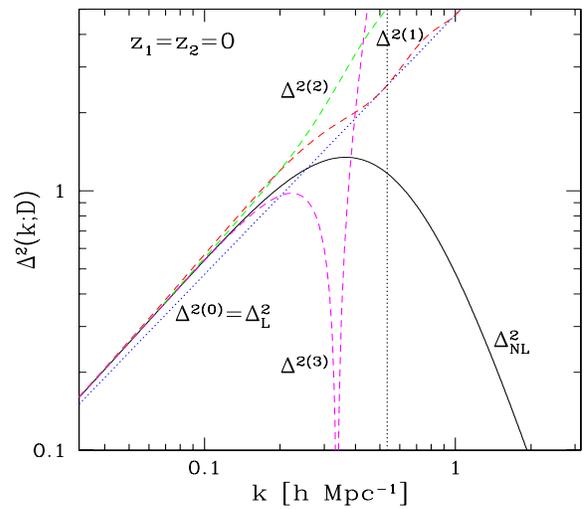}}
\end{center}
\caption{The expansion of the two-point correlation defined by the direct 
steepest-descent method, from Eq.(\ref{hPindef}). Higher-order terms display an
exponential growth at high $k$ for $p\geq 2$.}
\label{figlG_dsd}
\end{figure}

\begin{figure}[htb]
\begin{center}
\epsfxsize=8 cm \epsfysize=7 cm {\epsfbox{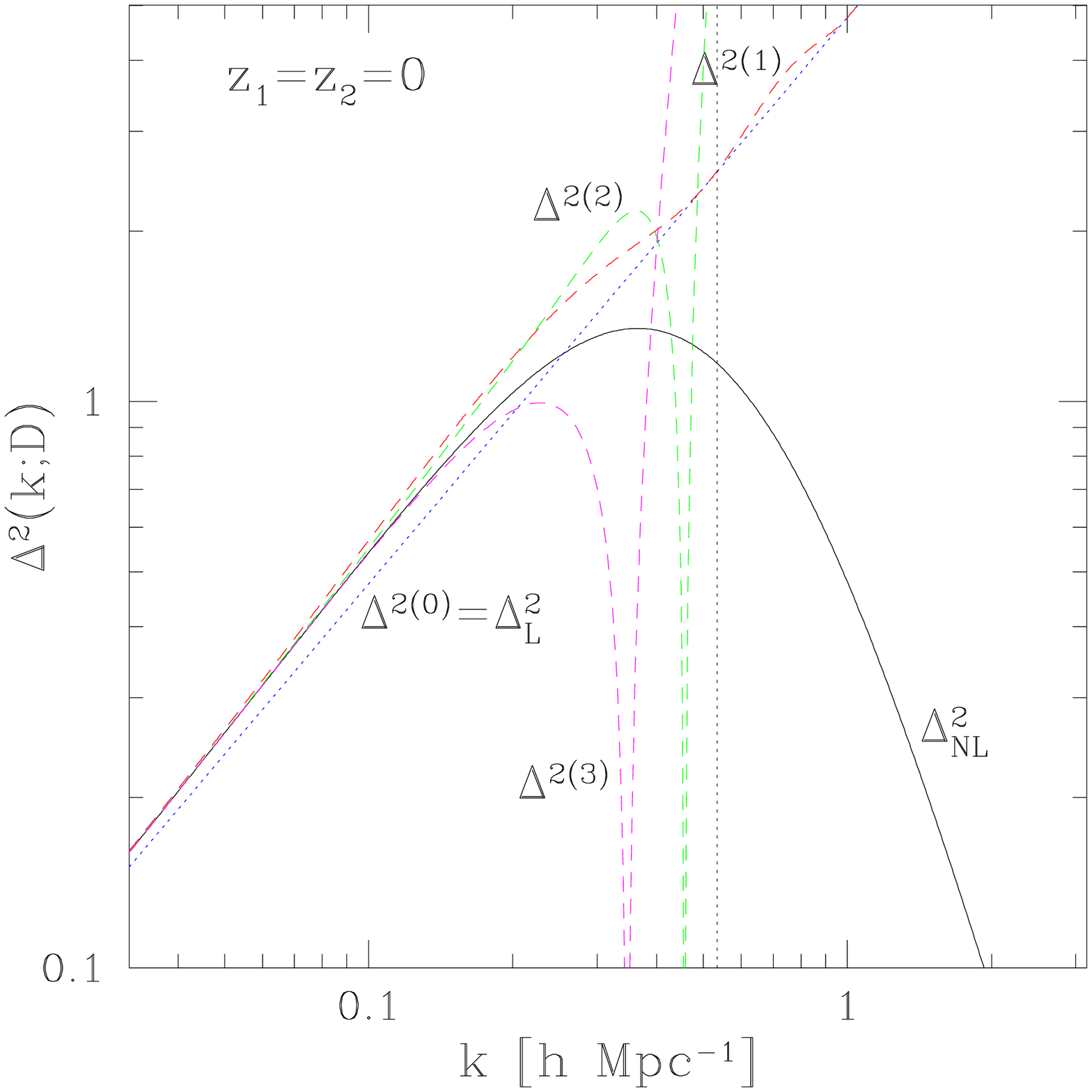}}
\end{center}
\caption{The expansion of the two-point correlation defined by the direct 
steepest-descent method, from Eq.(\ref{hPindef}), but using the Pad\'{e} 
approximants (\ref{r2t})-(\ref{r3t}) for the response function. Higher-order 
terms display a power-law growth at high $k$.}
\label{figlG_Pade}
\end{figure}

\begin{figure}[htb]
\begin{center}
\epsfxsize=8 cm \epsfysize=7 cm {\epsfbox{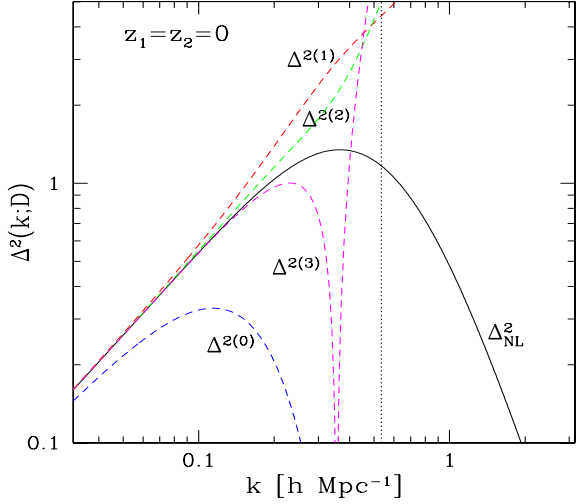}}
\end{center}
\caption{The expansion of the two-point correlation defined by the direct 
steepest-descent method, from Eq.(\ref{hPindef}), but using the exact response 
function (\ref{RNL}). Higher-order terms display a power-law growth at high $k$.}
\label{figlG_Gauss}
\end{figure}

Then, the direct-steepest descent method prediction at order $p$ is obtained by
applying Eq.(\ref{GPi}) using the response $R^{(p)}$ and the self-energy $\hPi^{(p)}$
at that order. Note that this expansion does not have the form of a standard 
perturbative expansion such as Eqs.(\ref{Dp1}) or (\ref{Dp2}), since all terms 
in the series get modified as we go to higher orders,
because the response $R^{(p)}$ has a different functional form, see (\ref{rpt}).
At linear order $p=0$ we have $\Pi=0$ and $\Delta^{2(0)}=\Delta_L^2$. At first order 
$p=1$, we obtain from Eq.(\ref{rpt})
\beqa
\lefteqn{ \Delta^{2(1)}(k;D_1,D_2) = \Delta_L^2 \cos(\omega_1) \cos(\omega_2) }
\nonumber \\
&& + \Delta_L^2 \left( \frac{3\pi^2}{64} \Delta_L^2 + \omega_1\omega_2 \right) 
\frac{\sin(\omega_1)\sin(\omega_2)}{\omega_1\omega_2} .
\label{D2dsd_1}
\eeqa
Of course, if we expand Eq.(\ref{D2dsd_1}) over powers of $\omega$, we recover
the usual perturbative result, and at equal time we recover Eq.(\ref{Dp1}) up to
order $\Delta_L^4$. Moreover, we already know that at equal times $D_1=D_2$, all
terms $\omega$ must cancel out up to order $P_{L0}^{p+1}$ because the exact power
(\ref{FtF}) does not depend on $\omega$. This is a general result that does not
depend on the shape of the linear power spectrum and can be seen from 
Eq.(\ref{G11Z4}).
This is related to the cancellation of IR divergences recalled in 
Sect.~\ref{Asymptotic-behavior} associated with Galilean invariance.
In a similar fashion, at order $p=2$, we obtain from Eq.(\ref{rpt})
\beqa
\lefteqn{ \Delta^{2(2)}= \Delta_L^2 r^{(2)}_1 r^{(2)}_2 + \Delta_L^2 
\left( \frac{3\pi^2}{64} \Delta_L^2 + \omega_1\omega_2 \right) 
r^{(2;1)}_1 r^{(2;1)}_2 } \nonumber \\
&& - \Delta_L^2 \left( \frac{3\pi^2}{64} \Delta_L^2 + \omega_1\omega_2 \right) 
\left(\omega_1^2 r^{(2;3)}_1 r^{(2;1)}_2+\omega_2^2 r^{(2;1)}_1 r^{(2;3)}_2\right) 
\nonumber \\
&& + \Delta_L^2 \left( \!\!  -\frac{\pi^2}{8} \Delta_L^4 \! + \! \frac{3\pi^2}{16} 
\Delta_L^2 \omega_1 \omega_2 \! + \! 2 \omega_1^2 \omega_2^2\right) 
r^{(2;2)}_1 r^{(2;2)}_2
\eeqa
where we have defined ($\ell\geq 1$)
\beq
r^{(p)}_i=r^{(p)}(\omega_i) , \;\; r^{(p;\ell)}_i= \int_0^1 \d t \; t^{\ell-1} \; 
r^{(p)}[\omega_i (1-t)] .
\label{rpi}
\eeq
We display our results for the first few terms in Fig.~\ref{figlG_dsd}.
We can see that the exponential terms associated with the response function 
(see Eq.(\ref{rpt})) make the higher-order approximations grow exponentially
at high $k$ (for $p\geq 2$). This very strong growth makes the series of little
practical value for $p\geq 2$, where the convergence is not better than for
the standard perturbative expansion displayed in Fig.~\ref{figlG_P0}.
We note that the equal-time power can become negative at high $k$, whereas
both $\Pi$ and $G$ should be positive matrices from Eq.(\ref{GPi}), which implies
that the components $G_{ii}$ and $\Pi_{ii}$ are positive at equal times.
This failure comes from the truncation of the self-energy $\hPi$ 
in Eq.(\ref{hPindef}), 
which breaks the positivity of $\hPi$. Indeed, from Eq.(\ref{hPi2}) and the scalings
$\Delta_L^2\propto k$ and $\omega\propto k$, we see that 
$\hPi_2 \sim -3\pi^2/32 \Delta_L^4 \omega^2$ at high $k$.

We show in Fig.~\ref{figlG_Pade} the results obtained from the expansion 
(\ref{hPindef}) for the self-energy $\hPi$ but using the Pad\'{e} approximants
(\ref{r2t})-(\ref{r3t}) for the response function. The higher-order terms no
longer grow as exponentials at high $k$ but as power laws, since the response 
functions only contain constants and cosines instead of exponentials.
However, this is not sufficient for significantly improving the convergence of the
series to the exact nonlinear power $\Delta^2$.

Finally, we show in Fig.~\ref{figlG_Gauss} the results obtained from the expansion 
(\ref{hPindef}) for the self-energy $\hPi$ when we use the exact response function
(\ref{RNL}). Note that in this case the lowest-order approximation $\Delta^{2(0)}$
is equal to the linear power multiplied by a Gaussian damping factor. It is
actually equal to the first term of Eq.(\ref{Dp2}). Higher-order terms do not
exhibit such a Gaussian damping because of the time-integrals involved in the last
term of Eq.(\ref{GPi}). Indeed, expansion (\ref{hPindef}) yields power laws
over $D_1,D_2$, which gives enough power generated at all times to build large
density fluctuations into the nonlinear regime. We can see from 
Fig.~\ref{figlG_Gauss} that using the exact response function is not enough
to significantly improve the convergence of the series obtained for the matter
power spectrum. Therefore, it appears that to improve the results 
one should use other expansion schemes for the self-energy $\hPi$: improving the
response function alone does not help much.

\subsection{Using a Gaussian decay for the self-energy $\Pi$}
\label{Gaussian-decay-for-the-self-energy-Pi}

\begin{figure}[htb]
\begin{center}
\epsfxsize=8 cm \epsfysize=7 cm {\epsfbox{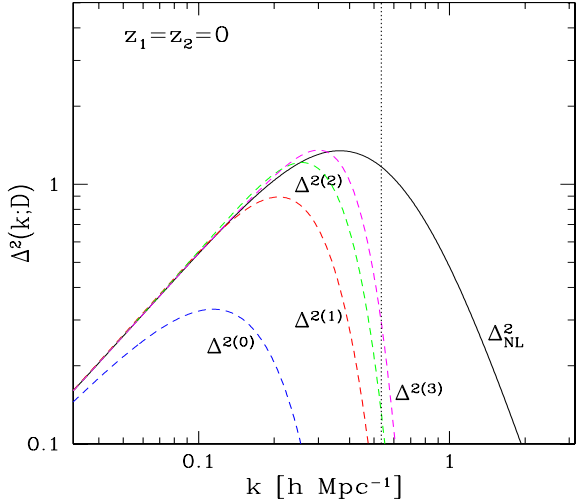}}
\end{center}
\caption{The expansion of the two-point correlation from Eq.(\ref{hPiexpdef}), 
using the exact response function (\ref{RNL}) and the expansion (\ref{hPiexpdef}) 
with Gaussian factors for the self-energy $\hPi$. All terms display a Gaussian 
decay at high $k$.}
\label{figlG_exp_Gauss}
\end{figure}

The previous discussion and the results obtained from expansion (\ref{Dp2}),
shown in Fig.~\ref{figlG_P0exp}, suggest that it may be useful to factor out
a Gaussian term of the form $e^{-D^2\omega^2}$ from the self-energy $\Pi$.
Therefore, we now replace expansion (\ref{hPindef}) by
\beq
\! \hPi(k;D_1,D_2) = \frac{e^{-(\omega_1^2+\omega_2^2)/2}}{4\pi k^3} \! 
\sum_{n=1}^{\infty} \! \hPi_n(k;D_1,D_2)  \left(\bea{cc} 1 & 1 \\ 1 & 1 \ea\right)
\label{hPiexpdef}
\eeq
where $\hPi_n$ is again of order $P_{L0}^{n+1}$ and $\omega_i=D_i\omega$ as in 
Eq.(\ref{omega12}). This expansion can be obtained as in 
Sect.~\ref{Correlation-G-and-self-energy-Pi} or it can be derived from the
expansion (\ref{hPindef}) by multiplying by a factor $e^{(\omega_1^2+\omega_2^2)/2}$
and again expanding over powers of $P_{L0}$. 
We display in Fig.~\ref{figlG_exp_Gauss}
the results obtained from Eq.(\ref{hPiexpdef}) using the exact response function
(\ref{RNL}) that also shows a Gaussian decay at high $k$. Of course, because of
the Gaussian damping factor introduced in Eq.(\ref{hPiexpdef}) all terms decay
as $e^{-k^2\sigma_v^2}$ at high $k$ and the expansion looks better behaved than
the one displayed in Fig.~\ref{figlG_Gauss}. Besides, the approximation obtained at
a given order seems to provide good accuracy over a slightly wider range than in 
both Figs.~\ref{figlG_Gauss} and \ref{figlG_P0exp}. Thus, decomposing the
two-point correlation in terms of response function and self-energy as in 
Eq.(\ref{GPi}) and using a Gaussian decay ansatz at high $k$ appears to be
a good scheme. However, this is somewhat artificial (see Sect.~\ref{High-k-limit}).

\begin{figure}[htb]
\begin{center}
\epsfxsize=8 cm \epsfysize=7 cm {\epsfbox{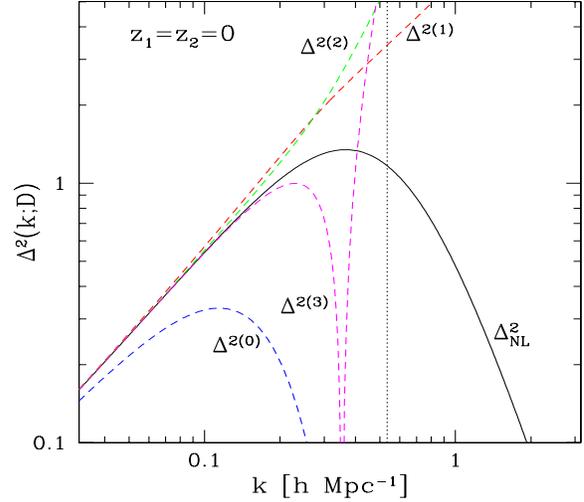}}
\end{center}
\caption{The expansion of the two-point correlation from Eq.(\ref{hPiexpdiffdef}), 
using the exact response function (\ref{RNL}) and the 
expansion (\ref{hPiexpdiffdef}) with Gaussian factors for the self-energy $\hPi$. 
This Gaussian factor only damps the contributions at different times, and 
high-order terms display a power-law growth at high $k$.}
\label{figlG_expdiff_Gauss}
\end{figure}

As explained in Sect.~\ref{High-k-limit} below, the Gaussian damping factors
$e^{-D^2k^2\sigma_v^2/2}$ merely correspond to the linear displacement
field that moves the location of  large-scale structures between different
times. However, this process does not affect the matter clustering, and in
particular the linear velocity variance $\sigma_v$ can show IR divergences for
$n\leq-1$, which cancel out for equal-time statistics (Vishniac 1983; 
Jain \& Bertschinger 1996). This suggests that it would make more sense to
factor out a Gaussian factor of the form
\beq
\! \hPi(k;D_1,D_2) = \frac{e^{-(\omega_1-\omega_2)^2/2}}{4\pi k^3} \! 
\sum_{n=1}^{\infty} \! \hPi_n(k;D_1,D_2)  \left(\bea{cc} 1 & 1 \\ 1 & 1 \ea\right) ,
\label{hPiexpdiffdef}
\eeq
which is equal to unity at equal times. This also agrees with the functional form
of the exact two-point function (\ref{G11Z7}) and of the simple approximation 
(\ref{D2kapprox}) below. We show the results of the expansion (\ref{hPiexpdiffdef})
in Fig.~\ref{figlG_expdiff_Gauss}, again using the exact response function 
(\ref{RNL}). The high-order terms now grow as power laws at high $k$ for
the equal-time power (clearly they would still decay as a Gaussian for unequal 
times), but the expansion does not fare better than the straightforward expansion
obtained from Eq.(\ref{hPindef}) displayed in Fig.~\ref{figlG_Gauss}.
This shows that using a reasonable Gaussian decay ansatz (which must disappear
at equal times) is not sufficient to bring a significant improvement over previous
expansions.

\section{High-$k$ limit ?}
\label{High-k-limit}

To improve the behavior of expansion schemes, Crocce \& Scoccimarro (2006b) 
suggest using a response function that matches both the low-$t$ behavior 
(obtained by perturbative expansions) and the Gaussian decay at high $t$. 
Indeed, for the case of the gravitational dynamics where the exact
response $R$ is not known, one can still obtain the low-$t$ behavior by perturbative
expansions, such as those described above, by summing over higher-loop diagrams;
see Crocce \& Scoccimarro (2006a) and Valageas (2007). On the other hand, 
Crocce \& Scoccimarro (2006b) manage to resum a subset of diagrams in the high-$k$
limit, which gives rise to the same Gaussian decay $e^{-t^2/2}$ as for the Zeldovich
dynamics. Then, one may hope that by using a ``fit'' for the response function
that closely follows the expected behavior, one has done
half the work needed to obtain a good prescription for the matter power spectrum,
and all that is left is to use a good recipe for the self-energy $\Pi$. In fact,
we have shown in Sect.~\ref{Correlation-G-and-self-energy-Pi} and 
Fig.~\ref{figlG_Gauss} that this is not so simple, because even using the exact
response $R$ is not enough to improve the predictions for the power spectrum
if we use a simple expansion over powers of $P_{L0}$ for the self-energy $\Pi$,
as in Eq.(\ref{hPindef}). Moreover, as shown in 
Sect.~\ref{Gaussian-decay-for-the-self-energy-Pi} using such a Gaussian cutoff
for $\Pi$ is not sufficient either.

\begin{figure}[htb]
\begin{center}
\epsfxsize=7 cm \epsfysize=5 cm {\epsfbox{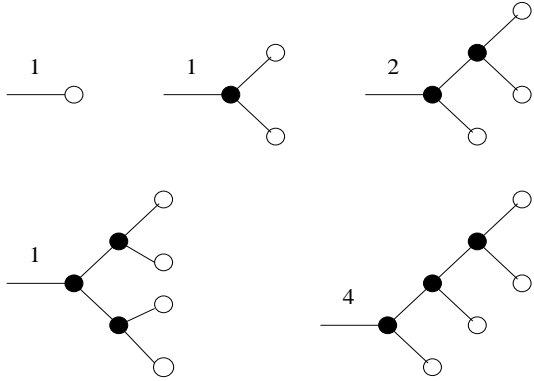}}
\end{center}
\caption{The expansion of the nonlinear field $\psi$ over the linear growing mode
$\psi_L$ from Eq.(\ref{OtKs}), up to order $\psi_L^4$. The filled circles are 
the vertex $\tKs$, whereas the white circles are the linear input $\psi_L$. 
The numbers are the multiplicity factor associated with each diagram.}
\label{figdiag1}
\end{figure}

\begin{figure}[htb]
\begin{center}
\epsfxsize=8.9 cm \epsfysize=1.46 cm {\epsfbox{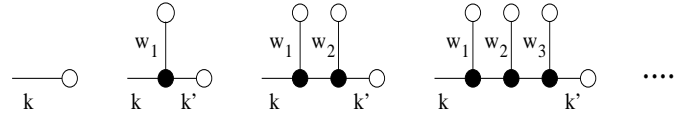}}
\end{center}
\caption{The diagrams assumed to dominate in the high-$k$ limit. 
We have $k'\simeq k$, and all intermediate wavenumbers $w_i$ are much smaller 
than $k$.}
\label{figdiag2}
\end{figure}

Nevertheless, this section revisits the approximate resummation performed
in Crocce \& Scoccimarro (2006b) to clarify its meaning and its shortcomings.
Let us start from the general integral equation of motion (\ref{OtKs}).
This equation can be solved perturbatively as an expansion over powers of $\psi_L$
of the form
\beq
\psi=\psi_L+\tKs\psi_L^2+2 \tKs^2\psi_L^3+5 \tKs^3\psi_L^4+...
\label{psipsiLexpand}
\eeq
which can also be written in a diagrammatic manner, see 
Crocce \& Scoccimarro (2006a) and Valageas (2001), as shown in Fig.~\ref{figdiag1}. 
In the high-$k$ limit, one assumes that all wavenumbers $w_i$ associated
with the linear fields $\psi_L$ are much smaller than $k$ except for one field
(because of the conservation of momentum associated with the Dirac factor 
$\delta_D(\bk_1+\bk_2-\bk)$ in the vertex $K_s$, see Eq.(\ref{Ksdef})).
This assumes that the power is generated over some finite range of wavenumbers 
(i.e. the dynamics is not governed by an extended UV tail). 
Crocce \& Scoccimarro (2006b) assume that the dominant contribution is provided
by the diagrams shown in Fig.\ref{figdiag2} where all low-$w_i$ fields 
$\psi_L(\bw_i)$ 
are directly connected to the ``principal path'' that joins the only one high-$k$ 
field $\psi_L$ to the root of the diagram. The idea is that in such diagrams
the ``principal path'' only interacts with other fields $\psi_L(\bw_i)$, which are
pure linear growing modes, as they have not already suffered nonlinear interactions
through the coupling vertex $\tKs$. Then, these diagrams should maximize the
cross-correlations since nonlinear interactions are expected to erase the
memory of initial conditions. Therefore, the response function $R$ may be
dominated by these diagrams in the high-$k$ limit.

On the other hand, the diagrams of Fig.\ref{figdiag2} are actually 
generated by the equation of motion
\beq
\hpsi(x)=\psi_L(x)+2\tKs(x;x_1,x_2) . \psi_L^<(x_1) \hpsi(x_2) ,
\label{OtKsinf}
\eeq
where $\psi_L^<(x_1)$ is the linear growing mode restricted to low wavenumbers
$w_1\ll k$ (the factor $2$ comes from the fact that we can associate $\psi_L^<$
to either $x_1$ or $x_2$ in Eq.(\ref{OtKs})), and we noted by a hat the 
approximate field $\hpsi$ obtained with this high-$k$ limit. 
Note that $\psi_L(x)$ and $\psi_L^<(x_1)$ are {\it independent} Gaussian 
fields (since $w_1\neq k$) and Eq.(\ref{OtKsinf}) is now a {\it linear} equation
for the field $\hpsi$. Indeed, we can check that, by solving Eq.(\ref{OtKsinf})
as a perturbative series over powers of $\psi_L$, we recover the diagrams of 
Fig.\ref{figdiag2}. In other words, by keeping only the diagrams of
Fig.\ref{figdiag2} we have actually approximated the nonlinear equation of 
motion (\ref{OtKs}) by the linear equation of motion (\ref{OtKsinf}). 
Next, following Crocce \& Scoccimarro (2006b) we note that the vertices $\gamma^s$
of Eqs.(\ref{gamma1})-(\ref{gamma2}) satisfy at the high-$k$ limit,
\beq
\!\!\frac{k}{w}\!\!\gg 1 \! :  \sum_m \! \gamma^s_{i;mj}(\bw,\bk) 
\psi_{Lm}^<(\bw,\eta') 
\simeq \delta_{i,j} \frac{\bk.\bw}{2w^2} e^{\eta'} \delta_{L0}(\bw) ,  
\label{highkgamma}
\eeq
where we have only kept the terms of order $k/w$ and neglected terms of order 
$1, w/k,..$. At this level, we can also replace the Dirac factor 
$\delta_D(\bw+\bk'-\bk)$ by $\delta_D(\bk'-\bk)$ in the vertex $K_s$, and  
using the second Eq.(\ref{tKs}) and Eq.(\ref{Ksdef}), we can write 
Eq.(\ref{OtKsinf}) as
\beq
\hdelta(\bk,\eta)=e^{\eta}\delta_{L0}(\bk)+ \!\int\!\d\bw\frac{\bk.\bw}{w^2} 
\delta_{L0}(\bw) e^{\eta} \! \int_{\eta_I}^{\eta} \! \d\eta' \hdelta(\bk,\eta') ,
\label{highkdelta}
\eeq
where we use the fact that $\psi_L$ is also the linear growing mode.
Here we consider the case where the initial conditions are set up at a finite
time $\eta_I$ as in Sect.~\ref{intetaI}. 
This linear equation can be solved through the expansion
\beqa
\lefteqn{ \hdelta(\bk,\eta)=e^{\eta}\delta_{L0}(\bk) + e^{\eta}\delta_{L0}(\bk) 
\sum_{p=1}^{\infty} \prod_{j=1}^p \int \! \d\bw_j \frac{\bk.\bw_j}{w_j^2} 
\delta_{L0}(\bw_j) } \nonumber \\
&& \times \int_{\eta_I}^{\eta} \d\eta_1 e^{\eta_1}  \int_{\eta_I}^{\eta_1} 
\d\eta_2 e^{\eta_2} .. \int_{\eta_I}^{\eta_{p-1}} \d\eta_p e^{\eta_p} ,
\label{highkseries}
\eeqa
which can be resummed as:
\beq
\hdelta(\bk,D)= D \, \delta_{L0}(\bk) \, e^{(D-D_I)\int\d\bw \frac{\bk.\bw}{w^2} 
\delta_{L0}(\bw)} ,
\label{highkexp}
\eeq
where we use the time-coordinate $D=e^{\eta}$.
Let us recall that $\delta_{L0}(\bk)$ and $\delta_{L0}(\bw)$ must be treated as 
independent Gaussian variables in Eq.(\ref{highkexp}). Then, if we define a response
$\hat{R}_I(\bk,\eta;\bk')$ in a fashion similar to Eq.(\ref{tcR}) by
\beq
\hat{R}_I(\bk,\eta;\bk') = \lag \frac{\cD\hdelta(\bk,\eta)}{\cD\delta_{LI}(\bk')} 
\rag = \frac{1}{D_I}  \lag \frac{\cD\hdelta(\bk,\eta)}{\cD\delta_{L0}(\bk')} \rag   ,
\label{highkRdef}
\eeq
where $\cD$ is the functional derivative, we obtain
\beqa
\hat{R}_I & = & \delta_D(\bk-\bk') \frac{D}{D_I} \lag e^{(D-D_I)\int\d\bw 
\frac{\bk.\bw}{w^2} \delta_{L0}(\bw)} \rag \nonumber \\
& = & \delta_D(\bk-\bk') \frac{D}{D_I} \, e^{-\frac{1}{2} (D-D_I)^2 k^2 \sigma_v^2} .
\label{highkR}
\eeqa
Thus we recover the exact Gaussian decay at high $k$ obtained by 
Crocce \& Scoccimarro (2006b). Note that since we have restricted the initial 
conditions
to the linear growing mode in Eq.(\ref{highkdelta}), the response $\hat{R}_I$ of
Eq.(\ref{highkRdef}) actually corresponds to the sum $\tcR_{11}+\tcR_{12}$ of the 
components of the response defined in Eq.(\ref{tcR}).  
The advantage of this formulation is that one may see more clearly through 
Eqs.(\ref{OtKsinf})-(\ref{highkexp}) the meaning of the assumptions involved in 
this high-$k$ limit. In particular, it is interesting to compare Eq.(\ref{highkexp})
with the exact result (\ref{deltakq}), which reads as
\beq
\delta(\bk,D) = \int\frac{\d\bq}{(2\pi)^3} \, e^{-i\bk.\bq} \, 
e^{D \int\d\bw \, e^{i\bw.\bq} \, \frac{\bk.\bw}{w^2} \, \delta_{L0}(\bw)} .
\label{deltakexact}
\eeq
Then, if we assume that we can split the integral over $\bw$ into low and high 
wavenumber parts $w<\Lambda$ and $w>\Lambda$, such that most of the power is
associated with $w<\Lambda$ and the high-wavenumber contribution is small,
we can expand the exponential over this high-wavenumber part as
\beqa
\lefteqn{ \delta(\bk,D) = \int\frac{\d\bq}{(2\pi)^3} \, e^{-i\bk.\bq} \,
e^{D \int_{w<\Lambda}\d\bw \, e^{i\bw.\bq} \, \frac{\bk.\bw}{w^2} \, 
\delta_{L0}(\bw)} } \nonumber \\
&& \times \left[ 1+D \int_{w'>\Lambda}\d\bw' \, e^{i\bw'.\bq} \frac{\bk.\bw'}{w'^2} 
\delta_{L0}(\bw') + .. \right] .
\label{deltakapprox1}
\eeqa
Next, if we assume that this expression is dominated by $q\sim 1/k$, we can neglect
the factor $e^{i\bw.\bq}$ in the exponent in the high-$k$ limit $k\gg\Lambda$.
Then the integration over $\bq$ yields the Dirac factor $\delta_D(\bw'-\bk)$ to
obtain at lowest order (the factor $1$ does not contribute)
\beq
\delta(\bk,D) \simeq D \, \delta_{L0}(\bk) \, 
e^{D \int_{w<\Lambda}\d\bw \, \frac{\bk.\bw}{w^2} \, \delta_{L0}(\bw)} .
\label{deltakapprox2}
\eeq
Thus we recover Eq.(\ref{highkexp}) with $D_I\rightarrow 0$ and letting 
$\Lambda\rightarrow\infty$. Of course, the assumptions involved 
in the derivation of Eq.(\ref{deltakapprox2}) from Eq.(\ref{deltakexact}) are
identical to those involved in the derivation of Eq.(\ref{highkexp}).
To check whether these assumptions are valid, we can compare the nonlinear
power spectrum predicted by Eq.(\ref{highkexp}) with the exact power studied in
Sect.~\ref{Two-point-correlation}. This gives (with $D_I=0$)
\beq
\hat{\Delta}^2(k;D_1,D_2)= \Delta_L^2 \, e^{-\frac{1}{2} (D_1-D_2)^2 k^2\sigma_v^2} .
\label{D2kapprox}
\eeq
We again recover the exact Gaussian decay at high $k$ for unequal 
times, which corresponds to the exponential term in the exact expression 
(\ref{G11Z7}); but for equal times, we merely get back the linear power 
$\Delta_L^2$. In fact, the analysis of Sect.~\ref{Asymptotic-behavior} shows that 
the assumptions underlying Eq.(\ref{deltakapprox2}) are not valid. For a 
power-law linear power spectrum
(\ref{P0kn}), the derivation leading to Eq.(\ref{F3}) shows that, in the highly 
nonlinear regime the nonlinear power at wavenumber $k$ is associated with scales
$q \sim k^{-1+(n+3)/(n+1)}$ and with linear wavenumbers $w\sim k^{1-(n+3)/(n+1)}$.
Thus, for $-3<n<-1$, where the system is well-defined, we find that the power
at a nonlinear wavenumber $k$ is generated by linear wavenumbers $w$, which 
actually grow faster\footnote{That the power at a nonlinear wavenumber $k$ mostly 
comes from higher
linear wavenumbers $w$, as shown by the explicit expression (\ref{G11Z5}) analyzed
in Sect.~\ref{Asymptotic-behavior}, may seem a bit counter-intuitive as one may
expect a ``direct cascade'' from larger to smaller scales. However, it appears that
the actual process is somewhat more complicated within the highly nonlinear regime.
In particular, as analyzed in Taylor \& Hamilton (1996) (see also 
Schneider \& Bartelmann 1995), the equal-time nonlinear power behaves as 
$\Delta^2(k,D) \propto D^{-3}$ (which is independent of $k$) in the highly nonlinear
regime if the spectral index is less than $-3$ on small scales (e.g. the linear 
power shows a high-$k$ cutoff $P_L(k) \propto k^n e^{-k^2/\Lambda^2}$). 
This scaling can be read from Eq.(\ref{G11Z4}) with $w$ fixed and $q \sim 1/(Dk)$,
which shows that the power comes from a fixed range of wavenumbers $w$ 
(e.g. $w<\Lambda$) associated with caustics (Schneider \& Bartelmann 1995).
However, for a spectral index larger than $-3$ on small scales the scaling
is quite different as shown by Eq.(\ref{Delta2asymp}) 
(see also Taylor \& Hamilton 1996),
and one cannot neglect smaller scales to describe the nonlinear evolution.} 
than $k$ instead of being restricted to a finite range $w<\Lambda$. 
Therefore, one cannot define a fixed cutoff $\Lambda$ beyond which nonlinear 
interactions are negligible so that we can expand the high-$w$ part as in 
Eq.(\ref{deltakapprox1}).
The latter integral is actually large in the nonlinear regime, and one needs
to take its full non-perturbative expression into account to obtain the non-trivial
scaling of Eq.(\ref{Delta2asymp}).

As was already clear from the linear 
Eqs.(\ref{OtKsinf}), (\ref{highkdelta}), the high-$k$ approximations described above
neglect the nonlinear interactions associated with high wavenumbers, which actually
govern the formation of large-scale structures on small scales, and only keep track 
of the overall
displacement associated with the large-scale velocity field. This is why the 
low-$k$ nonlinear interactions could be resummed as $e^{-(D_1-D_2)k^2\sigma_v^2}$,
which only involves the variance $\sigma_v$ of the linear displacement field.
This is clearly an important feature of the dynamics for unequal times, especially
for the simple Zeldovich dynamics studied in this paper where the exact trajectories
satisfy the simple law (\ref{xq}) and the response function (\ref{RNL}) happens to 
be fully determined by a similar advection process. In fact, Eq.(\ref{highkexp}) 
reads in real space (with $D_I=0$) as
\beq
\hdelta(\bx,D)= \delta_L(\bx-\bs_L(\bq=0,D)) ,
\label{advec}
\eeq
where $\bs_L=D_+\bs_{L0}$ is the linear displacement field of Eq.(\ref{xq}).
As expected, Eq.(\ref{advec}) explicitly shows that the approximation 
$\delta_D(\bw+\bk'-\bk) \simeq \delta_D(\bk'-\bk)$ in the vertex $K_s$
used to obtain Eq.(\ref{highkdelta}) and to simplify the diagrams of 
Fig.~\ref{figdiag2} has broken the invariance through translations of the system.
As seen above, this invariance is restored as in Eqs.(\ref{highkR}) and 
(\ref{D2kapprox}) by treating $\bs_L(\bq=0)$ as an independent Gaussian random 
variable. Equation (\ref{advec}) clearly shows that the effective dynamics 
associated with these high-$k$ approximations is simply the {\it uniform advection} 
of the linear density field by the linear velocity at $\bq=0$. After averaging 
over the Gaussian initial conditions, this random displacement of large-scale 
structures leads to an apparent ``diffusion'' in the form of a Gaussian decay 
$e^{-k^2\sigma_v^2}$ for both the response function and the different-time 
correlation, see Eqs.(\ref{highkR}) and (\ref{D2kapprox}). This apparent loss 
of memory happens to dominate the different-time behavior of two-point functions 
for the Zeldovich dynamics, as seen from Eqs.(\ref{G11Z7}) and (\ref{RNL}), 
but it is actually disconnected from the building of matter clustering as is 
obvious from Eq.(\ref{advec}).

It is clear that all steps going from Eqs.(\ref{psipsiLexpand}) to (\ref{highkR}),
and Eqs.(\ref{D2kapprox})-(\ref{advec}) can be applied identically to the exact 
gravitational dynamics, see  Crocce \& Scoccimarro (2006b). As seen above, the
high-$k$ approximations associated with Eq.(\ref{highkexp}) cannot aim at
capturing the physics of gravitational clustering but only the ``diffusion''
associated with the linear velocity variance, which affects different-time 
statistics. However, it is not obvious that the approximate response (\ref{highkR}) 
should again agree with the exact response in the high-$k$ limit, which may thus 
depart from such a Gaussian decay. The small-scale gravitational dynamics
is indeed quite different from the simple Zeldovich dynamics and other processes 
may come into play. Nevertheless, the apparent loss of memory due to the almost 
uniform random displacement of large-scale structures by low-$k$ modes, which is 
captured by the simple dynamics (\ref{advec}), clearly applies to the exact 
gravitational dynamics as well. Therefore, one can expect again a decay as fast 
as in Eq.(\ref{highkR}) although only wavelengths which are still linear would 
contribute to $\sigma_v$ (it is not clear whether nonlinear wavelengths would 
give rise to a stronger or weaker decay as compared with Eq.(\ref{highkR})).
On the other hand, for the exact gravitational dynamics
the fluid equations break down beyond shell-crossing so that the small-scale
limit associated with these hydrodynamical equations is not so well defined.

\section{2PI effective action method}
\label{2PI-effective-action-method2}

We now investigate the 2PI effective action method presented in 
Sect.~\ref{2PI-effective-action} and described in detail in Valageas (2007).
Thus, we need to solve the system of coupled equations (\ref{Geq})-(\ref{Req})
and (\ref{S2PI})-(\ref{P2PI}).
Thanks to causality, which leads to the Heaviside factor
$\theta(\eta_1-\eta_2)$ within both $R$ and $\Sigma$,
we solve this system by moving forward over time.
We refer the reader to Valageas (2007) for a description of the numerical scheme.
We consider the $\Lambda$CDM cosmology associated with Eq.(\ref{k0}) and we only 
investigate the one-loop predictions.

\begin{figure}[htb]
\begin{center}
\epsfxsize=9 cm \epsfysize=8 cm {\epsfbox{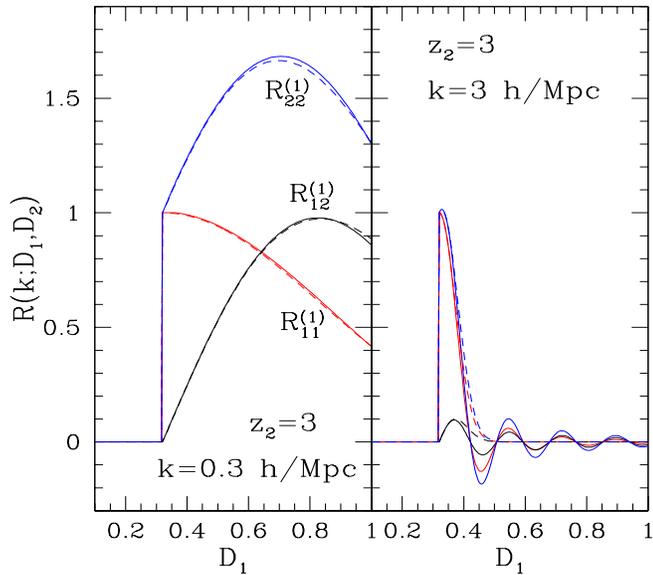}}
\end{center}
\caption{The nonlinear response $R(k;D_1,D_2)$ (solid lines) as a function 
of forward time $D_1$, for $D_2=0.32$ (i.e. $z_2=3$) and wavenumbers $k=0.3$ 
(left panel) and $3 \times h$ Mpc$^{-1}$ (right panel) for the 2PI effective 
action method at one-loop order. 
For comparison we also plot the exact response $R_{\rm NL}$ (dashed lines).}
\label{figRt1}
\end{figure}

\begin{figure}[htb]
\begin{center}
\epsfxsize=8 cm \epsfysize=7 cm {\epsfbox{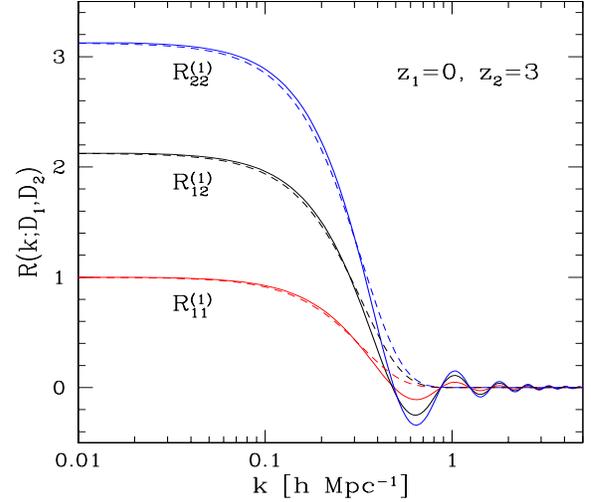}}
\end{center}
\caption{The nonlinear response function $R(k;D_1,D_2)$ (solid lines) 
as a function of wavenumber $k$, at times $z_1=0,z_2=3$, for the 2PI effective 
action method at one-loop order.  
We also plot the exact nonlinear response $R_{\rm NL}$ (dashed lines).}
\label{figRk}
\end{figure}

We first display in Fig.~\ref{figRt1} the evolution forward over time $D_1$
of the response $R(k;D_1,D_2)$. We can see that the nonlinear response
exhibits oscillations as for the steepest-descent result (\ref{rpt})
but its amplitude now decays as an inverse power-law at large times $D_1$ 
instead of following the linear envelope (at one-loop order). 
As shown in Sect.~6.1 of Valageas (2007) this behavior is due to the nonlinearity 
of the Schwinger-Dyson equation for the response $R$. Of course, in the weakly
nonlinear regime we also recover the exact response (\ref{RNL}) (dashed lines).
Next, we display in Fig.~\ref{figRk} the response function as a function of
wavenumber $k$. In agreement with Fig.~\ref{figRt1}, we again obtain damped
oscillations in the nonlinear regime. This is a clear improvement over both
the standard perturbative expansion displayed in Fig.~\ref{figR_P0} and the
steepest-descent result displayed in Fig.~\ref{figR_dsd}.
This behavior is identical to the one obtained for the gravitational dynamics
studied in Valageas (2007).

\begin{figure}[htb]
\begin{center}
\epsfxsize=8 cm \epsfysize=7 cm {\epsfbox{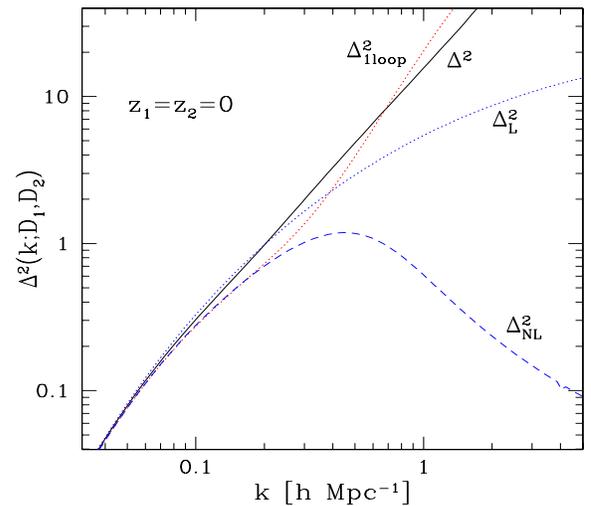}}
\end{center}
\caption{The logarithmic power $\Delta^2(k)$ (solid line) at redshift $z=0$, 
that is, at equal times $z_1=z_2=0$. We also display the linear power $\Delta^2_L$
(dotted line), the usual one-loop perturbative result $\Delta^2_{\rm 1 loop}$ 
of Eq.(\ref{P1loop}) (dotted line) and the exact nonlinear power $\Delta^2_{\rm NL}$
of Eq.(\ref{G11Z5}) (dashed line).}
\label{figlGkz0z0}
\end{figure}

\begin{figure}[htb]
\begin{center}
\epsfxsize=8 cm \epsfysize=7 cm {\epsfbox{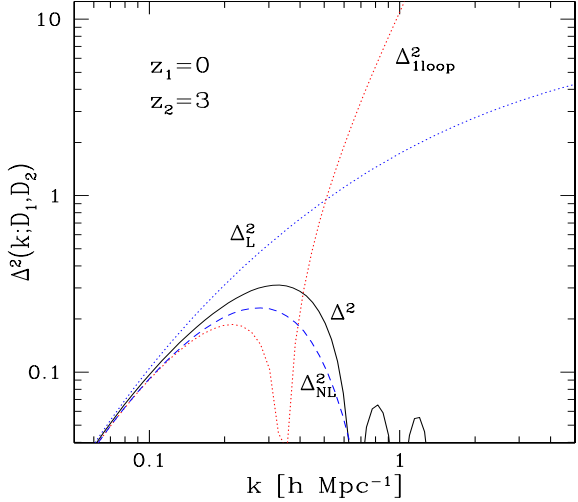}}
\end{center}
\caption{The logarithmic power $\Delta^2(k)$ at unequal times 
$(D_1=1,D_2=0.32)$ (i.e. $z_1=0,z_2=3$).}
\label{figlGkz0z3}
\end{figure}

Finally, we show the logarithmic power $\Delta^2(k;D_1,D_2)$ as a function of 
wavenumber $k$ in Figs.~\ref{figlGkz0z0} and \ref{figlGkz0z3}.
We compare the 2PI effective action prediction at one-loop order 
with the linear power, the exact nonlinear power obtained from Eq.(\ref{G11Z5}),
and the usual one-loop result obtained from standard perturbative analysis.
The latter may also be obtained by expanding Eq.(\ref{G11Z5}) up to order $P_{L0}^2$,
and it reads for the Zeldovich dynamics as,
\beq
P^{\rm 1loop}(k;D_1,D_2) = P_L + P_{22} + P_{13} ,
\label{P1loop}
\eeq
with
\beq
P_L= D_1 D_2 P_{L0}(k),
\label{PL12}
\eeq
\beq
P_{13}= - \frac{D_1^2+D_2^2}{2} D_1 D_2 P_{L0}(k) k^2 \sigma_v^2 ,
\label{P13}
\eeq
\beq
\!\!\!\!P_{22}\!= \!D_1^2 D_2^2 \!\!\int\!\!\d\bw 
\frac{(\bk.\bw)^2 [\bk.(\bk-\bw)]^2}{2w^4|\bk-\bw|^4} P_{L0}(w) P_{L0}(|\bk-\bw|) .
\label{P22}
\eeq
The results match the steepest-descent predictions, as well
as the usual one-loop power (\ref{P1loop}), on large scales.
On small scales, contrary to the usual one-loop power (\ref{P1loop}) and the
steepest-descent predictions, the 2PI effective action methods yields a logarithmic 
power $\Delta^2(k;D_1,D_2)$, which decays for different times $D_1\neq D_2$, see
Fig.~\ref{figlGkz0z3}.
This high-$k$ power-law damping is due to the decay of the response function
already shown in Figs.~\ref{figRt1} and \ref{figRk}, leading to a decorrelation 
on small scales and for large time separations; however, it is only
a power-law decay instead of the exact Gaussian damping seen in Eq.(\ref{G11Z7}).
On the other hand, for equal times $D_1=D_2=D$, we obtain a steady growth of the 
power $\Delta^2(k)$ in between the linear prediction $\Delta_L^2$ and the 
usual one-loop prediction $\Delta^2_{\rm 1 loop}$, see Fig.~\ref{figlGkz0z0}.
This is due to the contributions of nearby times $D_1'\simeq D_2'\simeq D$
in the last term of Eq.(\ref{GPi}), which are not damped because of their small 
time-difference. Note that the cancellation at equal times
of the damping associated with the decay of the response function is qualitatively
correct, as shown from the exact nonlinear solution studied in 
sects.~\ref{Two-point-correlation} and ~\ref{Asymptotic-behavior}, see 
Eq.(\ref{G11Z7}).
Therefore, at one-loop order, the 2PI effective action method shows a significant 
qualitative improvement over the standard perturbative expansions of 
Sect.~\ref{Standard-perturbative-expansions} and the direct steepest-descent 
method. However, it does not manage to predict the high-$k$ smooth power-law 
decay of the equal-time power $\Delta^2(k)$.

\section{Simple nonlinear schemes associated with the 2PI effective action method}
\label{nonlinear-schemes}

\subsection{Response function}
\label{nonlinear-schemes-response}

\begin{figure}[htb]
\begin{center}
\epsfxsize=4.34 cm \epsfysize=4.34 cm {\epsfbox{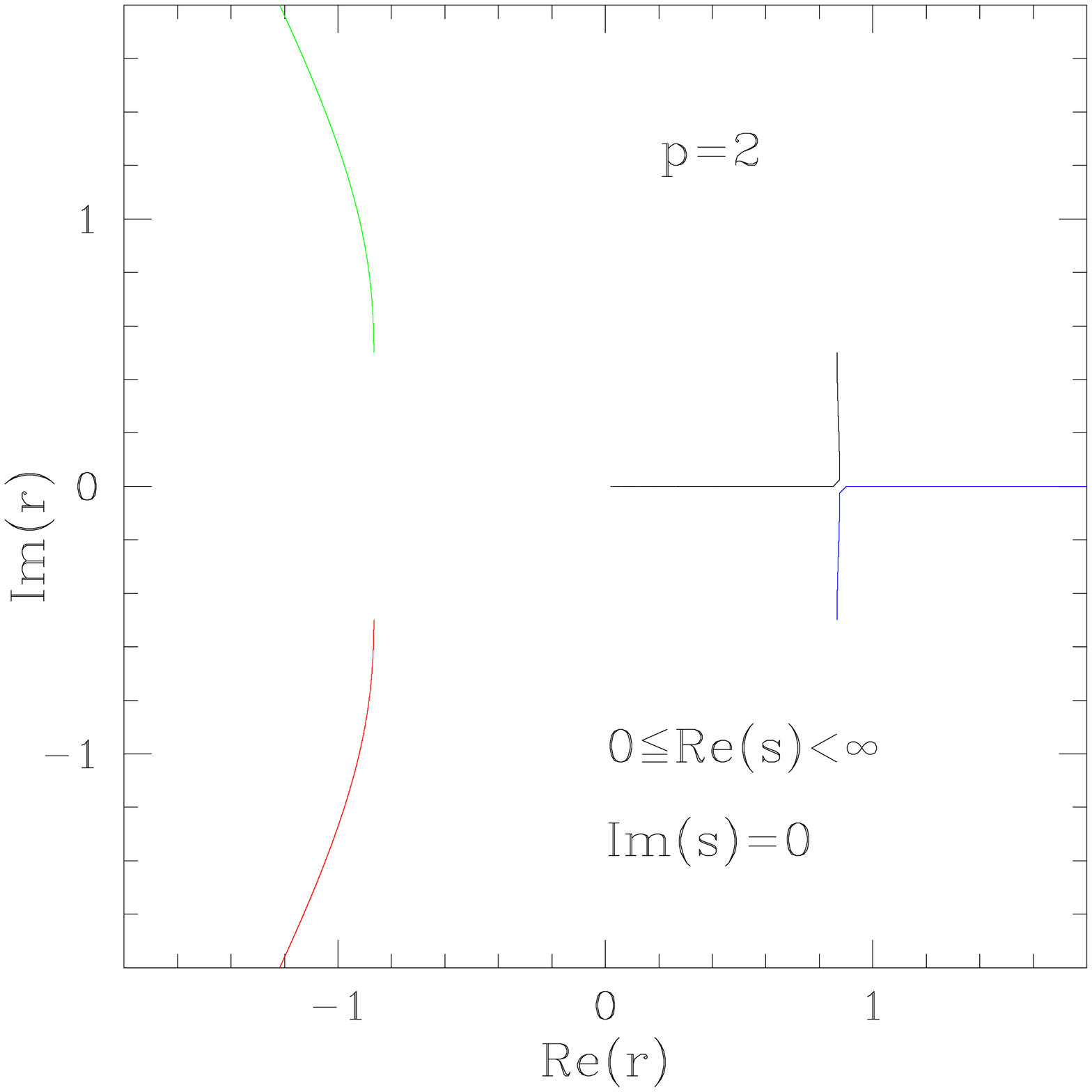}}
\epsfxsize=4.34 cm \epsfysize=4.34 cm {\epsfbox{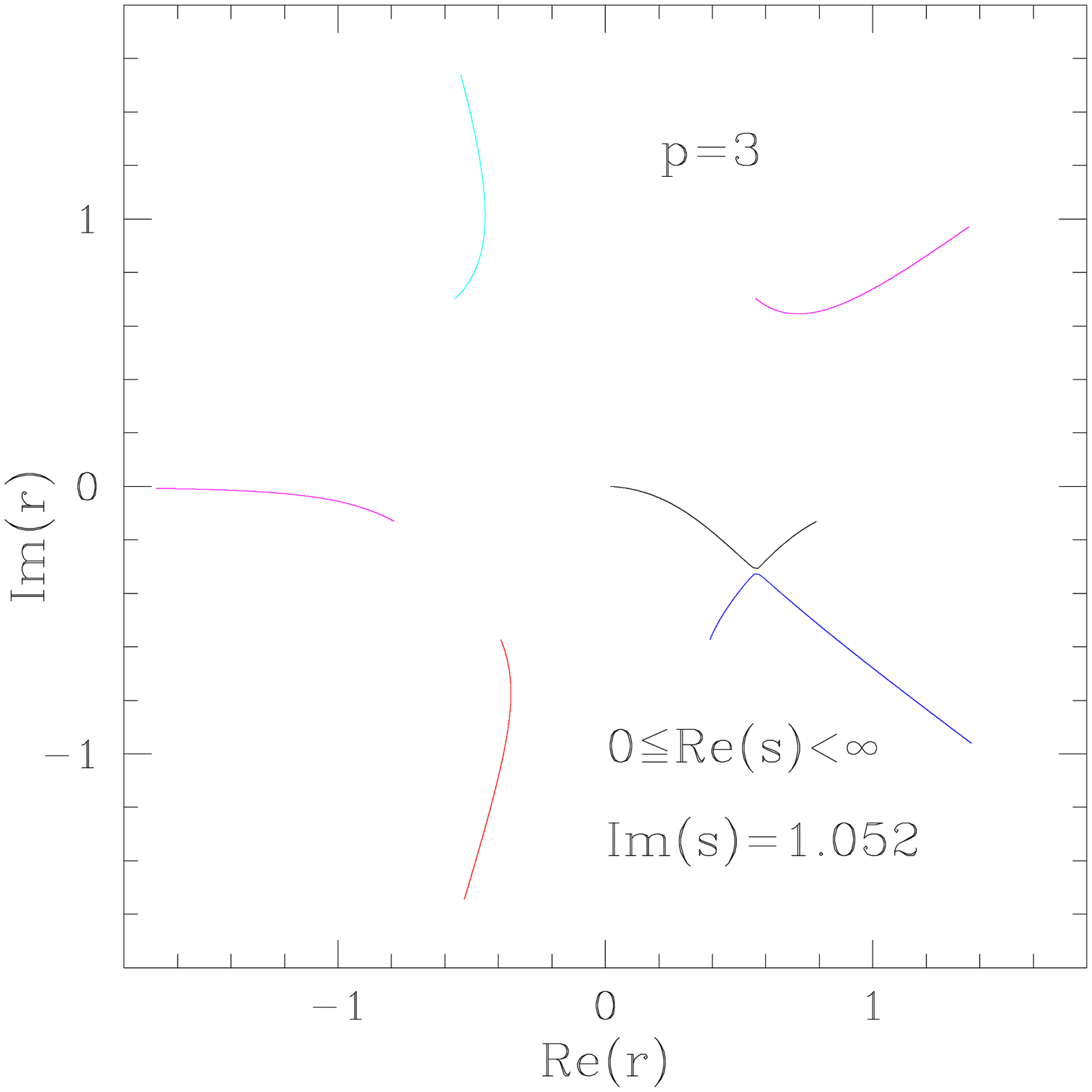}}
\end{center}
\caption{{\it Left panel}: the trajectories of the four roots of Eq.(\ref{rsp2})
as we follow $s$ along the real axis from $s=+\infty$ down to $s=0$.
{\it Right panel}: the trajectories of the six roots of Eq.(\ref{rsp3})
as we follow $s$ along the line ${\rm Im}(s)=1.052$ from ${\rm Re}(s)=+\infty$ 
down to ${\rm Re}(s)=0$.
The root of interest starting from $\tr=0$ ``collides'' with a second root and 
changes direction by $\pi/2$. This corresponds to a singularity for the
implicit function $\tr(s)$.}
\label{figrootsrs}
\end{figure}

For the 2PI effective action method, it is not straightforward to obtain the 
self-energy $\Sigma$ at a given order from the exact expressions (\ref{tsigNL}) or 
(\ref{sigtexp}). Indeed, for the steepest-descent approach investigated in 
Sect.~\ref{Direct-steepest-descent}, the self-energy at a given order simply
corresponds to the truncation of its expansion over powers of $P_{L0}$; therefore,
it could be directly obtained by expanding the exact result. By contrast, within the
2PI effective action scheme, the self-energy is obtained from a diagrammatic 
expansion
in terms of the nonlinear response $R$ and correlation $G$ (defined self-consistently
at this order). Then, the exact expressions of the response $R$ and correlation $G$
are not sufficient to fully define the equations associated with the 2PI effective 
action method at any order, so one must go back to its diagrammatic definition.
To avoid this complication, and to take advantage of the known expressions of the
exact two-point functions, which allowed us to bypass the computation of high-order
diagrams in Sect.~\ref{Direct-steepest-descent}, we investigate here a nonlinear
expansion that is not identical to the 2PI effective action but is expected to
show a similar behavior. Thus, we look for an expansion of the self-energy 
$\Sigma$ over the nonlinear response $R$. A simple way to build such an expansion
is to use the expansions over $1/s$ of the Laplace transforms $\tr(s)$ 
and $\tsig(s)$.
Thus, from Eq.(\ref{trsp}) and the Laplace transform of Eq.(\ref{sigtexp}), we have
\beqa
\tr(s) & = & \frac{1}{s} - \frac{1}{s^3} + \frac{3}{s^5} - \frac{15}{s^7} 
+ \frac{105}{s^9} + .. \label{rexps} \\ 
\tsig(s) & = & \frac{1}{s} - \frac{2}{s^3} + \frac{10}{s^5} - \frac{74}{s^7} 
+ \frac{706}{s^9} + ..  \label{tsigexps}
\eeqa
Then, the series (\ref{rexps}) may be inverted as
\beq
\frac{1}{s} = \tr+\tr^3+3\tr^7-20\tr^9+...
\label{sexpr}
\eeq
Composing this expansion with Eq.(\ref{tsigexps}), we obtain
\beq
\tsig = \tr-\tr^3+4\tr^5-27\tr^7+248\tr^9+...
\label{tsigexpr}
\eeq

This provides an expansion of the self-energy $\Sigma$ in terms of the nonlinear
response $R$. In real $t-$space this yields multiple integrals over $r(t)$,
\beq
\sigma(t) = r(t) - \int_0^t\d t_1\int_0^{t_1} \d t_2 r(t-t_1) r(t_1-t_2) r(t_2) +.. ,
\label{sigt2int}
\eeq
in a fashion similar to what would be obtained for the diagrammatic expansion
associated with the 2PI effective action. The main difference is that the
expansion (\ref{tsigexpr}) only involves the response $R$, whereas the 2PI effective 
action expansion involves both $R$ and $G$, as in Eq.(\ref{S2PI}). Note that this
shows that one can define several nonlinear expansion schemes. That it is
possible to write a simple expansion such as (\ref{tsigexpr}) is due to 
the exact response $R$ only depending on the linear power spectrum through 
the velocity dispersion $\sigma_v^2$. This is not the case for the gravitational 
dynamics where it may not be possible to write an expansion for $\Sigma$ only
in terms of $R$.
Next, truncating the expansion (\ref{tsigexpr}) at a given order and substituting
it into Eq.(\ref{rsig2}), we obtain at order $p=1$:
\beq
\tsig=\tr , \;\;\; \tr^2 + s\tr-1=0 , \;\;\; \tr^{(1)}(s)=\frac{\sqrt{s^2+4}-s}{2} .
\label{rsp1}
\eeq
The root of the polynomial of degree $2p$ which must be chosen, is the one that
is consistent with the expansion (\ref{rexps}) for $s\rightarrow\infty$.
Equation (\ref{rsp1}) is a well-known Laplace transform, and we obtain
\beq
r^{(1)}(t) = \frac{J_1(2t)}{t} .
\label{r1J1}
\eeq
Note that this expression is also obtained as a simple approximation for the
one-loop 2PI effective action approach for the gravitational dynamics in 
Valageas (2007).
We also recover the damped oscillations obtained within the 2PI effective action 
method
displayed in Figs.~\ref{figRt1} and \ref{figRk}. Thus, as expected this nonlinear
expansion and the 2PI effective action expansion show the same behavior at 
order $p=1$. At orders $p=2,3$, we obtain the polynomial equations:
\beqa
p=2 & : & -\tr^4 +\tr^2 + s\tr-1=0  \label{rsp2} \\
p=3 & : & 4\tr^6 - \tr^4 +\tr^2 + s\tr-1=0 . \label{rsp3}
\eeqa

However, we now find that the solutions $\tr^{(p)}(s)$ defined from these
equations have singularities in the right-hand half-plane ${\rm Re}(s)>0$.
For order $p=2$, this may be directly seen from the explicit solution of
Eq.(\ref{rsp2}) or from the behavior of the four roots $\tr_i(s)$ as a function 
of $s$
shown in left panel of Fig.~\ref{figrootsrs}. Indeed, as we follow
$s$ along the real axis from $+\infty$ to $0$, the root of interest
$\tr_1(s)$ that starts from $\tr_1=0$ at $s=+\infty$ ``collides'' with a second
root $\tr_2$ at $\tr_*\simeq 0.87$ ($s_*\simeq 0.94$) and afterwards forms a pair of
complex conjugates with $\tr_2$. This is associated with a square-root 
singularity for
$\tr(s)$. (A simple example is provided by the polynomial $\tr^2-s=0$ with a 
singularity at $\tr_*=0,s_*=0$.) For order $p=3$, the right panel of 
Fig.~\ref{figrootsrs}, where
we follow $s$ along the line ${\rm Im}(s)=1.052$ from ${\rm Re}(s)=+\infty$ down to
${\rm Re}(s)=0$, shows that we again have a singularity at the complex points
$s_*\simeq 1.17\pm 1.05 i, \tr_*\simeq 0.56\mp 0.31 i$. These singularities yield
exponential factors $e^{s_* t}$ for the response $r(t)$, which grow at large $t$.
Therefore, although the expansion (\ref{tsigexpr}) is much better than the 
steepest-descent approach (\ref{rpt}) at order $p=1$, since it exhibits a damping
in the nonlinear regime, it shows as for expansion (\ref{rpt}) growing exponentials
at higher orders. Thus, in this sense this nonlinear expansion is not well-behaved.
We can expect that a similar problem occurs for the 2PI effective action approach
at high orders.

\subsection{Correlation function}
\label{nonlinear-schemes-correlation}

We now investigate nonlinear schemes for the two-point correlation $G$. 
As for the response function, we look for a simple nonlinear expansion
that bypasses the need to compute high-order diagrams but that follows the
structure of the 2PI effective action method. 
This is not as straightforward as for the response $R$ because the two-point
correlation $G$ cannot be written in terms of a one-dimensional function such
as $r(t)$ for the response $R$. Thus, we focus on the equal-time nonlinear 
power for the case of a power-law linear power spectrum and we write
\beq
\Delta^2(k;D)= \Delta_L^2 \, g(t) \;\;\; \mbox{with} \;\;\; 
t= D \sqrt{\Delta_{L0}^2}=\sqrt{\Delta_L^2} ,
\label{gtdef}
\eeq
which defines the time-variable $t$ used in this section and the function $g(t)$.
From Eqs.(\ref{FtF}) and (\ref{Fp}), we have for $n=-2$
\beq
g(t)= \sum_{p=1}^{\infty} \frac{F_p}{p!} \, t^{2p-2} ,
\label{gtseries}
\eeq
which yields for the Laplace transform defined as in Eq.(\ref{Laplace}):
\beq
\tg(s)= \! \sum_{p=1}^{\infty} \frac{F_p}{p!} (2p-2)! \, s^{-2p+1} 
= \frac{1}{s} + \frac{3\pi^2}{32s^3} - \frac{3\pi^2}{4s^5} + ... 
\label{gsseries}
\eeq
This series can be inverted as:
\beq
\frac{1}{s}= \tg - \frac{3\pi^2}{32} \tg^3 + \frac{3(256\pi^2+9\pi^4)}{1024} \tg^5 
+... 
\label{sg}
\eeq
Next, the derivative of $g(t)$ verifies
\beq
g'(t) = \sum_{p=2}^{\infty} \frac{F_p}{p!} \, (2p-2) \, t^{2p-3} ,
\label{gpt}
\eeq
and the Laplace transform of this equation reads
\beqa
s \tg(s) - 1 & = & \sum_{p=2}^{\infty} \frac{F_p}{p!} \, (2p-2)! 
\, s^{-2p+2} \nonumber \\
& = & \frac{3\pi^2}{32 s^2} - \frac{3\pi^2}{4 s^4} - \frac{675\pi^4}{512 s^6} 
+ ... ,
\label{gps}
\eeqa
which could also be obtained from Eq.(\ref{gsseries}).
Then, substituting the series (\ref{sg}) into Eq.(\ref{gps}) gives
\beq
s \tg(s) - 1 = \frac{3\pi^2}{32} \tg^2 -\frac{3\pi^2(128+3\pi^2)}{512} \tg^4 +....
\label{gpg}
\eeq
Thus, as for the response function studied in 
Sect.~\ref{nonlinear-schemes-response},
we have obtained a simple nonlinear expansion scheme for the two-point correlation
$G$ (i.e. for the nonlinear power $\Delta^2(k;D)$). However, since the
functional dependence of the two-point correlation $G$ is not as simple as for
response $R$, the expansion (\ref{gpg}) only applies to the equal-time power, and
it is not built from the self-energy $\Pi$. Nevertheless, going back to real 
$t-$space, Eq.(\ref{gpg}) yields an integro-differential equation for $G$ 
such as Eq.(\ref{Geq}), with a fixed differential term on the left hand side 
and a series of multiple integrals over $g(t)$ on the right hand side. 
At order $p=0$, the right hand side is zero
and we recover the linear solution $g^{(0)}(t)=1$. At order $p=1$, we obtain
\beq
s\tg - 1 = \frac{3\pi^2}{32} \tg^2 , \;\;\; 
\tg^{(1)}(s)= \frac{16}{3\pi^2}\left[s-\sqrt{s^2-3\pi^2/8}\right] ,
\label{gsp1}
\eeq
which gives
\beq
g^{(1)}(t)= \sqrt{\frac{2}{3}} \frac{4}{\pi t} 
I_1\left(\sqrt{\frac{3}{2}} \frac{\pi t}{2}\right) ,
\label{gt1}
\eeq
where $I_1$ is the modified Bessel function of order 1.
Thus, the nonlinear power $\Delta^{2(1)}$ shows an exponential growth in the 
highly nonlinear regime. We can check that at order $p=2$ we again have an
exponential growth (with oscillations since ${\rm Im}(s_*)\neq 0$ where $s_*$
is the location of the singularity of $\tg^{(2)}(s)$).
Therefore, the nonlinear expansion (\ref{gpg})
is not better behaved than the ``linear'' expansions associated with the
steepest-descent approach studied in Sect.~\ref{Correlation-G-and-self-energy-Pi}.

\section{Simple nonlinear schemes associated with the running with the 
high-$k$ cutoff}
\label{nonlinear-schemes-2}

\subsection{Response function}
\label{Response-function-2}

In a fashion similar to Sect.~\ref{nonlinear-schemes}, we now investigate
some nonlinear schemes that may be built in the spirit of the method
outlined in Sect.~\ref{Running}, where we considered the dependence of the
system on a high-$k$ cutoff $\Lambda$. In order to separate the dependence
on $\Lambda$, we now write the response function as
\beq
R= R_L \, r(t,\omega^2) \;\;\;\; \mbox{with} \;\;\;\; t= D_1-D_2 ,
\label{rtLa1}
\eeq
\beq
\omega^2= k^2 \frac{4\pi}{3} \int_0^{\Lambda} \d w P_{L0}(w) .
\label{omegaLa}
\eeq
Thus, the exact response function $r(t,\omega^2)$ is, from Eq.(\ref{RNL}),
\beq
r(t,\omega^2) = e^{-\omega^2 t^2/2} .
\label{rtLa2}
\eeq
Defining the Laplace transform with respect to $t$ as in Eq.(\ref{Laplace}),
we obtain
\beq
\tr(s,\omega^2) = \sum_{p=0}^{\infty} (-1)^p \, (2p-1)!! \, \omega^{2p} 
\, s^{-2p-1} ,
\label{rsLa1}
\eeq
while the derivative with respect to $\Lambda$ is
\beq
\frac{\pl\tr}{\pl\Lambda}(s,\omega^2) = \frac{\d\omega^2}{\d\Lambda} 
\sum_{p=1}^{\infty} (-1)^p \, (2p-1)!! \, p \, \omega^{2(p-1)} \, s^{-2p-1} .
\label{drsLa1}
\eeq
Inverting the series (\ref{rsLa1}) and substituting into Eq.(\ref{drsLa1}) gives
the expansion
\beq
\frac{\pl\tr}{\pl\Lambda} = \frac{\d\omega^2}{\d\Lambda} \left[ -\tr^3 
+ 3 \omega^2 \tr^5 - 18 \omega^4 \tr^7 + ... \right] .
\label{drsrsLa1}
\eeq
At order $p=0$, we have $\pl\tr/\pl\Lambda=0$ and we recover the linear response
(since we impose $r(\Lambda=0)=r_L$). At order $p=1$ we obtain
\beq
\frac{\pl\tr}{\pl\Lambda} = - \frac{\d\omega^2}{\d\Lambda} \, \tr^3 ,
\label{drLa1}
\eeq
which is similar to Eq.(\ref{dRdLambdaRL0}) once we go back to real $t$-space,
which leads to a double integral over time (as in Eq.(\ref{sigt2int})).
We can recover Eq.(\ref{dRdLambdaRL}) by noting that the linear response is
$\tr_L(s)= 1/s$, whence $\tr_L^3=1/2 \d^2\tr_L/\d s^2$. Substituting this result
into Eq.(\ref{drLa1}) gives 
$\pl\tr/\pl\Lambda=-1/2 (\d\omega^2/\d\Lambda)\d^2\tr_L/\d s^2$. The inverse Laplace
transform of this equation gives back Eq.(\ref{dRdLambdaRL}). It is clear that
this procedure is not systematic and only applies to the linear response. This is 
why we could reduce the three response functions on the right hand side of 
Eq.(\ref{dRdLambdaRL0}) to the one response function on the right hand side of 
Eq.(\ref{dRdLambdaRL}). Then, replacing $R_L$ by $R$ in Eq.(\ref{dRdLambdaRL})
is clearly correct at the order $\omega^2$ at which the right hand side of 
Eq.(\ref{dRdLambda1}) was truncated, but this method cannot be extended to higher 
orders in a systematic fashion. Thus, let us consider Eq.(\ref{drLa1}) with
keeping the right hand side as it comes out from the systematic expansion 
obtained in (\ref{drsrsLa1}). Integrating over $\Lambda$ gives
\beq
\tr^{(1)}(s)=\frac{1}{\sqrt{s^2+2\omega^2}} \;\; \mbox{whence} \;\; 
r^{(1)}(t)= J_0(\sqrt{2}\omega t) .
\label{rLap1}
\eeq
Thus, we obtain decaying oscillations into the nonlinear regime, which are
damped as $[k(D_1-D_2)]^{-1/2}$. Note that the damping is smaller than for
the 2PI effective action method where the simplified expansion (\ref{tsigexpr})
gave Eq.(\ref{r1J1}), which decays as $[k(D_1-D_2)]^{-3/2}$. 
At next order $p=2$ we obtain
\beq
\frac{\pl\tr}{\pl\Lambda} \! = \! \frac{\d\omega^2}{\d\Lambda} \! [ -\tr^3 
+ 3 \omega^2 \tr^5 ] \;\; \mbox{hence} \;\; \frac{\pl\tr}{\pl\omega^2} 
= - \tr^3 + 3 \omega^2 \tr^5 .
\label{drp2La1}
\eeq
Then, looking for a solution of the form
\beq
r(t,\omega^2) = \rho(\omega t) , \;\;\; \tr(s,\omega^2)= \frac{1}{\omega} 
\trho(y) \;\; \mbox{with} \;\; y=\frac{s}{\omega} ,
\label{trtrho}
\eeq
we obtain for the Laplace transform $\trho(y)$ the equation
\beq
\trho+y\trho'=2\trho^3-6\trho^5 .
\label{trhoy}
\eeq
Note that Eq.(\ref{trhoy}) no longer involves an explicit dependence on $\omega$.
It can be solved in implicit form as
\beq
y= \frac{1}{\trho} (1-2\trho^2+6\trho^4)^{1/4} e^{\left[\arctan\sqrt{5}
-\arctan\left(\frac{\sqrt{5}}{1-6\trho^2}\right)\right]/(2\sqrt{5})} .
\label{ytrho}
\eeq
Going back to $s=\omega y$, we see that $\tr(s,\omega^2)$ is singular at the point
\beq
s_* = \omega y_* \;\;\; \mbox{with} \;\;\;
y_*=6^{1/4} e^{\frac{1}{2\sqrt{5}}\arctan\sqrt{5}} , \;\;\; s_*>0 ,
\label{ssingLa}
\eeq
where $|\tr|=\infty$. Therefore, the response function obtained at order $p=2$
grows into the nonlinear regime as $r^{(2)} \sim e^{s_*t}$, which gives
$R^{(2)} \sim e^{y_* k \sigma_v (D_1-D_2)}$. Thus, as for the nonlinear scheme
of Sect.~\ref{nonlinear-schemes-response} associated with the 2PI effective action
method, we find that at high orders, the expansion (\ref{drsrsLa1}) gives rise
to response functions that exhibit an exponential growth into the nonlinear
regime, even though at lowest-order $p=1$, it managed to provide a nonlinear
damping. Therefore, that the Gaussian decay could be recovered 
from Eq.(\ref{dRdLambdaRL})
is not due to the good convergence properties of the method outlined in 
Sect.~\ref{Running}. As discussed in Sect.~\ref{Running}, from Eq.(\ref{dRdalpha})
it merely comes from the successive steps that have been performed using the 
properties of the linear response until one gets the linear Eq.(\ref{dRdLambdaRL}), 
which is correct at lowest order and which has the desired solution.
However, these intermediate steps cannot be directly 
extended to high orders without ambiguities, and the systematic nonlinear expansion
(\ref{drsrsLa1}) does not recover this damping at high orders.

\subsection{Correlation function}
\label{Correlation-function}

For the correlation function $G$, following the procedure of 
Sects.~\ref{nonlinear-schemes-correlation} and \ref{Response-function-2},
we write
\beq
\Delta^2(k;D)= \Delta_L^2 \, g(t,\Delta_{L0}^2) \;\;\;\; \mbox{with} \;\;\;\;
t= D .
\label{gtdefLa}
\eeq
Then, again introducing the Laplace transform with respect to $t$, inverting the
series $\tg(s,\Delta_{L0}^2)=1/s+..$ and substituting into $\pl\tg/\pl\Lambda$
gives the expansion:
\beq
\frac{\pl\tg}{\pl\Lambda} = \frac{\d\Delta_{L0}^2}{\d\Lambda} \left[ 
\frac{3\pi^2}{32} \tg^3 - \frac{3\pi^2(512+9\pi^2)}{1024} \Delta_{L0}^2 \tg^5 +....
\right]
\label{gpgLa}
\eeq
At order $p=0$ we have the linear correlation $\pl\tg/\pl\Lambda=0$, $\tg=1/s$ and
$g(t)=1$. At order $p=1$ we obtain
\beq
\frac{\pl\tg}{\pl\Delta_{L0}^2}=\frac{3\pi^2}{32} \tg^3 , \;\;\;
\tg^{(1)}(s)= \frac{1}{\sqrt{s^2-3\pi^2\Delta_{L0}^2/16}} ,
\label{tg1La}
\eeq
which gives
\beq
g^{(1)}(t,\Delta_{L0}^2)= I_0\left(\frac{\sqrt{3}\pi}{4}\Delta_{L0}\,t\right) ,
\label{gtp1La}
\eeq
where $I_0$ is the modified Bessel function of order 0.
Thus, as for the simple nonlinear scheme associated with the 2PI effective action 
method of Sect.~\ref{nonlinear-schemes-correlation}, we obtain at order $p=1$
a nonlinear equal-time power $\Delta^2$ which shows an exponential growth in the
nonlinear regime. Following the method leading to Eq.(\ref{ytrho}), we can
actually integrate the nonlinear equation (\ref{gpgLa}) at any order, using
the scaling $\tg(s,\Delta_{L0}^2)=\tilde{\gamma}(s/\Delta_{L0})/\Delta_{L0}$,
which transforms Eq.(\ref{gpgLa}) into an implicit equation giving $y=s/\Delta_{L0}$
as the integral of a rational function of $\tilde{\gamma}$.
We can check that, at order $p=2$, we again have an exponential growth into the
nonlinear regime (with oscillations because ${\rm Im}(y_*)\neq 0$).

\section{Weakly nonlinear scales}
\label{Weakly-nonlinear-scales}

In the previous sections, we have investigated the convergence properties
of several expansion schemes that may be used to study gravitational clustering
in the expanding Universe, applied to the case of the Zeldovich dynamics
where exact results can be obtained. In this section, we complete this study
by a brief description of the results obtained at one-loop order on weakly
nonlinear scales. Indeed, an accurate prediction for the matter power spectrum
on these scales is of great practical interest for several cosmological probes,
such as baryonic acoustic oscillations (Eisenstein et al. 1998, 2005) and 
weak-lensing shear (Munshi et al. 2007).
For the exact gravitational dynamics, the various expansion schemes must be
compared with numerical simulations, which is not very convenient to evaluate
their power. Therefore, it is interesting to check the accuracy and the behavior 
of these expansion methods against the Zeldovich dynamics.

\subsection{Linear expansion schemes}
\label{Linear-expansion-schemes-2}

\begin{figure}[htb]
\begin{center}
\epsfxsize=8 cm \epsfysize=7 cm {\epsfbox{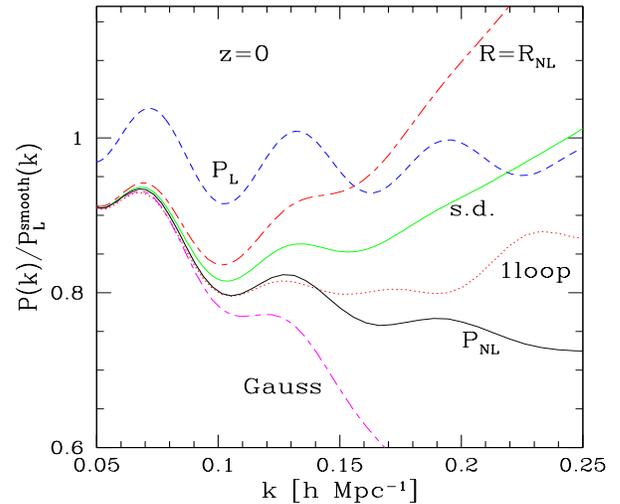}}
\end{center}
\caption{The power spectrum $P(k)$ divided by a smooth linear power 
$P_L^{\rm smooth}$ at redshift $z=0$. We display the linear power $P_L(k)$
(dashed line), the exact nonlinear power $P_{\rm NL}$ of Eq.(\ref{G11Z5}),
the standard 1-loop result (dotted line) of Eq.(\ref{P1loop}), and the 
steepest-descent result of Eq.(\ref{GPi})
(upper solid line ``s.d.''). We also show the results obtained by
adding a Gaussian factor to the standard perturbative result as in Eq.(\ref{Dp2})
(lower dot-dashed line) or by using the exact nonlinear response $R_{\rm NL}$
in Eq.(\ref{GPi}) within the steepest-descent scheme (upper dashed line), as in 
Fig.~\ref{figlG_Gauss}.}
\label{figGlink_z0}
\end{figure}

\begin{figure}[htb]
\begin{center}
\epsfxsize=8 cm \epsfysize=7 cm {\epsfbox{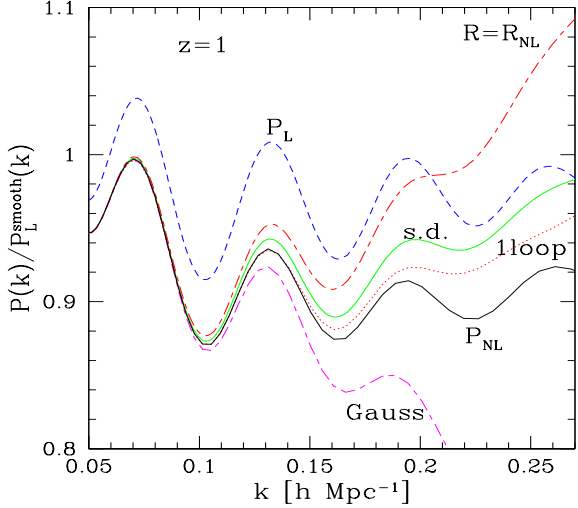}}
\end{center}
\caption{The power spectrum $P(k)$ divided by a smooth linear power 
$P_L^{\rm smooth}$ as in Fig.~\ref{figGlink_z0}, but at redshift $z=1$.}
\label{figGlink_z1}
\end{figure}

We first consider ``linear'' expansion schemes, that is, methods that give rise
to expansions in terms of the linear power spectrum, such as the standard
perturbative expansions of Sect.~\ref{Standard-perturbative-expansions} and
the direct steepest-descent methods of Sect.~\ref{Direct-steepest-descent}.
We focus on the equal-time power spectrum for the $\Lambda$CDM universe described
in the first paragraph of Sect.~\ref{Standard-perturbative-expansions}.
We used the CAMB Boltzmann code (Lewis et al. 2000) to obtain the linear 
power spectrum with the baryonic acoustic oscillations. In order to magnify the 
difference between various schemes, we show in 
Figs.~\ref{figGlink_z0} and \ref{figGlink_z1} the nonlinear 
power divided by the linear power $P_L^{\rm smooth}$ associated to
a smooth power spectrum without baryonic oscillations, taken from 
Eisenstein \& Hu (1998).
We compared the results of various expansion schemes with the exact nonlinear power
(solid line labeled $P_{\rm NL}$) obtained from the numerical integration
of Eq.(\ref{G11Z5}) for the $\Lambda$CDM power spectrum.
First, we see in Fig.~\ref{figGlink_z0} that all schemes agree with the exact 
power up to $k\simeq 0.1 h$ Mpc$^{-1}$ at redshift $z=0$ and follow the 
departure from the linear power $P_L$. On smaller scales, the various expansion
schemes deviate from one another and from the exact nonlinear power. It is
interesting to note that using the exact nonlinear response (\ref{RNL})
within the steepest-descent scheme does not improve the
agreement over the original steepest-descent scheme where
we use the response function predicted at the same order.
On the other hand, it appears that the standard one-loop expansion provides the 
best results at this order. 

Following Eq.(\ref{Dp2}) and Fig.~\ref{figlG_P0exp}, we also
consider the expansion defined from the standard perturbative series by factorizing
a Gaussian decay:
\beqa
P^{\rm 1loop}_{\rm Gauss.}(k) & = & e^{-D^2\omega^2} \left[ P_L + P_{22} + P_{13}
+ D^2 \omega^2 P_L \right] \nonumber \\
& = & e^{-D^2\omega^2} (P_L + P_{22}) .
\label{P1loopGauss}
\eeqa
In agreement with Sect.~\ref{Standard-perturbative-expansions} and 
Crocce \& Scoccimarro (2006a), this gives a positive power whatever 
the shape of the linear power spectrum; however, Fig.~\ref{figGlink_z0} shows that
this does not necessarily improve the accuracy as compared with the usual one-loop 
result (\ref{P1loop}). We show the power spectrum obtained at redshift $z=1$
in Fig.~\ref{figGlink_z1}. Of course the various expansion schemes agree more closely
on the same scales with the exact result, since we are closer to the linear regime.
We can see that we recover the same behaviors as in Fig.~\ref{figGlink_z0}.

The behavior of these various expansion schemes at higher orders was investigated
in sects.~\ref{Standard-perturbative-expansions}-\ref{Direct-steepest-descent}.
We found that no linear scheme provides a significant improvement over the 
standard perturbative expansion. Factorizing a Gaussian decaying term as in 
Eq.(\ref{P1loopGauss}) seemed to give a small improvement in Fig.~\ref{figlG_P0exp},
but as discussed below Fig.~\ref{figlG_P0exp}, this is not very robust and would not
apply to any power spectrum. On the other hand, Fig.~\ref{figlG_Gauss} shows that
even using the exact nonlinear response within the direct steepest-descent scheme
does not generically provide a significant improvement either.
Nevertheless, the most promising scheme within this framework is probably
to combine a good ansatz for the response function with such expansions
(see also Sect.~\ref{nonlinear-expansion-schemes-2} below).
Then, one could hope to gain from the higher orders, while the imposed decay of the
response function could tame the increasingly fast growth at higher orders
obtained in the standard perturbative expansion.
An application of such a strategy was presented in Crocce \& Scoccimarro (2007)
for a $\Lambda$CDM universe, and a significant improvement over the standard
perturbative expansion was obtained up to two-loop order.

\subsection{Nonlinear expansion schemes}
\label{nonlinear-expansion-schemes-2}

\begin{figure}[htb]
\begin{center}
\epsfxsize=8 cm \epsfysize=7 cm {\epsfbox{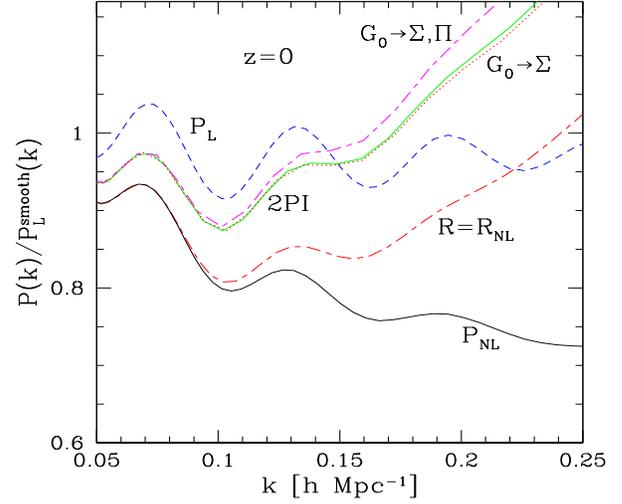}}
\end{center}
\caption{The power spectrum $P(k)$ divided by a smooth linear power 
$P_L^{\rm smooth}$ at redshift $z=0$. We display the linear power $P_L(k)$
(dashed line), the exact nonlinear power $P_{\rm NL}$ of Eq.(\ref{G11Z5}),
and the 2PI effective action result of Eq.(\ref{GPi}) (upper solid line ``2PI''). 
We also show the results obtained by using the exact nonlinear response 
$R_{\rm NL}$ in Eq.(\ref{GPi}) within the 2PI scheme (lower dashed line), or by 
using the linear two-point correlation to compute both $\Sigma$ and $\Pi$ 
(upper dot-dashed line) or only $\Sigma$ (dotted line).}
\label{figGnonlink_z0}
\end{figure}

\begin{figure}[htb]
\begin{center}
\epsfxsize=8 cm \epsfysize=7 cm {\epsfbox{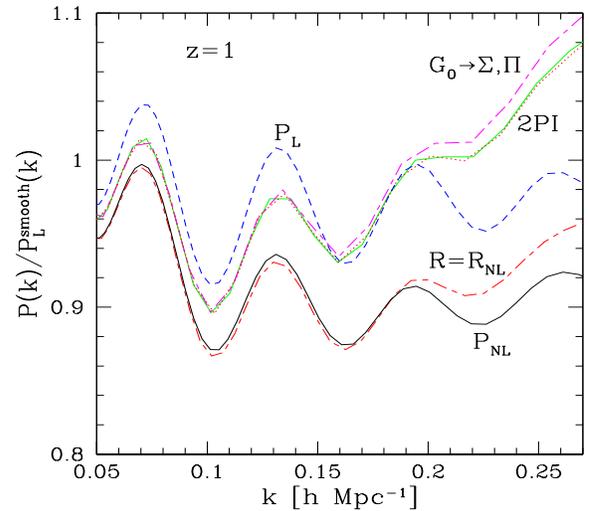}}
\end{center}
\caption{The power spectrum $P(k)$ divided by a smooth linear power 
$P_L^{\rm smooth}$ as in Fig.~\ref{figGnonlink_z0}, but at redshift $z=1$.}
\label{figGnonlink_z1}
\end{figure}

Finally, we study in this section ``nonlinear'' expansion schemes, that is, methods
that give rise to expansions in terms of the nonlinear two-point functions $R$ and
$G$, such as the 2PI effective action method of Sects.~\ref{2PI-effective-action}
and \ref{2PI-effective-action-method2}.
We show our results for the power spectrum (divided again by $P_L^{\rm smooth}$) 
at redshifts $z=0,1$ in Figs.~\ref{figGnonlink_z0} and \ref{figGnonlink_z1}.
First, we note that the 2PI effective action result at one-loop order overestimates
the power spectrum on weakly nonlinear scales. Both the
standard one-loop result and the direct steepest-descent method actually work
better in this range. This behavior can be traced back to the damping self-energy
$\Sigma$. Indeed, since $\Sigma \propto RG$ at one-loop order (see Eq.(\ref{S2PI})), 
the decay of the response and
of the two-point correlation at high $k$ (shown in Figs.~\ref{figRk} and 
\ref{figlGkz0z3})) leads to a smaller $\Sigma(k)$ at high $k$ as compared with the
$\Sigma_0$ obtained for the direct steepest-descent method. This in turns yields
a response $R$ that is somewhat larger than for the steepest-descent method
in the weakly nonlinear regime (but smaller at high $k$ where it decays) whence
a two-point correlation $G$ that is somewhat larger from Eq.(\ref{GPi}). Of course,
this slight overestimate of $R$ and underestimate of $\Sigma$ will be corrected by 
higher-order terms. For the Zeldovich dynamics, we know that $R$ and
$\Sigma$ actually depend only on $\sigma_v$, so that terms that depend on other
integrals over $P_{L0}(k)$ must cancel out in the full resummation. However, this
discrepancy makes the one-loop 2PI effective action result insufficient for
practical purposes (at least at the one-loop order). 

We also display the results obtained when both the self-energies $\Sigma$ and $\Pi$
are obtained from the linear correlation $G_0$ (so that
only the response $R$ coupled to $\Sigma$ is obtained from nonlinear equations)
and when only the self-energy $\Sigma$ is obtained from $G_0$ (so that 
the nonlinear systems for the pairs $\{\Sigma,R\}$ and $\{\Pi,G\}$ are
decoupled). We see that these two methods give results very close to the
original 2PI effective action prediction where all two-point functions
$\{\Sigma,\Pi,R,G\}$ are coupled. 

On the other hand, we also show the power spectrum
obtained from the coupled equations (\ref{P2PI}) and (\ref{GPi}), when we use
the exact nonlinear response (\ref{RNL}). We can see that this significantly
improves the
agreement with the exact power spectrum, and the result is slightly better than
the steepest-descent prediction shown in Fig.~\ref{figGlink_z0}, but it is still
a bit less accurate than the standard perturbative result at $z=0$.
However, Fig.~\ref{figGnonlink_z1} shows that at $z=1$ the power obtained in
this fashion is slightly more accurate than the standard perturbative result.
Therefore, contrary to the case of the steepest-descent method studied in 
Sect.~\ref{Linear-expansion-schemes-2}, using the exact nonlinear response
improves the prediction for the two-point correlation and provides a scheme
that can be competitive with the usual perturbative expansion on weakly nonlinear
scales.

At higher orders, the nonlinear 2PI expansion scheme -- especially if it uses the
exact nonlinear response or a reliable ansatz (as for the lower dashed lines of
Figs.~\ref{figGlink_z0}-\ref{figGnonlink_z1}) -- may provide an even greater
improvement as compared with the standard perturbative expansion. The analysis
of Sect.~\ref{nonlinear-schemes}-\ref{nonlinear-schemes-2} showed that nonlinear
schemes can give unwanted exponential growths at higher orders for both the
response function and the power spectrum. However, this does not rule out a good
convergence on weakly nonlinear scales. Nevertheless, this behavior and
Figs.~\ref{figGlink_z0}-\ref{figGnonlink_z1} suggest that the most promising scheme
would be to combine a good ansatz for the response with such nonlinear expansions.
As for the linear schemes of Sect.~\ref{Linear-expansion-schemes-2},
one could hope to benefit from the higher orders, while the imposed decay of the
response function could partly restrain their increasingly fast growth. This
could make the improvement over the standard perturbative expansion even more
significant at higher orders, but such a study is left for future work.

\section{Conclusion}
\label{Conclusion}

In this article we have applied to the Zeldovich dynamics various expansion 
schemes that may also be used for the gravitational dynamics in the expanding 
Universe. We derived the path-integral formalism that describes this system,
starting either from the differential or the integral form of the equations of 
motion, and we obtained the relationship between the associated response functions.
These response functions describe the response of the system to a small perturbation
applied at any time and they also encode the memory of initial conditions.
Next, we briefly described how to build various expansion schemes from these
path integrals, such as large-$N$ expansions or running with a high-$k$ cutoff.
All these results apply almost identically to the case of the gravitational 
dynamics. 

Then, we have derived the exact nonlinear two-point functions associated
with the Zeldovich dynamics, taking advantage of the well-known exact solution
of the equations of motion. Whereas the equal-time nonlinear power decays as a 
power law in the highly nonlinear regime, different-time two-point functions,
such as the response $R(k;t_1,t_2)$ or the correlation $G(k;t_1,t_2)$, show a
Gaussian decay $e^{-k^2\sigma_v^2(D_1-D_2)^2/2}$, where $\sigma_v$
is the variance of the linear velocity. 
This damping is associated with the uniform random displacement of particles between
different times by long wavelength modes. This leads to an effective decorrelation
but it is not directly related to the matter clustering. 
In particular, $\sigma_v^2$ can be made very large, and even infinite,
through low-$k$ divergences, without affecting the equal-time matter power 
spectrum. This also means that
the matter power spectrum $P(k;t)$ may still be very close to linear 
on scale $k$ even though two-point functions such as $R(k;t_1,t_2)$ may have already
shown large deviations from their linear values on the same scale at previous times
$t_2<t_1<t$. Therefore, departures from linearity of different two-point functions
are not necessarily related.

Next, we have studied the standard perturbative expansions for both the response $R$
and the logarithmic power $\Delta^2(k;t)\propto k^3 P(k;t)$, focussing on a 
power-law linear power spectrum $n=-2$ for the latter. The usual perturbative 
expansion being equivalent to a Taylor expansion it cannot capture the 
different-time Gaussian decay of $R$ and $\Delta^2$, nor the equal-time power-law 
decay of $\Delta^2$, since it gives polynomial approximations of increasing order. 
In the spirit of Crocce \& Scoccimarro (2006a),
we noticed that for the power $\Delta^2$, reorganizing the perturbative series
by factorizing a simple Gaussian term $e^{-D^2k^2\sigma_v^2}$ gives an expansion
which looks better behaved, as all terms become positive and there is a Gaussian
damping. However, this procedure only works for linear power-spectra such that the
scales which govern $\sigma_v^2$ and $\Delta^2$ are close (as for typical 
$\Lambda$CDM power-spectra).

Then, we have studied the steepest-descent method derived from a large-$N$ 
expansion. At first order $p=1$, there is some improvement for the 
response function, which remains bounded (as a cosine) instead of growing as $k^2$. 
(The same behavior is obtained for the gravitational dynamics (Valageas 2007).) 
However, at higher orders the expansion worsens as it exhibits an exponential 
growth in the nonlinear regime. We showed that this could be cured by using 
Pad\'{e} approximants, which remain bounded at all orders as a sum of cosines 
(they do not decay but after integration the oscillations should produce some 
effective damping). 
The power $\Delta^2$ displays an exponential growth at orders $p\geq 2$
for the plain steepest-descent method and a polynomial growth using the
Pad\'{e} approximants. As for the standard perturbative expansion, one can
reorganize the series by factorizing a Gaussian term 
$e^{-(D_1^2+D_2^2)k^2\sigma_v^2/2}$ in the self-energy $\Pi$ that describes the
generation of power by nonlinear interactions. This appears to give slightly
better results than with the standard perturbative expansion, but this procedure
obviously suffers from the same restrictions. Alternatively, one can factorize
a Gaussian term such as $e^{-(D_1^2-D_2^2)k^2\sigma_v^2/2}$ into the self-energy
$\Pi$, to reproduce the fact that the Gaussian damping must disappear for 
equal-time statistics. Then, one again obtains a polynomial growth into the
highly nonlinear regime for the power $\Delta^2(k;t)$, because of the contribution
of mode couplings at recent times $t_1\simeq t_2\simeq t$. Therefore, none of
these methods shows very satisfactory global convergence properties.

Next, we have discussed a high-$k$ resummation proposed by 
Crocce \& Scoccimarro (2006b) to improve the behavior of such expansion schemes. 
We showed that the partial resummation involved in this procedure is equivalent 
to approximating the nonlinear equation of motion by a linear equation. Using 
the high-$k$ asymptotic of the coupling kernel, as in 
Crocce \& Scoccimarro (2006b), we derived the explicit expression of the nonlinear 
density field $\delta(\bx,t)$ associated with these approximations. Then,
both the nonlinear response $R$ and power $\Delta^2$ are equal to their linear
counterpart multiplied by the same different-time Gaussian decay. 
Thus, these approximations manage to capture the different-time Gaussian decay 
(associated with the advection by long wavelengths) but they fail to capture the 
equal-time
properties of the system. A close analysis of this procedure shows that the
underlying assumptions are not valid because one cannot define a high-$k$ limit
in a simple manner. (For a linear power spectrum with $-3<n<-1$ at high $k$ 
the nonlinear power at wavenumber $k$ is
generated by the highly nonlinear couplings of modes $k'\gg k$ instead of being
produced by small wavenumbers restricted to some finite range $k'<\Lambda$.) 
The same caveats should apply to the gravitational dynamics, although it is not
totally obvious whether the Gaussian decay at different times is exact in this case
(but this is not necessarily important for practical purposes).

Then, we have turned to nonlinear schemes, that is, expansions over powers of 
nonlinear two-point functions, such as the 2PI effective action method built 
from a large$-N$ expansion. At one-loop order ($p=1$), we obtain damped 
oscillations (with a power-law decay) for the two-time response function and 
power $\Delta^2(k;t_1,t_2)$, but the equal-time power still grows on small scales. 
(One obtains the same behavior for the gravitational dynamics (Valageas 2007)). 
Then, from the exact two-point functions we have built
simple nonlinear expansion schemes that are similar to the 2PI effective action
expansion, but that can be more easily handled analytically. We recover damped 
oscillations for the response function at order $p=1$, but we find an exponential 
growth at higher orders; the equal-time power also shows an exponential growth.
Next, we investigated simple nonlinear expansion schemes associated with the
evolution of the system with a high-$k$ cutoff $\Lambda$. We find a similar behavior
as we again obtain damped oscillations at order $p=1$ and exponential growth at 
higher orders for the response function; the equal-time power also exhibits
an exponential growth. To recover the Gaussian decay from this expansion at 
order $p=1$, one must introduce some further approximations, as in 
Matarrese \& Pietroni (2007a,b), which are correct at this order but which do 
not give a systematic procedure. Therefore, it appears that at high orders 
$p\geq 2$ nonlinear methods do not fare much better than linear schemes.

Finally, we have studied the quantitative predictions of these various schemes
at one-loop order on weakly nonlinear scales for the equal-time matter 
power spectrum. All linear expansions agree with the exact results on 
quasi-linear scales ($k<0.1 h$ Mpc$^{-1}$ at $z=0$) where there is already 
a deviation of $10\%$ from linear theory.  
On smaller scales they depart from each other and the standard
perturbation theory actually works best for the case studied in this article.
Moreover, factorizing a Gaussian damping factor or using the exact response
function does not improve the predictions obtained within this framework.
On the other hand, the nonlinear 2PI effective action method overestimates 
the nonlinear power spectrum on these scales because of the nonlinear feedback 
involved by the coupled system obeyed by the response $R$. However, using the 
exact nonlinear response within this method now improves
the agreement with the exact result and can be competitive with the standard
perturbation expansion (but this depends on the exact shape of the linear
power spectrum as in our case it is slightly better at $z=1$ but slightly 
worse at $z=0$). 

Since the equations of motion associated with the gravitational and the Zeldovich
dynamics are very close we can expect these results to apply to the gravitational
case. (This is the case at one-loop order as shown by the comparison with 
Valageas (2007).) We have found that none of these schemes shows good global 
convergence properties at high orders. 
Indeed, they all lead to polynomial or exponential
growth into the nonlinear regime for the response function, except for the
use of Pad\'{e} approximants which gives bounded response functions
with fast oscillations in the highly nonlinear regime. On the other hand, nonlinear
schemes manage to reproduce the damping at one-loop order $p=1$ but fail at higher 
orders. Next, no scheme manages to recover the power-law damping of the nonlinear
matter power spectrum. They either display increasing growth at higher order or
a Gaussian decay which is also somewhat artificial. 

Nevertheless, an expansion
may still be very useful on weakly nonlinear scales even if it converges badly (or
even diverges) on highly nonlinear scales. There, we found that the best methods
seem to be the standard perturbation theory or a nonlinear expansion where one
uses the exact nonlinear response (with the Gaussian decay).
These results are somewhat disappointing, since it appears to be difficult
to build systematic expansion schemes that significantly improve over the
standard expansion. One may still obtain some improvement as in 
Crocce \& Scoccimarro (2007) or Matarrese \& Pietroni (2007a,b), but this requires
some additional ingredients, such as the use of an ansatz, that shows a Gaussian decay,
for the response function and in some cases for other two-point functions as well. 
On the other hand, the use of several expansion schemes can be of interest by 
itself, since they should be accurate at least over the range where they all agree.
This allows one to obtain an estimate of their range of validity without the need
to perform numerical simulations.

In order to make progress, it appears that it may be advantageous for 
observational purposes to be guided by the expected behavior of two-point
functions and to combine such systematic expansions with reasonable ansatze
(e.g. Crocce \& Scoccimarro 2007). From a theoretical perspective, one may also
look for different approaches. For instance, one could try to work directly
with the Vlasov equation (Valageas 2004). However, this would make the
computations significantly more difficult, and it is not clear whether it is not
more efficient to stick to the hydrodynamical approach and to simply compute
higher order terms, especially if one is mostly interested in weakly nonlinear 
scales. On the other hand, one may consider simpler effective dynamics that attempt
to go beyond shell crossing (based for instance on a Schroedinger equation,
Widrow \& Kaiser 1993).

\end{document}